\documentclass[12pt,preprint]{aastex}
\usepackage{amsmath}
\usepackage{amssymb}

\def\phantfrac{\vphantom{\frac{1}{2}}}
\def\underorder#1#2{\underset{\vphantom{\bigl(} #1}{\phantfrac #2}}

\def\pseudofigureone#1#2#3{
\begin{figure}
\plotone{#2}
\caption{#3
\label{#1}}
\end{figure}}

\def\pseudofiguretwot#1#2#3#4{
\begin{figure}
\plottwo{#2}{#3}
\caption{#4
\label{#1}}
\end{figure}}

\def\pseudofiguretwob#1#2#3#4{
\begin{figure}
\plottwo{#2}{#3}
\caption{#4
\label{#1}}
\end{figure}}

\def\pseudotable#1#2#3#4{
\begin{deluxetable}{#3}
\tablewidth{0pt}
\tablecaption{#2\label{#1}}
\startdata
#4
\enddata
\end{deluxetable}
}

\usepackage[numberedappendix]{emulateapj5}

\usepackage{apjfonts}

\journalinfo{The Astrophysical Journal, submitted
{\rm [e-print~{\tt astro-ph/0511812}]}}

\def\pseudofigureone#1#2#3{{
\refstepcounter{figure}
\label{#1}
\vskip5mm
\plotone{#2}
\footnotesize\def\baselinestretch{1.0}
\begin{minipage}{\columnwidth}
{\scshape ~~Fig.}\space\thefigure.--- #3
\end{minipage}
\vskip5mm
}}

\def\pseudofiguretwot#1#2#3#4{{
\begin{figure*}[t!]
\epsscale{2.2}
\plottwo{#2}{#3}
\epsscale{1.0}
\caption{#4
\label{#1}}
\end{figure*}
}}

\def\pseudofiguretwob#1#2#3#4{{
\begin{figure*}[b!]
\epsscale{2.2}
\plottwo{#2}{#3}
\epsscale{1.0}
\caption{#4
\label{#1}}
\end{figure*}
}}

\def\pseudotable#1#2#3#4{{
\begin{table*}[t]
\centering
\caption{#2}
\label{#1}
\begin{tabular}{#3}
#4
\end{tabular}
\end{table*}
}}

\renewcommand{\eqref}[1]{equation~(\ref{#1})}
\newcommand{\eqsref}[2]{equations~(\ref{#1}) and (\ref{#2})}
\newcommand{\eqsmoreref}[2]{equations~(\ref{#1})--(\ref{#2})}
\newcommand{\Order}{O}
\newcommand{\appref}[1]{Appendix~\ref{#1}}
\newcommand{\figref}[1]{Figure~\ref{#1}}
\newcommand{\tabref}[1]{Table~\ref{#1}}

\newcommand{\secref}[1]{\S\ref{#1}}
\newcommand{\appsref}[2]{Appendices~(\ref{#1}) and (\ref{#2})}
\newcommand{\E}{E} 
 
\newcommand{\zhat}{\mbox{$\hat{\mathbf{z}}$}} 
\newcommand{\xhat}{\mbox{$\hat{\mathbf{x}}$}} 
\newcommand{\yhat}{\mbox{$\hat{\mathbf{y}}$}} 
 
\newcommand{\V}[1]{\mathbf{#1}}

\shorttitle{ASTROPHYSICAL GYROKINETICS}
\shortauthors{HOWES, COWLEY, DORLAND, HAMMETT, QUATAERT, AND SCHEKOCHIHIN}

\begin{document}

\title{Astrophysical Gyrokinetics: Basic Equations and Linear Theory}
\author{
Gregory~G.~Howes,\altaffilmark{1}
Steven~C.~Cowley,\altaffilmark{2,3} 
William ~Dorland,\altaffilmark{4}
Gregory~W.~Hammett,\altaffilmark{5}
Eliot~Quataert\altaffilmark{1}
and Alexander~A.~Schekochihin\altaffilmark{6}
}
\altaffiltext{1}{Department of Astronomy, 601 Campbell Hall, 
University of California, Berkeley, CA 94720.} 
\altaffiltext{2}{Department of Physics and Astronomy, 
University of California, Los Angeles, CA~90095-1547.}
\altaffiltext{3}{Department of Physics, Imperial College London, 
Blackett Laboratory, Prince Consort Road, 
London~SW7~2BW, UK.}
\altaffiltext{4}{Department of Physics, 
University of Maryland, College Park, MD~20742-3511.} 
\altaffiltext{5}{Princeton University Plasma Physics Laboratory, 
P.~O.~Box 451, Princeton, NJ~08543.}
\altaffiltext{6}{Department of Applied Mathematics and 
Theoretical Physics, University of Cambridge, 
Wilberforce Road, Cambridge~CB3~0WA, UK.}

\email{ghowes@astro.berkeley.edu}

\begin{abstract}

Magnetohydrodynamic (MHD) turbulence is encountered in a wide variety
of astrophysical plasmas, including accretion disks, the solar wind,
and the interstellar and intracluster medium.  On small scales, this
turbulence is often expected to consist of highly anisotropic fluctuations 
with frequencies small compared to the ion cyclotron frequency.  For a number of
applications, the small scales are also collisionless, so
a kinetic treatment of the turbulence is necessary.  We show that this
anisotropic turbulence is well described by a low frequency expansion
of the kinetic theory called gyrokinetics.  This paper is the first in
a series to examine turbulent astrophysical plasmas in the gyrokinetic
limit.  We derive and explain the nonlinear gyrokinetic equations and
explore the linear properties of gyrokinetics as a prelude to
nonlinear simulations.  The linear dispersion relation for
gyrokinetics is obtained and its solutions are compared to those of
hot-plasma kinetic theory. These results are used to validate the
performance of the gyrokinetic simulation code {\tt GS2} in the
parameter regimes relevant for astrophysical plasmas. New results on
global energy conservation in gyrokinetics are also derived. We
briefly outline several of the problems to be addressed
by future nonlinear simulations, including particle heating by
turbulence in hot accretion flows and in the solar wind, the magnetic and
electric field power spectra in the solar wind, and the origin of
small-scale density fluctuations in the interstellar medium.

\end{abstract}

%==============================================================================
\section{Introduction} 
The \citet{gol95} (hereafter GS) theory of MHD turbulence \citep[see
also][]{sri94,gol97,lit01} in a mean magnetic field predicts 
that the energy cascades primarily by developing 
small scales perpendicular to the local field, with
$k_\perp \gg k_\parallel$, as schematically shown in
\figref{fig:gk_fig} 
\citep[cf.\ earlier work by][]{mon81,she83}.
Numerical simulations of magnetized turbulence with a 
dynamically strong mean field
support the idea that such a turbulence is strongly anisotropic 
\citep{mar01,cho02}. \emph{In situ} measurements of turbulence in the
solar wind \citep{bel71,mat90} and observations of interstellar
scintillation \citep{wil94,tro98,ric02,den03} also provide evidence
for significant anisotropy.

In many astrophysical environments, small-scale perturbations in the
MHD cascade have (parallel) wavelengths much smaller than, or at least
comparable to, the ion mean free path, 
implying that the turbulent dynamics should be calculated using
kinetic theory.  As a result of the intrinsic anisotropy of MHD
turbulence, the small-scale perturbations also have frequencies well
below the ion cyclotron frequency, $\omega \ll \Omega_i$, even when
the perpendicular wavelengths are comparable to the ion gyroradius
(see \figref{fig:gk_fig}). Anisotropic MHD turbulence in plasmas
with weak collisionality can be described by a system of
equations called gyrokinetics.

Particle motion in the small-scale turbulence is dominantly the
cyclotron motion about the unperturbed field lines.  Gyrokinetics
exploits the time-scale separation ($\omega \ll \Omega_i$) for the
electromagnetic fluctuations to eliminate one degree of freedom in the
kinetic description, thereby reducing the
problem from 6 to 5 dimensions (three spatial plus two in velocity space).  
It does so by averaging over the
fast cyclotron motion of charged particles in the mean magnetic field.
The resulting ``gyrokinetic'' equations describe charged ``rings''
moving in the ring-averaged electromagnetic fields. The removal of one
dimension of phase space and the fast cyclotron time-scale achieves a more
computationally efficient treatment of low frequency turbulent
dynamics. The gyrokinetic system is completed by 
electromagnetic field equations that are obtained by applying the
gyrokinetic approximation to Maxwell's equations. The gyrokinetic
approximation orders out the fast MHD wave and the cyclotron resonance, 
but retains finite-Larmor-radius effects, collisionless 
dissipation via the parallel Landau resonance, and collisions. Both
the slow MHD wave and the Alfv\'en wave are captured by the gyrokinetics,
though the former is damped 
when its wavelength along the magnetic field is smaller than
the ion mean free path \citep{bar66,foo79}.

Linear \citep{rut68,tay68,cat78,ant80,cat81} and nonlinear gyrokinetic
theory \citep{fri82,dub83,lee83,lee87,hah88,bri92} has proven to be a valuable tool in
the study of laboratory plasmas.  It has been extensively employed to
study the development of turbulence driven by microinstabilities,
\emph{e.g.,} the ion- and electron-temperature-gradient instabilities 
\citep[e.g.,][]{dim96,dor00,jen00,rog00,jen01,jen01b,jen02,can04,par04}. 
For these applications, the structure of the mean equilibrium field and
the gradients in mean temperature and density are critical. The
full gyrokinetic equations allow for a spatially varying mean 
magnetic field, temperature, and density. In an 
astrophysical plasma, the microinstabilities associated with 
the mean spatial gradients are unlikely to be as
important as the MHD turbulence produced by violent large scale events
or instabilities. Our goal is to use gyrokinetics to
describe this MHD turbulence on small scales (see Fig. \ref{fig:gk_fig}). 
For this purpose, the variation of the
large-scale field, temperature, and density is unimportant.  We,
therefore, consider gyrokinetics in a uniform equilibrium field with no 
mean temperature or density gradients. 

\pseudofigureone{fig:gk_fig}{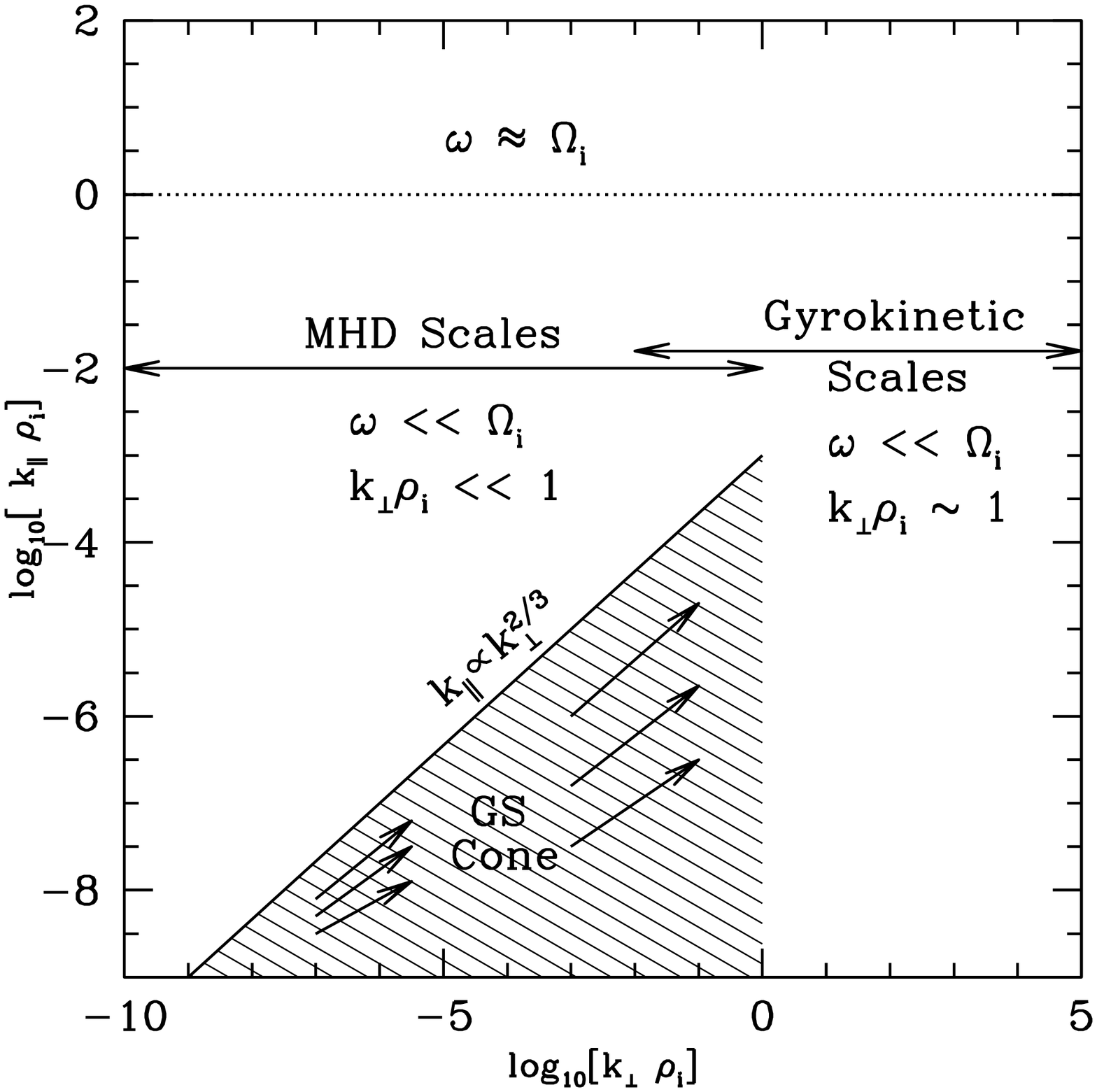}{Schematic diagram 
of the low-frequency, anisotropic Alfv\'en-wave cascade in wavenumber
space: the horizontal axis is perpendicular wavenumber; the vertical
axis is the parallel wavenumber, proportional to the frequency.  MHD
\citep[or, rather, strictly speaking, its Alfv\'en-wave part;
see][]{sch06b} is valid only in the limit $\omega \ll
\Omega_i$ and $k_\perp \rho_i \ll 1$; gyrokinetic theory remains valid
when the perpendicular wavenumber is of the order of the ion Larmor
radius, $k_\perp \rho_i \sim 1$.  Note that 
$\omega\to\Omega_i$ only when $k_\parallel \rho_i \to 1$, so gyrokinetics is
applicable for $k_\parallel \ll k_\perp$.}

This is the first in a series of papers to apply gyrokinetic theory to
the study of turbulent astrophysical plasmas.  In this paper, we
derive the equations of gyrokinetics in a uniform equilibrium field
and explain their physical meaning.  We also derive and analyze the
linear gyrokinetic dispersion relation in the collisionless regime, 
including the high-beta regime, which is of particular interest in astrophysics. 
Future papers will present analytical reductions of the
gyrokinetic equations in various asymptotic limits \citep{sch06b} and
nonlinear simulations applied to specific astrophysical problems
These simulations use the gyrokinetic simulation code {\tt GS2}. As part
of our continuing tests of {\tt GS2}, we demonstrate here that it 
accurately reproduces the frequencies and damping rates of the linear
modes.

The remainder of this paper is organized as follows.  In \S\ref{sec:gk}, we
present the gyrokinetic system of equations, giving a physical
interpretation of the various terms and a detailed discussion of the
gyrokinetic approximation. A detailed derivation of the gyrokinetic equations
in the limit of a uniform, mean magnetic field with no mean
temperature or density gradients is presented in
\appref{app:gk_deriv}. \appref{app:gk_heat} contains the first
published derivation of the long-term transport and heating equations
that describe the evolution of the equilibrium plasma --- these are 
summarized in \S\ref{sec:heating}. In \S\ref{sec:disprel}, 
we introduce the linear collisionless dispersion relation of the gyrokinetic 
theory (detailed derivation given in \appref{app:disprel}, various analytical limits 
worked out in \appref{app:limits}).
The numerics are presented in \S \ref{sec:numres}, 
where we compare the gyrokinetic dispersion relation 
and its analytically tractable  
limits with the full dispersion relation of the hot-plasma theory and with 
numerical results from the gyrokinetic simulation code {\tt GS2}. 
We also discuss the effect of collisions on collisionless damping rates
(\S\ref{sec:collisions}) and illustrate the limits of applicability of
the gyrokinetic approximation (\S\ref{sec:limappl}).
Finally, \S\ref{sec:discuss} summarizes our results and discusses
several potential astrophysical applications of gyrokinetics.
Definitions adopted for plasma parameters in this paper are given in
\tabref{tab:defs1}.

\pseudotable{tab:defs1}{Definitions of plasma parameters.}{cl}{
\hline
$s$ & particle species \\
$m_s$ & particle mass\\
$q_s$ & particle charge (we take $q_i=-q_e=e$)\\
$n_{0s}$ & number density\\
$T_{0s}$ & temperature (in units of energy)\\
$\beta_s = 8\pi n_{0s} T_{0s}/B_0^2$ & plasma beta \\
$v_{th_s}= (2 T_{0s}/m_s)^{1/2}$ & thermal speed \\
$c_s = (T_{0e}/m_i)^{1/2}$ & sound speed \\
$v_A = B_0/(4\pi m_in_{0i})^{1/2}$ & Alfv\'en speed \\
$c$ & speed of light \\
$\Omega_s= q_s B_0/(m_s c)$ & cyclotron frequency (carries sign of $q_s$) \\
$\rho_s = v_{th_s}/\Omega_s$ & Larmor radius \\
\hline
}

%==============================================================================
\section{Gyrokinetics}
\label{sec:gk}
In this section we describe the gyrokinetic approximation and present
the gyrokinetic equations themselves in a simple and physically
motivated manner (the details of the derivations are given in the
Appendices). To avoid obscuring the physics with the complexity of
gyrokinetics in full generality, we treat the simplified case of
a plasma in a straight, uniform mean magnetic
field, $\V{B}_0=B_0 \zhat$ and with a spatially uniform equilibrium
distribution function,$\nabla F_0 =0$ (the slab limit). This case is
also of most direct astrophysical relevance because the mean gradients 
in turbulent astrophysical plasmas are generally dynamically unimportant on
length scales comparable to the ion gyroradius.

%==============================================================================
\subsection{The Gyrokinetic Ordering}
\label{sec:ordering}
The most basic assumptions that must be satisfied for the gyrokinetic equations
to be applicable are weak coupling, strong magnetization, low frequencies and small
fluctuations. The weak coupling is the standard assumption of plasma physics: $n_{0e}
\lambda_{De}^3 \gg 1$, where $n_{0e}$ is the mean electron number density and
$\lambda_{De}$ is the electron Debye length. This approximation
allows the use of the Fokker--Planck equation to describe the kinetic
evolution of all plasma species.

The conditions of strong magnetization and low frequencies 
in gyrokinetics mean that 
the ion Larmor radius $\rho_i$ must be much smaller than  
the macroscopic length scale $L$ of the equilibrium plasma and
that the frequency of fluctuations $\omega$ must be small compared to the ion
cyclotron frequency $\Omega_i$,
\begin{equation}
\rho_i = \frac{v_{th_i}}{\Omega_i} \ll L,\qquad 
\omega \ll \Omega_i. 
\end{equation}
The latter assumption allows one to average all quantities 
over the Larmor orbits of particles, one of the key
simplifications allowed by the gyrokinetic theory. Note that the assumption of
strong magnetization does not require the plasma beta 
(the ratio of the thermal to the magnetic pressure, $\beta = 8 \pi p/B^2$)  
to be small. A high-beta plasma can satisfy this constraint as
long as the ion Larmor radius is small compared to the gradients of
the equilibrium system. In most astrophysical contexts, even a very
weak magnetic field meets this requirement.

To derive the gyrokinetic equations, we order the time and
length scales in the problem to separate fluctuating and equilibrium
quantities. The remainder of this section defines this formal
ordering and describes some simple consequences that follow from it.

Two length scales are characteristic of gyrokinetics: 
the small length scale, the ion Larmor radius $\rho_i$, 
and the larger length scale $l_0$, which is here introduced 
formally and will be argued below 
to be the typical parallel wavelength of the fluctuations. 
Their ratio defines the fundamental expansion parameter $\epsilon$ 
used in the formal ordering:
\begin{equation}
\epsilon = \frac{\rho_i}{l_0} \ll 1.
\label{eq:epsilon}
\end{equation}

There are three relevant time scales, or frequencies, of interest.  The
fast time scale is given by the ion cyclotron frequency
$\Omega_i$. The distribution function and the electric and magnetic 
fields are assumed to be stationary on this time scale. The intermediate 
time scale corresponds to the frequency of the turbulent fluctuations, 
\begin{equation}
\omega \sim \frac{v_{th_i}}{l_0} \sim \Order(\epsilon \Omega_i).
\end{equation}  
The slow time scale is connected to the rate of heating in the system, 
ordered as follows 
\begin{equation}
\frac{1}{t_{heat}} \sim \epsilon^2 \frac{v_{th_i}}{l_0} \sim \Order( \epsilon^3\Omega_i).
\end{equation}

The distribution function $f$ of each species $s$ ($=e,i$, the species 
index is omitted unless necessary) and magnetic and electric fields $\V{B}$
and $\V{E}$ are split into equilibrium parts (denoted with a subscript
$0$) that vary at the slow heating rate and fluctuating parts (denoted
with $\delta$ and a subscript indicating the order in $\epsilon$) that
vary at the intermediate frequency $\omega$:
\begin{eqnarray}
f(\V{r},\V{v},t) &=& F_0(\V{v},t) + \delta f_1 (\V{r},\V{v},t) + 
\delta f_2 (\V{r},\V{v},t) + \cdots,\\
\label{eq:dist_split}
\V{B}(\V{r}, t) &=& \V{B}_0 + \delta \V{B}(\V{r}, t) 
= B_0\zhat + \nabla\times\V{A},\\
\V{E}(\V{r}, t) &=& \delta \V{E}(\V{r}, t) = -\nabla\phi 
- \frac{1}{c}\frac{\partial \V{A}}{\partial t}.
\end{eqnarray}

Let us now list the gyrokinetic ordering assumptions.
\begin{itemize}
\item {\em Small fluctuations about the equilibrium.}  
Fluctuating quantities are formally of order $\epsilon$ in the
gyrokinetic expansion, 
\begin{equation}
\frac{\delta f_1}{F_0} \sim \frac{\delta \V{B}}{B_0}
\sim \frac{\delta \V{E}}{(v_{th_i}/c) B_0} \sim \Order( \epsilon).
\label{eq:order_fluct}
\end{equation}  
Note that although fluctuations are small, the 
theory is fully nonlinear (interactions are strong). 
\item {\em Slow-time-scale variation of the equilibrium.} 
The equilibrium varies on the heating time-scale, 
\begin{equation}
\frac{1}{F_0}\frac{\partial F_0}{\partial t} \sim \Order\left(\frac{1}{t_{heat}}\right) 
\sim \Order\left(\epsilon^2\frac{v_{th_i}}{l_0}\right).
\end{equation}
Derivations for laboratory plasmas 
\citep{fri82} have included a large-scale [$\sim\Order(1/l_0)$] spatial
variation of the equilibrium ($F_0$ and $ \V{B}_0$)---this we omit.
The slow-time-scale evolution of the equilibrium, however, is treated
for the first time here. 
\item {\em Intermediate-time-scale variation of the fluctuating quantities.}  
The fluctuating quantities vary on the intermediate time scale
\begin{equation} 
\omega\sim\frac{1}{\delta f}\frac{\partial \delta f}{\partial t} \sim 
\frac{1}{|\delta \V{B}|}\frac{\partial \delta \V{B}}{\partial t} \sim
\frac{1}{|\delta \V{E}|}\frac{\partial \delta \V{E}}{\partial t} \sim
\Order\left(\frac{v_{th_i}}{l_0}\right).
\end{equation}  
\item {\em Intermediate-time-scale collisions.}  
The collision rate in gyrokinetics is ordered to be the same as the 
intermediate time-scale 
\begin{equation}
\nu \sim \Order\left(\frac{v_{th_i}}{l_0}\right)
\sim \Order(\omega).
\end{equation} 
Collisionless dynamics with $\omega > \nu$
are treated correctly as long as $\nu > \epsilon\omega$.
\item {\em Small-scale spatial variation of fluctuations across the mean field.} 
Across the mean magnetic field, the fluctuations occur on the
small length scale 
\begin{equation}
k_\perp\sim\frac{{\bf{\hat z}} \times \nabla \delta f}{\delta f} \sim 
\frac{{\bf{\hat z}} \times \nabla \delta \V{B}}{|\delta \V{B}|} \sim 
\frac{{\bf{\hat z}} \times \nabla \delta \V{E}}{|\delta \V{E}|} \sim 
\Order\left(\frac{1}{\rho_i}\right).
\end{equation}  
\item {\em Large-scale spatial variation of fluctuations along the mean field.}  
Along the mean magnetic field the fluctuations occur on the
larger length scale
\begin{equation}
k_\parallel\sim\frac{{\bf{\hat z}} \cdot \nabla \delta f}{\delta f} \sim 
\frac{{\bf{\hat z}} \cdot \nabla \delta \V{B}}{|\delta \V{B}|} \sim
\frac{{\bf{\hat z}} \cdot \nabla \delta \V{E}}{|\delta \V{E}|} \sim 
\Order\left(\frac{1}{l_0}\right).
\label{eq:order_kpar}
\end{equation}  
\end{itemize}

\pseudofigureone{fig:gk_cartoon}{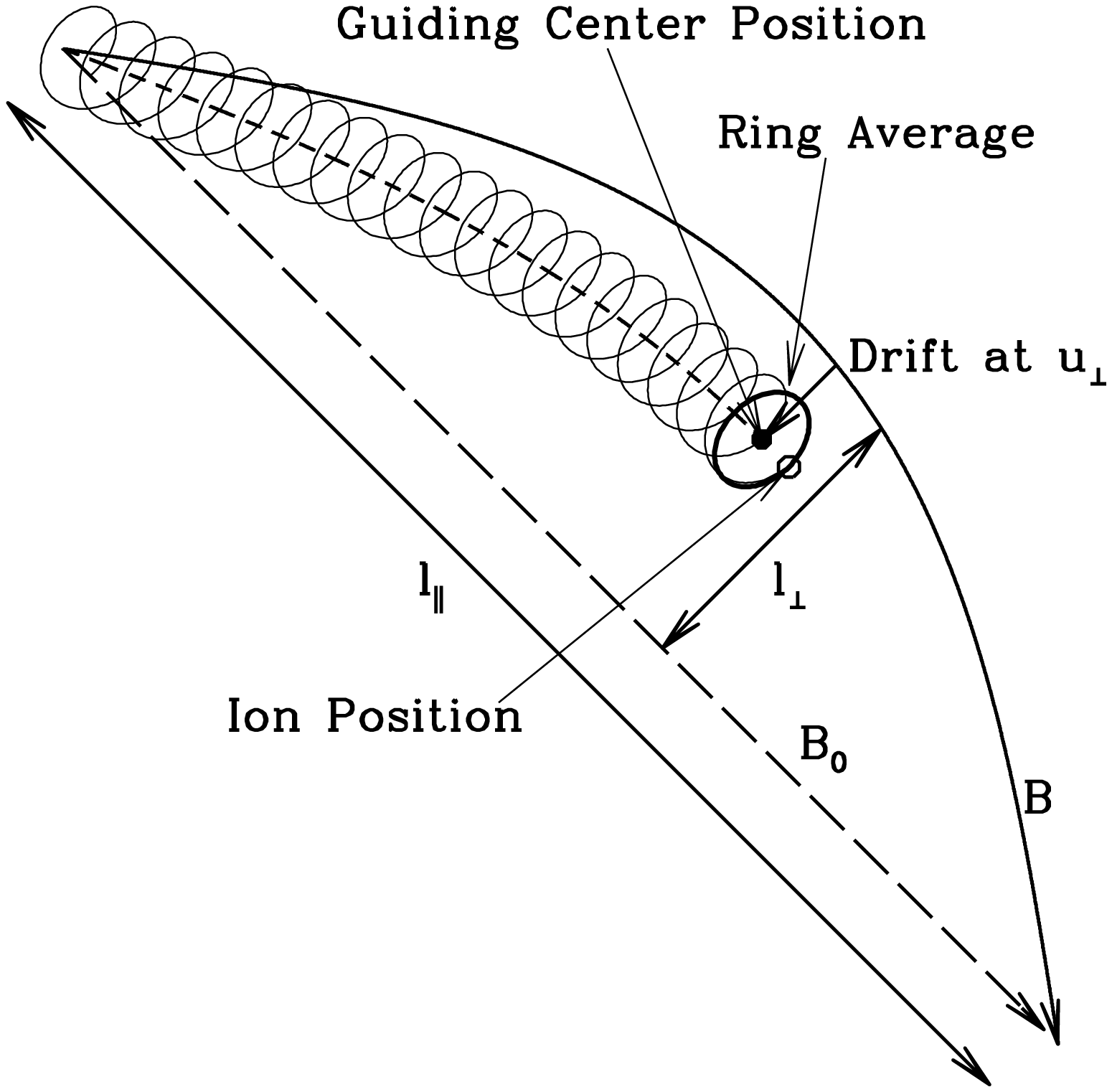}{Diagram illustrating the ring average in gyrokinetics. 
The ion position is given by the open circle, the guiding center
position by the filled circle; the ring average, centered on the
guiding center position, is denoted by the thick-lined circle passing
through the particle's position. The characteristic perpendicular and
parallel length scales in gyrokinetics are marked $l_\perp$ and
$l_\parallel$; here the perpendicular scale is exaggerated for
clarity.  The unperturbed magnetic field $\V{B}_0$ is given by the
long-dashed line and the perturbed magnetic field $\V{B}$ by the solid
line. The particle drifts off of the field line at $\V{u}_\perp$,
roughly the $\V{E} \times \V{B}$ velocity.}

With a small length scale across the field and a large
length scale along the field, the typical gyrokinetic
fluctuation is highly anisotropic:
\begin{equation}
\frac{k_\parallel}{k_\perp}
\sim \frac{\rho_i}{l_0} \sim \Order(\epsilon).
\label{eq:wavenumbers}
\end{equation}
\figref{fig:gk_cartoon} presents a schematic diagram that depicts 
the length-scales associated with the gyrokinetic ordering.  
The typical perpendicular flow velocity, roughly the $\V{E}
\times \V{B}$ velocity, is 
\begin{equation}
\V{u}_\perp \sim \frac{c\delta \V{E} \times\V{B}}{B_0^2} 
\sim \Order(\epsilon v_{th_i}).
\end{equation} 
The typical perpendicular 
fluid displacement is $\sim u_\perp /\omega \sim 1/k_\perp\sim\Order(\rho_i)$, 
as is the field line displacement. Since displacements are of order 
the perpendicular wavelength or eddy size, the fluctuations are
fully nonlinear.

\subsection{The Gyrokinetic Ordering and MHD Turbulence}

The GS theory of incompressible MHD turbulence conjectures that, on
sufficiently small scales, fluctuations at all spatial scales 
always arrange themselves in
such a way that the Alfv\'en time scale and the nonlinear
decorrelation time scale are similar, 
$\omega\sim k_\parallel v_A\sim k_\perp u_\perp$. This is known
as the critical balance. A modification of \citet{kol41} dimensional
theory based on this additional assumption then leads to the scaling
$u_\perp\sim U(k_\perp L)^{-1/3}$, where $U$ and $L$ are the velocity
and the scale at which the turbulence is driven. 
For a detailed discussion of these results, we refer the
reader to Goldreich and Sridhar's original papers or to a review by
\citet{sch05}. 

Here, we show that the gyrokinetic ordering is
manifestly consistent with, and indeed can be constructed on the
basis of, the GS critical balance conjecture. Using the critical balance
and the GS scaling of $u_\perp$, we find that the ratio of the
turbulent frequency $\omega\sim k_\perp u_\perp$ to the ion
cyclotron frequency is
\begin{equation}
\frac{\omega}{\Omega_i} \sim \frac{U}{v_{th_i}}
\left(k_\perp \rho_i \right)^{2/3} \left(\frac{\rho_i}{L} \right)^{1/3}.
\label{eq:gs_freq}
\end{equation}
The ratio of parallel and perpendicular wave numbers~is
\begin{equation}
\frac{k_\parallel}{k_\perp} \sim \frac{U}{v_A}
\left(k_\perp \rho_i \right)^{-1/3} \left(\frac{\rho_i}{L} \right)^{1/3}.
\label{eq:gs_length}
\end{equation}
Both of these ratios have to be order $\epsilon$ in the gyrokinetic 
expansion. Therefore, for the GS model of
magnetized turbulence, we define the expansion parameter 
\begin{equation}
\epsilon = \left(\frac{\rho_i}{L}\right)^{1/3}.
\label{eq:eps_GS} 
\end{equation} 
Comparing this with \eqref{eq:epsilon}, we may formally define the
length scale $l_0$ used in the gyrokinetic ordering as $l_0 =
\rho_i^{2/3}L^{1/3}$. Physically, this definition means that 
$l_0$ is the characteristic parallel
length scale of the turbulent fluctuations in the context of GS
turbulence. Note that our assumption of no spatial
variation of the equilibrium is, therefore, equivalent to assuming that the
variation scale of $F_0$ and $\V{B}_0$ is $\gg l_0$---this is
satisfied, e.g., for the injection scale $L$.

One might worry that the power of $1/3$ in \eqref{eq:eps_GS} 
means that the expansion is 
valid only in extreme circumstances. For astrophysical plasmas,
however, $\rho_i/L$ is so small that this is probably
not a significant restriction.  To take the interstellar medium as a
concrete example, $\rho_i \sim 10^8$ cm and $L \sim
100$ pc $\simeq 3 \times 10^{20}$ cm (the supernova scale), so
$\left(\rho_i/L \right)^{1/3} \sim 10^{-4}$. For
galaxy clusters, $\left(\rho_i/L\right)^{1/3}\sim10^{-5}$
\citep{sch06a},
for hot, radiatively inefficient accretion flows around
black holes, $\left(\rho_i/L \right)^{1/3} \sim 10^{-3}$
\citep{qua98}, and for the solar wind, $\left(\rho_i/L \right)^{1/3}
\sim 10^{-2}$.

In the gyrokinetic ordering, all cyclotron frequency effects 
(such as the cyclotron resonances) 
and the fast MHD wave are ordered out
\citep[for a more general approach using a kinetic description of 
plasmas in the gyrocenter coordinates that retains the 
high-frequency physics, see the gyrocenter-gauge theory of][]{qin00}. 
The slow and Alfv\'en waves are retained, however, and
collisionless dissipation of fluctuations occurs via the Landau
resonance, through Landau damping and transit-time, or \citet{bar66}, 
damping. The slow and Alfv\'en waves are accurately described
for arbitrary $k_\perp\rho_i$, as long as they are anisotropic ($k_\parallel \sim \epsilon
k_\perp$). Subsidiary ordering of collisions (as long as it does not
interfere with the primary gyrokinetic ordering) allows for a
treatment of collisionless and/or collisional dynamics.  Similarly,
subsidiary ordering of the plasma beta allows for both low- and high-beta plasmas. 

The validity of GS turbulence theory 
for compressible astrophysical plasmas is an
important question. Direct numerical simulations of compressible MHD
turbulence \citep{cho03} demonstrate that spectrum and anisotropy
of slow and Alfv\'en waves are consistent with the GS
predictions. A recent work exploring weak compressible MHD turbulence
in low-beta plasmas \citep{cha06} shows that interactions of
Alfv\'en waves with fast waves produce only a small amount of energy at
high $k_\parallel$ (weak turbulence theory for incompressible MHD 
predicts no energy at high $k_\parallel$), but this is unlikely to alter the
prediction of anisotropy in strong turbulence.  Both of these works
demonstrate that a small amount of fast-wave energy 
cascades to high frequencies, 
but the dynamics in this regime are energetically dominated
by the low-frequency Alfv\'en waves. 

%==============================================================================
\subsection{Coordinates and Ring Averages}
\label{sec:ring_ave}
Gyrokinetics is most naturally described in guiding center coordinates, 
where the position of a particle $\V{r}$ and the position of its guiding
center $\V{R}_s$ are related by
\begin{equation}
\V{r} = \V{R}_s-\frac {\V{v}\times\zhat}{\Omega_s}.
\label{eq:posguid}
\end{equation}
The particle velocity can be decomposed 
in terms of the parallel velocity
$v_\parallel$, perpendicular velocity $v_\perp$, and 
the gyrophase angle~$\theta$:
\begin{equation}
\V{v} = v_\parallel \zhat + v_\perp ( \cos \theta \xhat + \sin \theta \yhat).
\end{equation}

Gyrokinetics averages over the Larmor motion of particles and 
describes the evolution of a distribution of rings rather than
individual particles. The formalism requires defining two 
types of ring averages: the 
ring average at fixed guiding center $\V{R}_s$, 
\begin{equation}
\langle a(\V{r},\V{v},t)\rangle_{\V{R}_s} = 
\frac{1}{2 \pi} \oint d \theta\, 
a(\V{R}_s-\frac{\V{v} \times \zhat}{\Omega_s},\V{v},t),
\label{eq:avgR_def}
\end{equation}
where the $\theta$ integration is done keeping $\V{R}_s$, $v_\perp$
and $v_\parallel$ constant, and the ring average at fixed position
$\V{r}$, 
\begin{equation}
\langle a(\V{R}_s,\V{v},t)\rangle_\V{r} = 
\frac{1}{2 \pi} \oint d \theta\, 
a(\V{r}+\frac{\V{v} \times \zhat}{\Omega_s},\V{v},t),
\label{eq:avgr_def}
\end{equation}
where the integration is at constant $\V{r}$, $v_\perp$, $v_\parallel$. 

%==============================================================================
\subsection{The Gyrokinetic Equations}
\label{sec:gk_eqs}

The detailed derivation of the gyrokinetic equations is given in
\appref{app:gk_deriv}. Here we summarize the results and their
physical interpretation. 
The full plasma distribution function is expanded as follows
\begin{equation}
f_s=F_{0s}(v,t) \exp \left[-\frac{q_s \phi(\V{r},t)}{T_{0s}} \right] 
+ h_s(\V{R}_s,v, v_\perp, t) + \delta f_{2s} + \cdots,
\label{eq:distfunc2}
\end{equation}
where $v=(v_\perp^2+v_\parallel^2)^{1/2}$ and 
the equilibrium distribution function is Maxwellian:
\begin{equation}
F_{0s}= \frac{n_{0s}}{\pi^{3/2} v_{th_s}^3} \exp \left( -
\frac{v^2}{v_{th_s}^2}\right).
\label{eq:eqdistfunc}
\end{equation}
The first-order part of the distribution function is composed 
of a term that comes from the Boltzmann factor, 
$\exp [-q_s \phi(\V{r},t)/T_{0s}] \simeq 1 - q_s\phi(\V{r},t)/T_{0s}$, 
and the ring distribution $h_s$. The ring distribution $h_s$
is a function of the guiding center position $\V{R}_s$ (not the
particle position $\V{r}$) and two velocity coordinates, $v$ and $v_\perp$.\footnote{Note 
that in the inhomogeneous case, it is more convenient to use the
energy $m_s v^2/2$ and the first adiabatic invariant (the magnetic
moment) $\mu_s = (1/2)m_sv_\perp^2/B_0$ instead of $v$ and $v_\perp$.}
It satisfies the {\em gyrokinetic equation}: 
\begin{equation}
\frac{\partial h_s}{\partial t} 
+v_\parallel \zhat \cdot \frac{\partial h_s}{\partial \V{R}_s} +
\frac{c}{B_0} \left[ \langle \chi \rangle_{\V{R}_s} ,h_s \right] -
\left(\frac{\partial h_s}{\partial t} \right)_{\rm coll}
= {q_s} \frac{\partial \langle
\chi \rangle_{\V{R}_s}}{\partial t} \frac{F_{0s}}{T_{0s}},
\label{eq:gkequation1}
\end{equation}
where the electromagnetic field enters via the ring average 
of the gyrokenetic potential $\chi = \phi - \V{v}\cdot\V{A}/c$.
The Poisson bracket is defined by 
$[U,V]=\zhat\cdot[(\partial U/\partial\V{R_s})\times(\partial V/\partial\V{R_s})]$. 
The scalar potential $\phi$ and the vector potential $\V{A}$ are 
expressed in terms of $h_s$ via Maxwell's equations: 
the Poisson's equation, which takes the form of 
the quasineutrality condition
\begin{equation}
\sum_s q_s \delta n_s = 
\sum_s \left( - \frac{q_s^2 n_{0s}}{T_{0s}} \phi + q_s \int d^3 \V{v}
\langle h_s \rangle_\V{r} \right)= 0.
\label{eq:quasi1}\end{equation}
the parallel component of Amp\`ere's law,
\begin{equation}
-\nabla_\perp^2 A_\parallel
= \sum_s \frac{ 4 \pi}{c} q_s
\int d^3 \V{v}  v_\parallel \langle h_s \rangle_\V{r},
\label{eq:amp_par1}
\end{equation}
and the perpendicular component of Amp\`ere's law,
\begin{equation}
\nabla_\perp \delta B_\parallel
= \sum_s \frac{ 4 \pi}{c} q_s
\int d^3 \V{v} \langle(\zhat \times \V{v}_\perp) h_s \rangle_\V{r}.
\label{eq:amp_perp1}
\end{equation}

The gyrokinetic \eqref{eq:gkequation1} can be written in the following, 
perhaps more physically illuminating, form, 
\begin{equation}
\frac{\partial h_s}{\partial t} 
+ \left\langle \frac{d \V{R}_s}{dt}\right\rangle_{\V{R}_s} \cdot
\frac{\partial h_s}{\partial \V{R}_s} 
- \left(\frac{\partial h_s}{\partial t} \right)_{\rm coll}
= \left\langle \frac{d{\cal E}_s}{dt}\right\rangle_{\V{R}_s}\frac{F_{0s}}{T_{0s}}
\label{eq:gkequation_vel}
\end{equation}
where 
\begin{eqnarray}
\left\langle\frac{d \V{R}_s}{dt} \right\rangle_{\V{R}_s} &  =&   v_\parallel  \zhat  
-\frac{c}{B_0}\frac{\partial \langle \chi \rangle_{\V{R}_s}}{\partial
\V{R}_s} \times
\zhat \nonumber \\ 
\nonumber
 &=& v_\parallel \zhat - \frac{c}{B_0}\frac{\partial \langle \phi
\rangle_{\V{R}_s}}{\partial \V{R}_s} \times \zhat \\ 
&& + 
\frac{\partial \langle v_\parallel A_\parallel \rangle_{\V{R}_s}}{\partial \V{R}_s} 
\times \frac{\zhat}{B_0} +
\frac{\partial \langle \V{v}_\perp \cdot \V{A}_\perp \rangle_{\V{R}_s}}{\partial \V{R}_s} 
\times \frac{\zhat}{B_0}  
\label{eq:expanded}
\end{eqnarray}
is the ring velocity, 
${\cal E}_s = (1/2)m_s v^2 + q_s\phi$ is the total energy of the particle 
and 
\begin{equation}
\left\langle \frac{d{\cal E}_s}{dt}\right\rangle_{\V{R}_s}
= q_s \frac{\partial \langle \chi \rangle_{\V{R}_s}}{\partial t}.
\label{eq:energy1}
\end{equation}
Note that the right-hand side of \eqref{eq:gkequation_vel} is 
\begin{equation} 
\left\langle \frac{d{\cal E}_s}{dt}\right\rangle_{\V{R}_s}\frac{F_{0s}}{T_{0s}}
= -\left\langle \frac{d{\cal E}_s}{dt} \frac{\partial f_s}{\partial{\cal E}_s} 
\right\rangle_{\V{R}_s} 
\end{equation}
written to lowest order in $\epsilon$. 
Using this and the conservation of the first adiabatic invariant, 
$\langle d\mu_s/dt\rangle_{\V{R}_s} = 0$, where $\mu_s = m_sv_\perp^2/2B_0$, 
it becomes clear that \eqref{eq:gkequation_vel} is simply the gyroaveraged 
Fokker--Planck equation 
\begin{equation}
\left\langle\frac{d f_s}{dt} - 
\left(\frac{\partial f_s}{\partial t}\right)_{\rm coll}\right\rangle_{\V{R}_s} = 0 
\end{equation}
where only the lowest order in $\epsilon$ has been retained. 

A simple physical interpretation can now be given for each term in 
\eqref{eq:gkequation_vel}. 
It describes the evolution of a distribution of
rings $h_s$ that is subject to a number of physical influences: 

\begin{itemize}

\item motion of the ring along the ring-averaged 
total (perturbed) magnetic field: 
since $\nabla A_\parallel\times\zhat = \delta\V{B}_\perp$, 
\begin{equation}
\left(v_\parallel\zhat + \frac{\partial \langle v_\parallel A_\parallel \rangle_{\V{R}_s}}{\partial \V{R}_s} 
\times \frac{\zhat}{B_0}\right)\cdot \frac{\partial h_s}{\partial \V{R}_s} 
= \left\langle v_\parallel \frac{\V{B}}{B_0}\right\rangle_{\V{R}_s}
\cdot \frac{\partial h_s}{\partial \V{R}_s}; 
\end{equation}

\item the ring averaged  $\V{E} \times \V{B}$ drift: 
\begin{equation}
\left(- \frac{c}{B_0}\frac{\partial \langle \phi
\rangle_{\V{R}_s}}{\partial \V{R}_s} \times \zhat \right)\cdot \frac{\partial h_s}{\partial \V{R}_s} 
= \left\langle c\frac{\V{E}\times\V{B}_0}{B_0^2}\right\rangle_{\V{R}_s}
\cdot \frac{\partial h_s}{\partial \V{R}_s}; 
\end{equation}

\item the $\nabla B$ drift:  
\begin{equation}
\left(\frac{\partial \langle \V{v}_\perp \cdot \V{A}_\perp \rangle_{\V{R}_s}}{\partial \V{R}_s} 
\times \frac{\zhat}{B_0}\right)\cdot \frac{\partial h_s}{\partial \V{R}_s} 
= -\left\langle\frac{\delta B_\parallel}{B_0} \V{v}_\perp \right\rangle_{\V{R}_s}
\cdot \frac{\partial h_s}{\partial \V{R}_s}, 
\end{equation}
where, if we expand the ring average [\eqref{eq:avgR_def}] 
in small $\V{v}\times\zhat/\Omega_s$, we get, to lowest order, 
the familiar drift velocity: 
\begin{equation}
-\left\langle\frac{\delta B_\parallel}{B_0} \V{v}_\perp \right\rangle_{\V{R}_s}
\simeq -c\frac{\mu_s\nabla B\times\V{B}_0}{q_s B_0^2},
\end{equation}
where $\mu_s=m_s v_\perp^2/2B_0$ is the first adiabtic invariant 
(magnetic moment of the ring) and $\nabla B = \nabla\delta B_\parallel$ is 
taken at the center of the ring: $\V{r}=\V{R}_s$; 

\item the (linearized) effect of collisions on the perturbed ring distribution function:
$-(\partial h_s/\partial t)_{\rm coll}$ \citep[the gyrokinetic collision operator is 
discussed in detail in][]{sch06b};

\item the effect of collisionless work done on the rings by the fields 
(the wave-ring interaction): the right-hand side of \eqref{eq:gkequation_vel}. 

\end{itemize}

We have referred to the ring averaged versions of the more
familiar guiding center drifts. \figref{fig:gk_cartoon} shows the
drift of the ring along and across the magnetic field.

%==============================================================================
\subsection{Heating in Gyrokinetics} 
\label{sec:heating}
The set of equations given in the previous section determines the
evolution of the perturbed ring distribution and the field
fluctuations on the intermediate time-scale characteristic of the
turbulent fluctuations. To obtain the evolution of the distribution
function $F_0$ on the slow (heating) time-scale, we must
continue the expansion to order $\epsilon^2$. 
This is done in \appref{app:gk_heat}, where 
the derivations of the particle transport and heating equations
for our homogeneous equilibrium, including the equation defining the
conservation of energy in externally driven systems (e.g., ``forced''
turbulence), are given for the first time. 
In an inhomogeneous plasma, turbulent diffusion, or
transport, also enters at this order and proceeds on the 
slow time scale. Let us summarize the main results on heating. 

In the homogeneous case, there is no particle transport on the slow 
time-scale,
\begin{equation}
\frac{d  n_{0s}}{d  t}= 0.
\end{equation}
The evolution of the temperature $T_{0s}$ of species $s$ on this 
time scale is given by the {\em heating equation}
\begin{eqnarray}
\frac{3}{2} n_{0s} \frac{d  T_{0s}}{d t} &= &
\int d^3\V{v} \int \frac{d^3\V{R}_s}{V} 
q_s \overline{\frac{\partial \langle \chi \rangle_{\V{R}_s}}{\partial t}
h_s}  + n_{0s} \nu_{\E}^{sr} (T_{0r}-T_{0s})\nonumber \\ 
\nonumber
&= & - \int \frac{d^3\V{r}}{V} \int d^3\V{v} \frac{T_{0s}}{F_{0s}} 
\overline{\left\langle h_s\left(\frac{\partial h_s}{\partial t}\right)_{\rm coll}\right\rangle_{\V{r}}}\\
&& + n_{0s} \nu_{\E}^{sr} (T_{0r}-T_{0s}).
\label{eq:heat}
\end{eqnarray}
The overbar denotes the medium-time average over time $\Delta t$ 
such that $ 1/\omega \ll \Delta t \ll 1/(\epsilon^2\omega)$ [see \eqref{eq:timeave}].  
The second term on the right-hand side
(proportional to $\nu_{\E}^{sr}$) corresponds to the collisional
energy exchange \citep[see, e.g.,][]{hel02} between species $r$ and $s$\footnote{We
have been cavalier about treating the collision operator up to this
point.  The characteristic time scale of the interspecies collisional
heat exchange is $\sim \nu^{ii} (m_e/m_i)^{1/2}
(T_{0i}-T_{0e})/T_{0e}$. For the two terms on the right-hand
side of \eqref{eq:heat} to be formally of the same order, we must 
stipulate $\nu^{ii} (m_e/m_i)^{1/2}(T_{0i}-T_{0e})/T_{0e} \sim \Order(\epsilon^2\omega)$.  
This ordering not only ensures that the zeroth-order distribution
function is a Maxwellian but also provides greater flexibility in
ordering the collisionality relative to the intermediate time-scale of
the fluctuations. We ignore this technical detail here---in most
cases, the second term on the right-hand side of \eqref{eq:heat} is
small compared to the first term, allowing a relatively large temperature
difference between species to be maintained.}.  
It is clear from the second equality in \eqref{eq:heat} 
that the heating is ultimately always collisional, as it must be, because 
entropy can only increase due to collisions.  When the
collisionality is small, $\nu \ll \omega$, the heating is due to the
collisionless Landau damping in the sense that the distribution
function $h_s$ develops small-scale structure in velocity space, with
velocity scales $\Delta\V{v}\sim \Order(\nu^{1/2})$. Collisions smooth
these small scales at the rate $\nu v_{th_i}^2/(\Delta {v})^2 \sim
\omega$, so that the heating rate [given by the second expression in
\eqref{eq:heat}] becomes asymptotically independent of $\nu$ in the 
collisionless limit \citep[see related discussion of 
collisionless dissipation by][]{kro94,kro99}.
We stress that 
it is essential for any kinetic code, such as {\tt GS2}, to have some
collisions to smooth the velocity distributions at small scales and
resolve the entropy production. 
The numerical demonstration of the 
collisional heating and its independence of the 
collision rate is given in \secref{sec:collisions}. 

In the homogeneous case, turbulence will damp away unless driven. 
In our simulations, we study the steady-state homogeneous turbulence 
driven via an external antenna current $\V{j}_a$ introduced into Amp\`ere's
law---\emph{i.e.}, the parallel and perpendicular components of
$\V{j}_a$ are added to the right-hand sides of
\eqsmoreref{eq:amp_par1}{eq:amp_perp1}. The work done by the antenna
satisfies the \emph{power-balance equation}
\begin{equation}
%\sum_s \frac{3}{2} n_{0s} \frac{d  T_{0s}}{d t}
\int \frac{d^3\V{r}}{V} \overline{\V{j}_a  \cdot \V{E}} = 
\sum_s\int \frac{d^3\V{r}}{V} \int d^3\V{v} \frac{T_{0s}}{F_{0s}} 
\overline{\left\langle h_s\left(\frac{\partial h_s}{\partial t}\right)_{\rm coll}\right\rangle_{\V{r}}}
\label{eq:pb}
\end{equation}
(see Appendix \ref{app:driven}). 
Thus, the energy input from the driving antenna is
dissipated by heating the plasma species. The lesson of
\eqref{eq:heat} is that this heat is always produced by 
entropy-increasing collisions.

%==============================================================================
\subsection{Linear Collisionless Dispersion Relation} 
\label{sec:disprel}

The derivation of the linear dispersion relation from the 
gyrokinetic \eqsmoreref{eq:gkequation1}{eq:amp_perp1} 
is a straightfoward linearization procedure. 
In \appref{app:disprel}, it is carried out step by step. 
A key technical fact in this derivation is that once
the electromagnetic fields and the gyrokinetic
distribution function are expanded in plane waves, the ring
averages appearing in the equations can be written as multiplications
by Bessel functions. The resulting dispersion relation for linear,
collisionless gyrokinetics can be written in the following form
\begin{equation}
\left(\frac{\alpha_i A}{\overline{\omega}^2} -A B + B^2 \right)
\left(\frac{2A}{\beta_i}- AD +  C^2 \right)
 =  \left( AE +  BC \right)^2,
\label{eq:disprel}
\end{equation}
where $\overline{\omega}= \omega/|k_\parallel| v_A$ and, 
taking $q_i=-q_e=e$, $n_{0i}=n_{0e}$, 
\begin{equation}
A=1 + \Gamma_0(\alpha_i)\xi_i Z(\xi_i) 
+ \frac{T_{0i}}{T_{0e}}\left[1 + \Gamma_0(\alpha_e)\xi_e Z(\xi_e) \right],
\label{eq:defa}
\end{equation}
\begin{equation}
B=1-\Gamma_0(\alpha_i) + \frac{T_{0i}}{T_{0e}}\left[1-\Gamma_0(\alpha_e)\right],
\label{eq:defb}
\end{equation}
\begin{equation}
C=\Gamma_1(\alpha_i)\xi_i Z(\xi_i) - \Gamma_1(\alpha_e)\xi_e Z(\xi_e),
\label{eq:defc}
\end{equation}
\begin{equation}
D=2\Gamma_1(\alpha_i)\xi_i Z(\xi_i) + 2\frac{T_{0e}}{T_{0i}}\Gamma_1(\alpha_e)\xi_e Z(\xi_e),
\label{eq:defd}
\end{equation}
\begin{equation}
E=\Gamma_1(\alpha_i)-\Gamma_1(\alpha_e),
\label{eq:defe}
\end{equation}
where $\xi_s = \omega/|k_\parallel| v_{th_s}$, 
$Z(\xi_s)$ is the plasma dispersion function, 
$\alpha_s=k_\perp^2 \rho_s^2/2$, and 
$\Gamma_0(\alpha_s)=I_0(\alpha_s)e^{-\alpha_s}$, 
$\Gamma_1(\alpha_s)=[I_0(\alpha_s)-I_1(\alpha_s)]e^{-\alpha_s}$ 
($I_0$ and $I_1$
are modified Bessel functions). These functions 
arise from velocity-space integrations 
and ring averages: see \appref{app:disprel} for details. 

The complex eigenvalue solution
$\overline{\omega}$ to the
\eqref{eq:disprel} depends on three dimensionless parameters: the
ratio of the ion Larmor radius to the perpendicular wavelength,
$k_\perp \rho_i$; the ion plasma beta, or the ratio of ion thermal
pressure to magnetic pressure, $\beta_i$; and the ion to electron
temperature ratio, $T_{0i}/T_{0e}$. 
Thus, $\overline{\omega} =
\overline{\omega}_{GK}(k_\perp \rho_i, \beta_i, T_{0i}/T_{0e})$.

%==========================================================
\subsubsection{Long-Wavelength Limit} 
\label{sec:lowk}

Let us first consider the linear physics at 
scales large compared to the ion Larmor radius,
where the comparison to MHD is more straightforward. These are
not new results, but they are an important starting point for the more
general results to follow. First, recall the MHD waves in the
anisotropic limit $k_\parallel \ll k_\perp$:
\begin{eqnarray}
\omega &=& \pm k_\parallel v_A 
\qquad\qquad\quad \text{Alfv\'en waves}, \\
\omega &\simeq& \pm \frac{k_\parallel v_A}{\sqrt{1+v_A^2/c_s^2}} 
\qquad \text{slow waves}, \\
\omega &\simeq& \pm k_\perp \sqrt{c_s^2 + v_A^2} 
\qquad \text{fast magnetosonic waves}, \\
\omega &=& 0 
\qquad\qquad\qquad\quad \text{entropy mode},
\end{eqnarray}
where $c_s$ is the sound speed. The fast magnetosonic
waves have been ordered out of gyrokinetics because, when $k_\perp
\rho_i \sim 1$, their frequency is of order the cyclotron frequency
$\Omega_i$.  The removal of the fast waves is achieved by balancing
the perpendicular plasma pressure with the magnetic field pressure
[see \eqref{eq:amp_perp}].  Here we are concerned with the Alfv\'en
and slow waves in the collisionless limit. Note that the entropy mode
is mixed with the slow-wave mode when the parallel wavelength is below
the ion mean free path. In this paper, whenever we refer to the
``slow-wave'' part of the dispersion relation, we sacrifice
terminological precision to brevity.  Strictly speaking, the ``slow
waves,'' as understood below, are everything that is not Alfv\'en
waves---namely, modes involving fluctuations of the magnetic-field
strength, which can also be aperiodic (have zero real frequency)
\citep[for further discussion of this component 
of gyrokinetic turbulence, see][]{sch06b}. 

The left-hand side of \eqref{eq:disprel} contains two factors.  We
will see that the first factor corresponds to the Alfv\'en-wave
solution, the second to the slow-wave solution. The right-hand side of
\eqref{eq:disprel} represents the coupling between
the Alfv\'en and slow waves that is only important at finite ion Larmor
radius.

In the long-wavelength limit $k_\perp \rho_i \ll 1$, or $\alpha_i
\ll 1$, we can expand $\Gamma_{0}(\alpha_s) \simeq 1 -
\alpha_s$ and $\Gamma_{1}(\alpha_s) \simeq 1 - 3\alpha_s/2$.
We can also neglect terms that multiply powers of the electron-ion mass 
ratio, $m_e/m_i$, a small parameter. 
In this limit, $B\simeq\alpha_i$, $E\simeq-(3/2)\alpha_i$ and the
dispersion relation simplifies to
\begin{equation}
\left(\frac{1}{\overline{\omega}^2} - 1 \right)
\left(  \frac{2 A}{\beta_i}- AD +  C^2 \right) = 0.
\label{eq:disprel_all1}
\end{equation}

The first factor leads to the familiar Alfv\'en-wave dispersion
relation:
\begin{equation}
\omega = \pm k_\parallel v_A.
\label{eq:disprel_alf}
\end{equation}
It is not hard to verify that this branch corresponds to fluctuations 
of $\phi$ and $A_\parallel$, but not of $\delta B_\parallel$.
Thus, the Alfv\'en-wave dispersion relation in the $k_\perp\rho_i \to0$ 
limit is unchanged from the MHD result.  This is expected (and well
known) since this wave involves no motions or forces parallel to the
mean magnetic field. 
The wave is undamped and the plasma dispersion
function (which contains the wave-particle resonance effects) does not
appear in this branch of the dispersion relation.  To higher order in
$k_\perp \rho_i$, however, the Alfv\'en wave is weakly damped; taking
the high-beta result ($\beta_i \gg 1$), derived in \appref{app:weak},
the damping of the Alfv\'en wave in the limit $k_\perp\rho_i\ll1$ is 
\begin{equation}
\gamma = -|k_\parallel|v_A \frac{9}{16}
\frac{k_\perp^2 \rho_i^2 }{2}  
\sqrt{\frac{\beta_i}{\pi}}. 
\label{eq:damp_alf}
\end{equation}

The second factor in \eqref{eq:disprel_all1} represents the slow-wave
solution of the dispersion relation.  This involves motions
and forces along the magnetic field line (perturbations of 
$\delta B_\parallel$, but not of $A_\parallel$) and, unlike in the MHD
collisional limit, is damped significantly \citep{bar66,foo79}. 
The plasma dispersion function enters through $A$, $C$, and $D$; to further
simplify the expression for the slow wave, we consider the 
high- and low-beta limits.  

In the high-beta limit, $\beta_i \gg 1$, 
the argument of
the plasma dispersion function for the ion terms will be small, 
$\xi_i = \overline{\omega}/\sqrt{\beta_i} \sim \Order(1/\beta_i)$ (verified by the outcome), 
and we can use the power-series expansion \citep{fri61}
\begin{equation}
\xi_i Z(\xi_i) \simeq i \sqrt{\pi} \xi_i  - 2 \xi_i^2
\end{equation}
to solve for the complex frequency analytically. The electron terms 
may be dropped because $\xi_e=\xi_i(T_{0i}/T_{0e})^{1/2}(m_e/m_i)^{1/2}\ll\xi_i$.
Then we can approximate $A\simeq1+T_{0i}/T_{0e}$, $C\simeq i\sqrt{\pi}\xi_i$, 
$D\simeq 2\sqrt{\pi}\xi_i$. The dispersion relation reduces to 
\begin{equation}
{2\over \beta_i} - D = 0,
\end{equation}
whose solution is $\xi_i = -i/\sqrt{\pi}\beta_i$, or 
\begin{equation}
\omega = -i\frac{|k_\parallel| v_A}{\sqrt{\pi\beta_i}}.
\label{eq:sw}
\end{equation}
This frequency is purely imaginary, so the mode does
not propagate and is strictly damped, in agreement with \citet{foo79}. 
Note that $\xi_i\sim \Order(1/\beta_i)$, confirming the 
{\em a priori} assumption used to derived this result.

In the low beta limit, $\beta_i\ll1$, we shall see that the
phase velocity of the slow wave is of the order of the sound speed
$c_s=(T_{0e}/m_i)^{1/2}$.  The electrons then move faster than the wave 
and we can drop all terms involving electron plasma dispersion functions
because $\xi_e\sim c_s/v_{th_e}=(m_e/2m_i)^{1/2}\ll1$. If we
further assume that $T_{0e} \gg T_{0i}$, then the ions are moving slower
than the sound speed, so we have
$\xi_i \sim c_s/v_{th_i} = (T_{0e}/2T_{0i})^{1/2} \gg 1$. 
Expanding the plasma dispersion function in this limit
gives \citep{fri61} 
\begin{equation}
\xi_i Z(\xi_i) \simeq i \sqrt{\pi} \xi_i e^{-\xi_i^2} - 1 - 
\frac{1}{2 \xi_i^2}.
\end{equation}
Using this expansion in $A\simeq 1 + \xi_i Z(\xi_i) + T_{0i}/T_{oe}$, 
$C\simeq\xi_i Z(\xi_i)$ and $D\simeq2\xi_i Z(\xi_i)$, we find that the
slow-wave part of \eqref{eq:disprel_all1} now reduces to 
$A=0$, or 
\begin{equation}
\frac{T_{0i}}{T_{0e}} - 
\frac{1}{2}\left(\frac{k_\parallel v_{th_i}}{\omega}\right)^2 +
i \sqrt{\pi} \frac{\omega}{|k_\parallel| v_{th_i}}\,
e^{-(\omega/k_\parallel v_{th_i})^2}=0.
\label{eq:sweq} 
\end{equation}
Assuming weak damping, to be checked later, we can solve for the real
frequency and damping rate by expanding this equation about the 
real frequency.  Solving for the real frequency from the
real part of \eqref{eq:sweq} gives
\begin{equation}
\omega = \pm k_\parallel c_s.
\label{eq:disprel_sw_r}
\end{equation}
This is the familiar ion acoustic wave. Solving for the damping gives
\begin{equation}
\gamma = -|k_\parallel|c_s\sqrt{\frac{\pi}{8}} 
\left( \frac{T_{0e}}{T_{0i}} \right)^{3/2}
e^{-T_{0e}/2T_{0i}}.
\label{eq:disprel_sw_i}
\end{equation}
This solution agrees with the
standard solution for ion acoustic waves \citep[see, e.g.,][ \S
8.6.3]{kra86} in the limit $k^2 \lambda_{De}^2 \ll 1$.
Note that the \emph{a priori} assumptions we made above 
are verified by this result.

In summary, the gyrokinetic dispersion relation in the long-wavelength
limit, $k_\perp\rho_i \ll 1$, separates neatly into an Alfv\'en-wave
mode and a slow-wave mode, while the fast wave is ordered out by the
gyrokinetic approximation.  We have seen here that slow waves are
subject to collisionless Landau damping, even in the long-wavelength
limit, $k_\perp\rho_i \ll 1$. Therefore, if the scale of turbulent
motions falls below the mean free path, the slow mode should be
effectively damped out, particularly for high-beta plasmas.  In
contrast, the Alfv\'en waves are undamped down to scales around the
ion Larmor radius. The linear damping of the Alfv\'en waves at these
scales is worked out in \appsref{app:weak}{app:strong}, where we
present the high- and low-beta limits of the gyrokinetic dispersion
relation including the effects associated with the finite Larmor
radius.  The nature of the turbulent cascades of Alfv\'en and slow
waves at collisionless scales is discussed in more detail in
\citet{sch06b}.

%==========================================================
\subsubsection{Short-Wavelength Limit} 
\label{sec:highk}

At wavelengths small compared to the ion Larmor radius, $k_\perp
\rho_i \gg 1$, the low-frequency dynamics are those of kinetic Alfv\'en
waves. It is expected that, while the Alfv\'en-wave 
cascade is damped around $k_\perp\rho_i\sim1$, some fraction of the 
Alfv\'en-wave energy seeps through 
to wavelengths smaller than the ion Larmor radius and is 
channeled into a cascade of kinetic Alfv\'en waves.
This cascade extends to yet smaller wavelengths until
the electron Larmor radius is reached, $k_\perp\rho_e\sim1$,
at which point the kinetic Alfv\'en waves Landau-damp on the electrons.

In the limit $k_\perp\rho_i\gg1$, $k_\perp\rho_e\ll1$, 
we have $\Gamma_0(\alpha_i),\Gamma_1(\alpha_i)\to0$ 
and $\Gamma_0(\alpha_e)\simeq\Gamma_1(\alpha_e)\simeq1$,
whence $B\simeq1$, $E\simeq -1$. We assume \emph{a priori} and will 
verify later that $\xi_e\sim \Order(k_\perp\rho_e) \ll1$, so the electron 
plasma dispersion functions may be dropped to lowest order in $k_\perp\rho_e$. 
The gyrokinetic dispersion relation is then
\begin{equation}
\left(\frac{\alpha_i A}{\overline{\omega}^2} - A + 1\right)\frac{2}{\beta_i} 
= A,
\end{equation}
where $A\simeq 1 + T_{0i}/T_{0e}$. The solution is
\begin{equation}
\omega = \pm \frac{k_\parallel v_A k_\perp\rho_i}{\sqrt{\beta_i + 2/(1+ T_{0e}/T_{0i})}}.
\label{eq:kaw}
\end{equation}
This agrees with the kinetic-Alfv\'en-wave dispersion 
relation derived in the general plasma setting 
\citep[see, e.g.,][]{kin90}. Note that, for this solution, 
$\xi_e \sim \Order(k_\perp\rho_e)$ as promised. 

In order to get the (small) damping decrement of these waves, we retain 
the electron plasma dispersion functions: 
these are approximated by $Z(\xi_e)\simeq i\sqrt{\pi}$. 
Then $A\simeq 1 + (T_{0i}/T_{0e})(1+i\sqrt{\pi}\xi_e)$, 
$C\simeq -i\sqrt{\pi}\xi_e$ and $D\simeq i2(T_{0e}/T_{0i})\sqrt{\pi}\xi_e$. 
Expanding the resulting dispersion relation 
around the lowest-order solution [\eqref{eq:kaw}], we get
\begin{eqnarray}
\nonumber
\gamma &=& -i|k_\parallel| v_A \frac{k_\perp^2\rho_i^2}{2}
\left(\frac{\pi}{\beta_i} \frac{T_{0e}}{T_{0i}}\frac{m_e}{m_i}\right)^{1/2}\\
&&\times\left\{1 - \frac{1}{2}\frac{1+(1+ T_{0e}/T_{0i})\beta_i}{\left[1 + (1+T_{0e}/T_{0i})\beta_i/2\right]^2}\right\}.
\label{eq:kaw_damp}
\end{eqnarray}

The transition between the long-wavelength solutions of 
the previous section and the short-wavelength ones of this 
section is treated (in the anlytically tractable limits 
of high and low $\beta_i$) in \appsref{app:weak}{app:strong}.

%==============================================================================
\section{Numerical Tests} 
\label{sec:numres}
Gyrokinetic theory is a powerful tool for investigating
nonlinear, low-frequency kinetic physics. 
This section presents the
results of a suite of linear tests over a wide range of the three
relevant parameters: the ratio of the ion Larmor radius to the
perpendicular wavelength $k_\perp \rho_i$; the ion plasma beta, or the
ratio of ion thermal pressure to magnetic pressure, $\beta_i$; and the
ion to electron temperature ratio $T_{0i}/T_{0e}$. We compare the results
of three numerical methods: the gyrokinetic simulation code {\tt GS2}, the
linear collisionless gyrokinetic dispersion relation, and the linear
hot-plasma dispersion relation. 

For a wide range of parameters, we present three tests of the code
for verification: \S \ref{sec:damping} presents the frequency and
damping rate of Alfv\'en waves; \S \ref{sec:power} compares the ratio
of ion to electron heating due to the linear collisionless damping of
Alfv\'en waves; and \S \ref{sec:density} examines the density
fluctuations associated with the Alfv\'en mode when it couples to the
compressional slow wave around $k_\perp \rho_i \sim \Order(1)$.  The
effect of collisions on the collisionless damping rates is discussed
in \S \ref{sec:collisions}.  The breakdown of gyrokinetic theory in
the limit of weak anisotropy $k_\parallel \sim k_\perp$ and high
frequency $\omega \sim \Omega_i$ is demonstrated and discussed in \S
\ref{sec:limappl}.

\pseudofiguretwot{fig:wg_t_01}{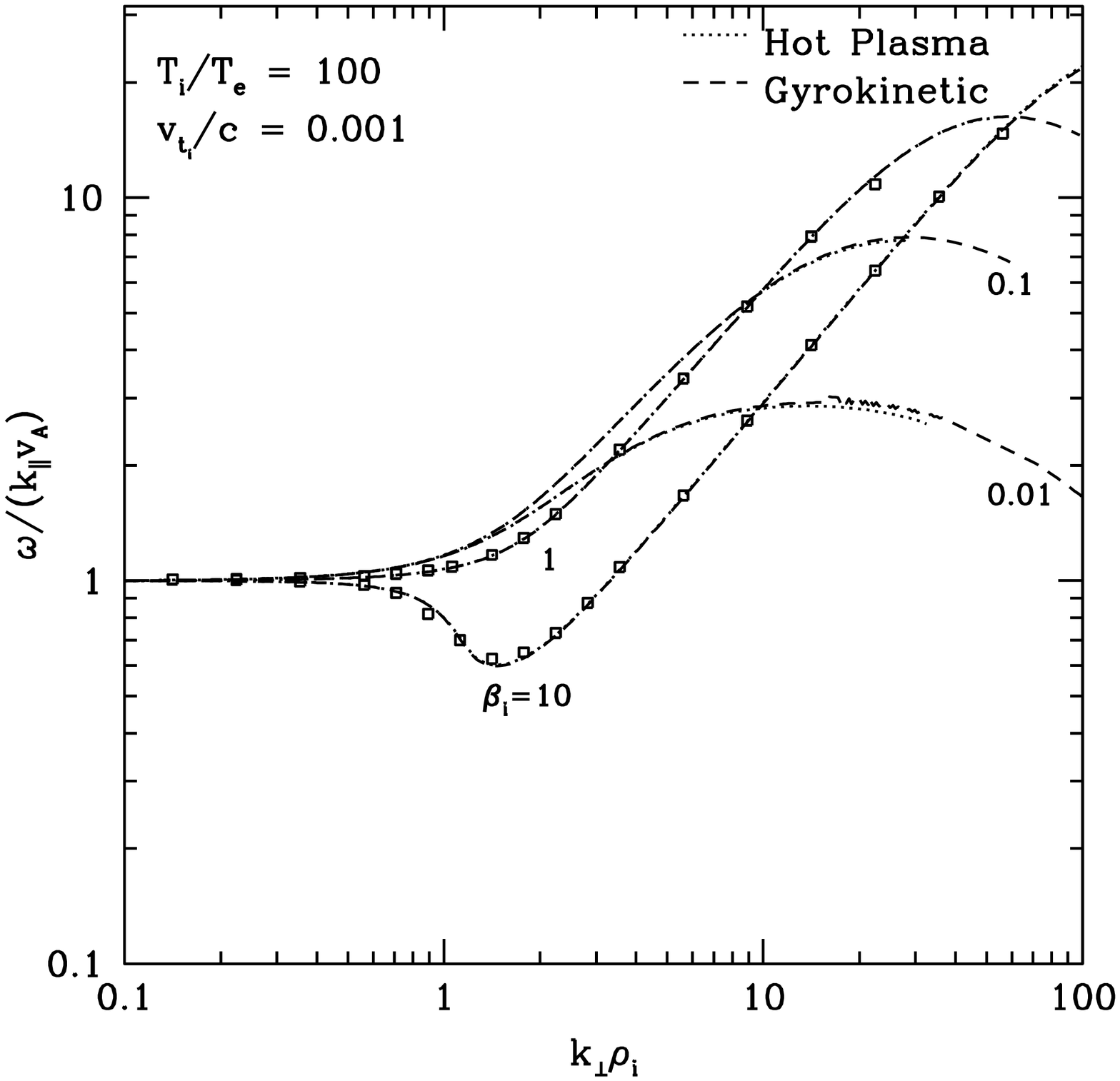}{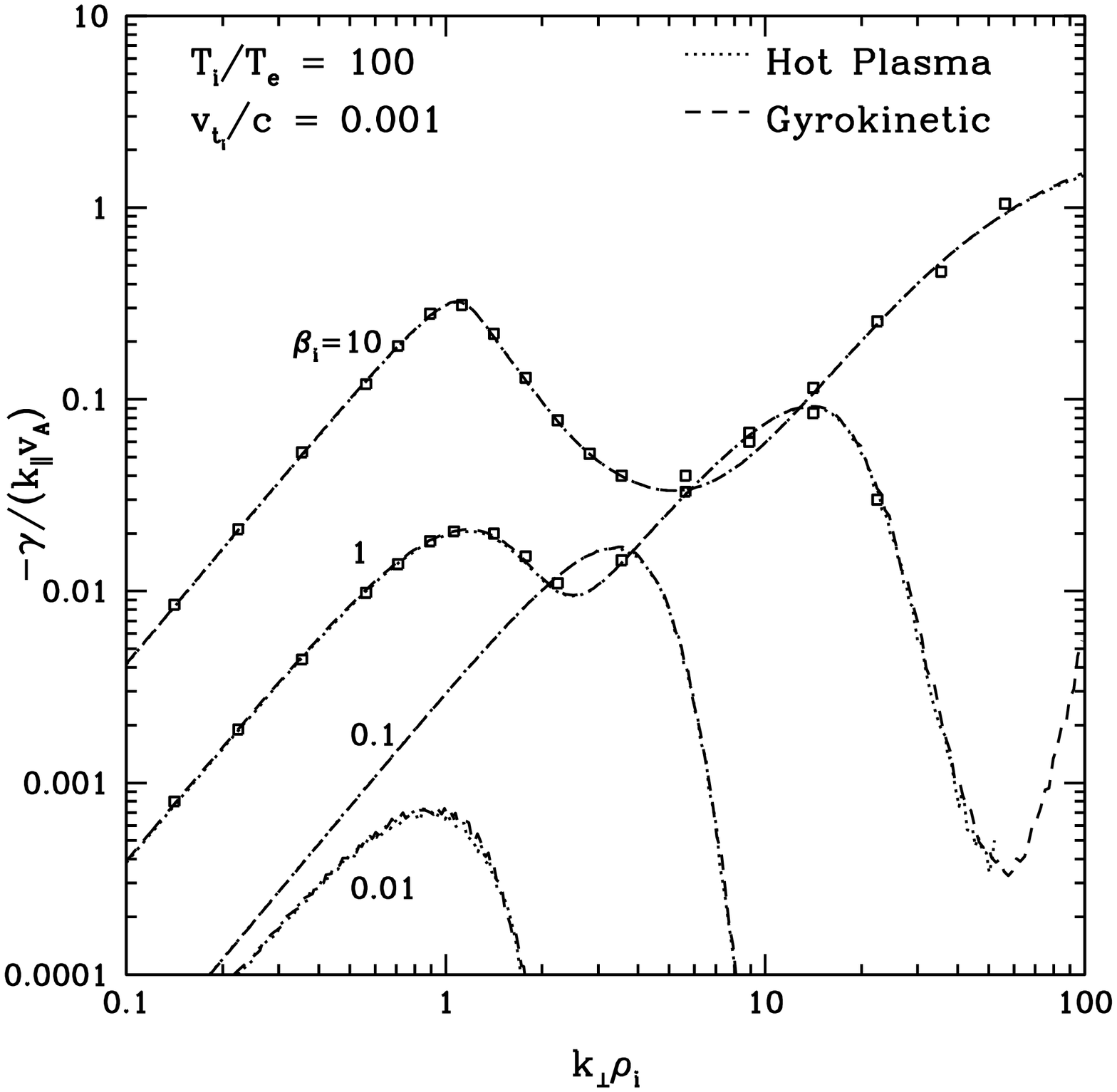}{The normalized real 
frequency $\omega/k_\parallel v_A$ (left) and 
damping rate $\gamma/k_\parallel v_A$ (right) vs.\  $k_\perp \rho_i$ for a
temperature ratio $T_{0i}/T_{0e}=100$ and ion plasma beta
values $\beta_i=10,1,0.1,0.01$. Plotted are numerical solutions to
the gyrokinetic dispersion relation (dashed lines), numerical
solutions to the hot-plasma dispersion relation (dotted lines), and
results from the {\tt GS2} gyrokinetic code (open squares).}

\pseudofiguretwob{fig:wg_t1}{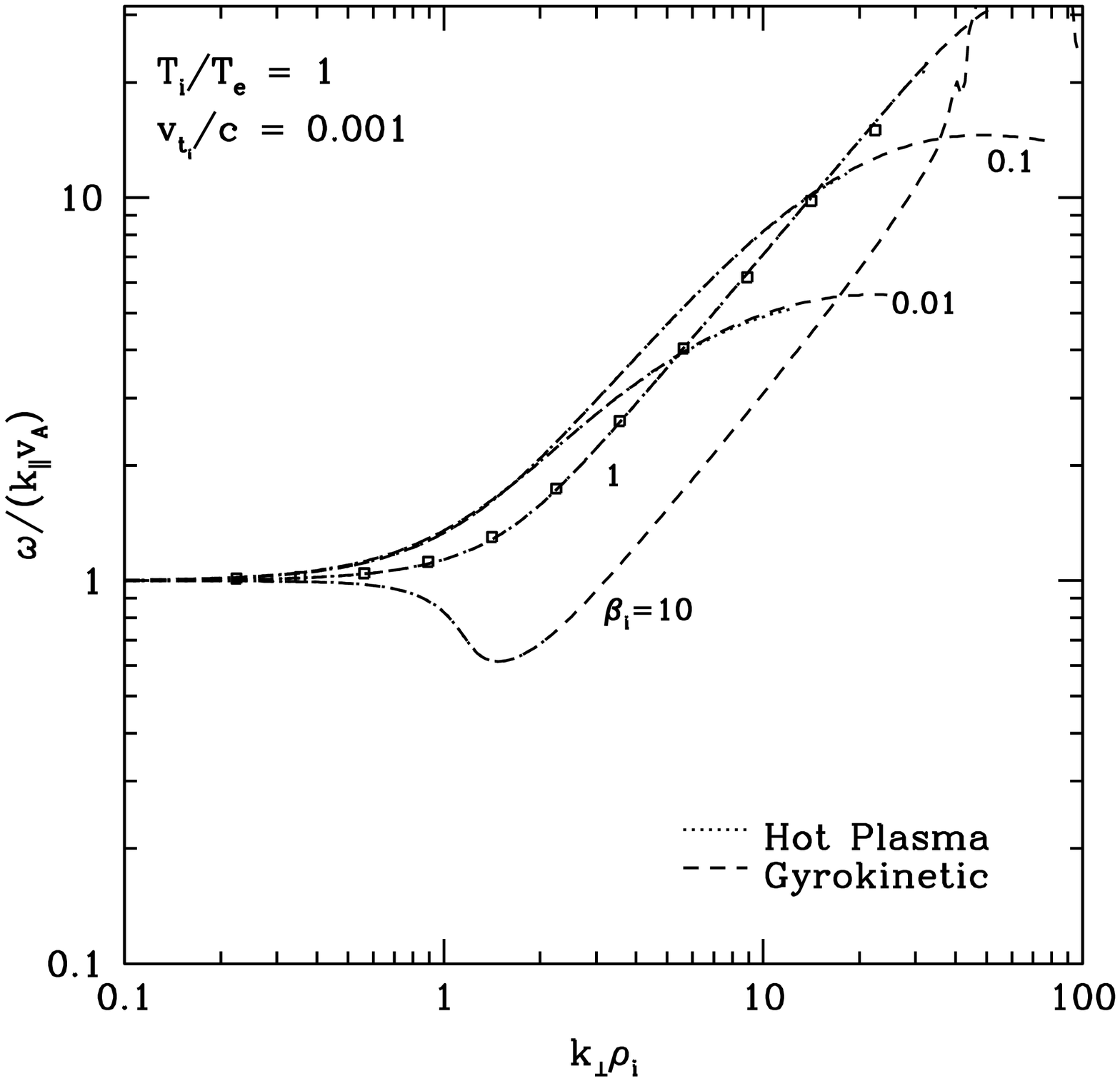}{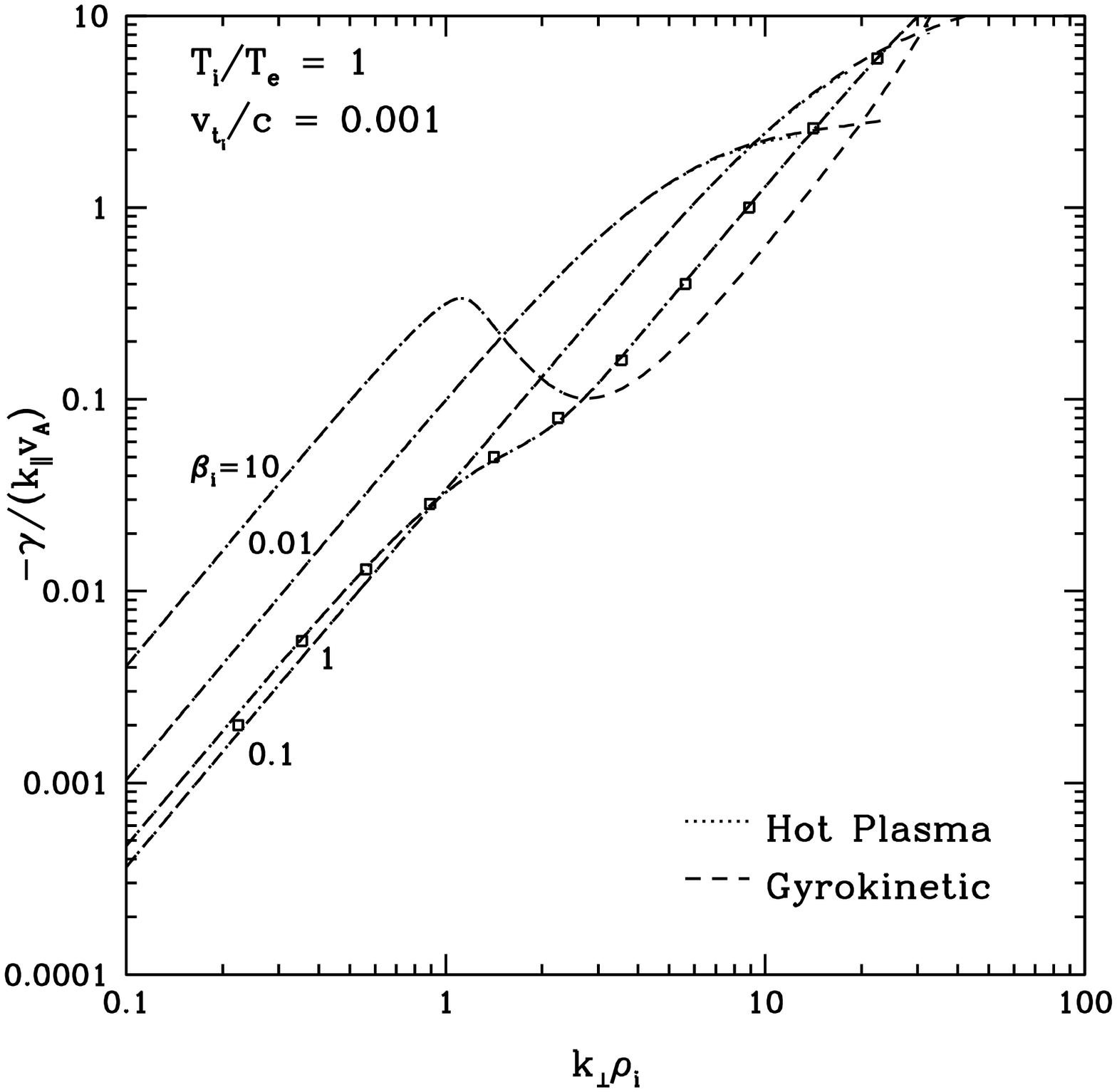}{Same as \figref{fig:wg_t_01}, 
but for $T_{0i}/T_{0e}=1$.}

\subsection{Technical Details}

{\tt GS2} is a publicly available, widely used gyrokinetic simulation
code, developed\footnote{Code development continues with support from
the Department of Energy's Center for Multiscale Plasma Dynamics.} to
study low-frequency turbulence in magnetized plasmas
\citep{kot95,dor00}.  The basic algorithm is Eulerian;
\eqsmoreref{eq:gkequation1}{eq:amp_perp1} are solved for the
self-consistent evolution of 5D distribution functions (one for each
species) and the electromagnetic fields on fixed spatial grids. All
linear terms are treated implicitly, including the field equations. 
The nonlinear terms are advanced with an explicit,
second-order accurate Adams-Bashforth scheme. 

Since turbulent structures in gyrokinetics are highly elongated along
the magnetic field, {\tt GS2} uses field-line-following Clebsch
coordinates to resolve such structures with maximal efficiency, in a
flux tube of modest perpendicular extent \citep{bch}.  Pseudo-spectral
algorithms are used in the spatial directions perpendicular to the
field and for the two velocity space coordinate grids (energy
$v^2/2$ and magnetic moment $v_\perp^2/2B$) for high accuracy
on computable 5D grids. The code offers wide flexibility in
simulation geometry, allowing for optimal representation of the
complex toroidal geometries of interest in fusion research.  For the
astrophysical applications pursued here, we require only the simple,
periodic slab geometry in a uniform equilibrium magnetic field with no
mean temperature or density gradients.

The linear calculations of collisionless wave damping and particle
heating presented in this section employed an antenna driving the
parallel component of the vector potential $A_\parallel$ 
(this drives a perpendicular perturbation of the magnetic field). 
The simulation was
driven at a given frequency $\omega_a$ and wavenumber $\V{k}_a$ by
adding an external current $j_{\parallel a}$ into the 
parallel Amp\`ere's law [\eqref{eq:amp_par1}]. 
To determine the mode frequency $\omega$ and
damping rate $\gamma$, the driving frequency was swept slowly
($\dot{\omega}_a/\omega_a \ll \gamma$) through the resonant frequency
to measure the Lorentzian response.   Fitting the curve of the
Lorentzian recovers the mode frequency and damping rate.  These
damping rates were verified in decaying runs: the plasma was
driven to steady state at the resonant frequency $\omega_a=\omega$;
then the antenna was shut off and the decay rate of the wave energy
measured. 

The ion-to-electron heating ratio was determined by driving the plasma
to steady state at the resonant frequency $\omega_a=\omega$ and
calculating the heating of each species using diagnostics based on
both forms of the heating \eqref{eq:heat}. In all methods, a realistic
mass ratio is used assuming a hydrogenic plasma. The linear {\tt GS2}
runs used a single $k_\perp$ mode and 16 points in the parallel
direction.  For most runs a velocity space resolution of $20 \times
16$ points was adequate; the ion and electron heating ratio runs
required higher velocity space resolution to resolve the heating of
the weakly damped species, with extreme cases requiring up to $80
\times 40$ points in velocity space.

In what follows, 
the linear results obtained from {\tt GS2} are compared to two sets of
analytical solutions:

1. Given the three input parameters $k_\perp \rho_i$, $\beta_i$, and
$T_{0i}/T_{0e}$, the linear, collisionless gyrokinetic dispersion relation
[\eqref{eq:disprel}] is solved numerically using a two-dimensional
Newton's method root search in the complex frequency plane, obtaining
the solution $\overline{\omega}=\overline{\omega}_{GK}(k_\perp \rho_i,
\beta_i, T_{0i}/T_{0e})$.

2. The hot-plasma dispersion relation \citep[see, e.g.,][]{sti92} is solved numerically
(also using a two-dimensional Newton's method root search) for an
electron and proton plasma characterized by an isotropic Maxwellian
with no drift velocities \citep{qua98}.  To obtain accurate results at
high $k_\perp \rho_i$, it is necessary that the number of terms kept
in the sums of Bessel functions appearing in the 
hot-plasma dispersion relation is
about the same as $k_\perp \rho_i$.  The linear hot-plasma dispersion
relation depends on five parameters: $k_\perp \rho_i$, the ion
plasma beta $\beta_i$, the ion to electron temperature ratio
$T_{0i}/T_{0e}$, the ratio of the parallel to the perpendicular wavelength
$k_\parallel/k_\perp$, and the ratio of the ion thermal velocity to
the speed of light $v_{th_i}/c$.  Hence, the solution 
may be expressed as $\overline{\omega} =
\overline{\omega}_{HP}(k_\perp \rho_i, \beta_i, T_{0i}/T_{0e}, 
k_\parallel/k_\perp, v_{th_i}/c)$.  The hot-plasma theory must reduce
to gyrokinetic theory in the limit of $k_\parallel \ll k_\perp$ and
$v_{th_i}/c \ll 1$, \emph{i.e.}, $\overline{\omega}_{HP}(k_\perp
\rho_i,\beta_i, T_{0i}/T_{0e}, 0, 0) = 
\overline{\omega}_{GK}(k_\perp \rho_i, \beta_i, T_{0i}/T_{0e})$.  

%==============================================================================
\pseudofigureone{fig:pipe_t_01}{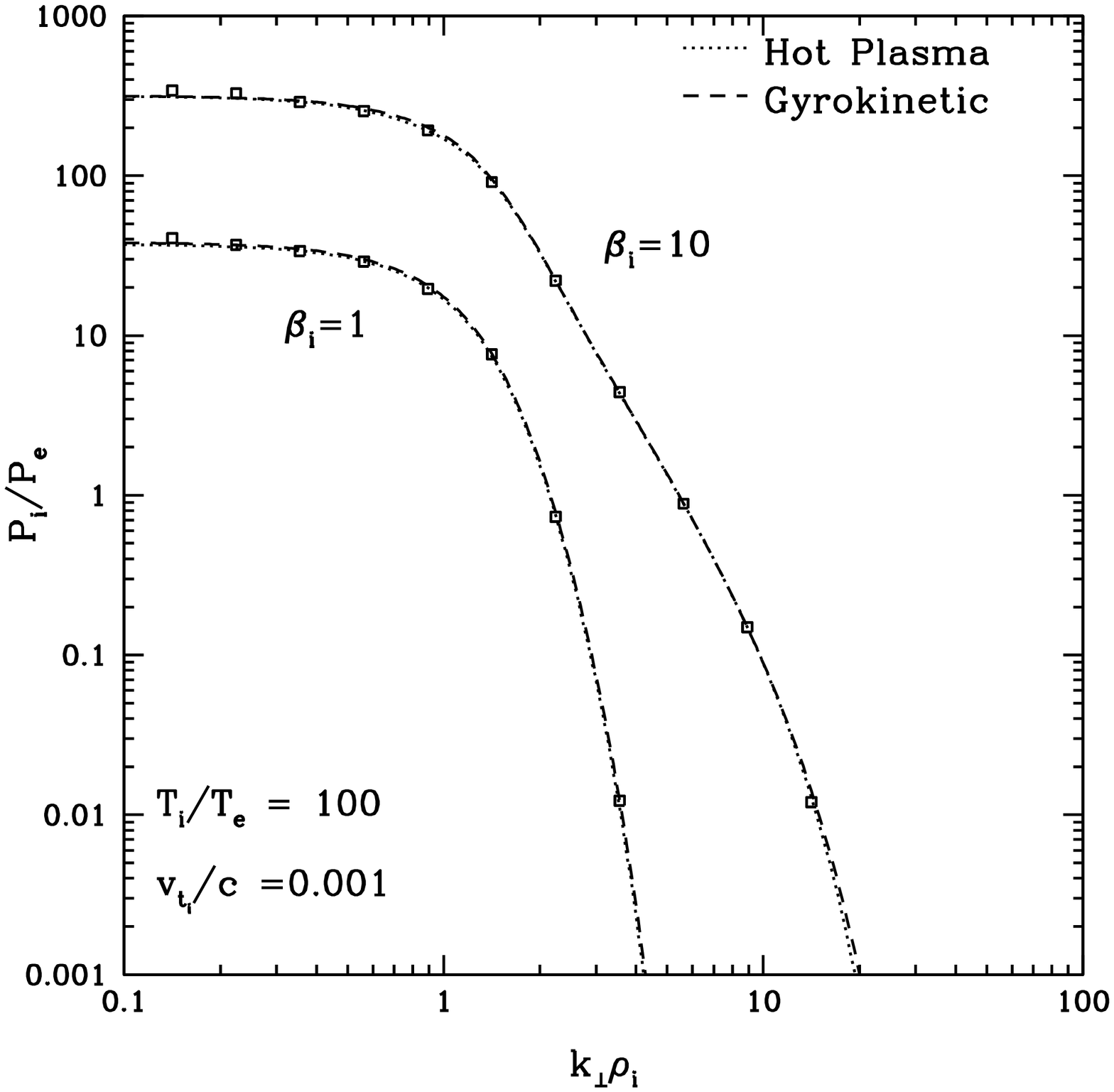}{Ratio of the ion to electron heating $P_i/P_e$. 
Results are shown for a temperature ratio of $T_{0i}/T_{0e}=100$ and values of ion
plasma beta $\beta_i=1,10$. Plotted are derived values using the
gyrokinetic dispersion relation (dashed lines) and the hot-plasma
dispersion relation (dotted lines); results from the {\tt GS2} gyrokinetic
initial-value code show good agreement over nearly five orders of
magnitude (open squares).}

\subsection{Frequency and Damping Rates} 
\label{sec:damping}
The frequency and collisionless damping rates for the three methods
are compared for temperature ratios $T_{0i}/T_{0e}=100$ and $T_{0i}/T_{0e}=1$ and
for ion plasma beta values $\beta_i=10,1,0.1,0.01$ over a range of
$k_\perp \rho_i$ from $0.1$ to $100$.  The temperature ratio
$T_{0i}/T_{0e}=100$ is motivated by accretion disk physics and the
temperature ratio $T_{0i}/T_{0e}=1$ is appropriate for studies of the
interstellar medium and the solar wind.  The real frequency $\omega$ 
and the damping rate $\gamma$ are normalized to the Alfv\'en frequency
$k_\parallel v_A$. The hot-plasma calculations in this section
all have $k_\parallel/k_\perp= 0.001$ and
$v_{th_i}/c=0.001$. The number of Bessel
functions used in the sum for these results was 100, so the results
will be accurate for $k_\perp \rho_i \lesssim 100$.

\figref{fig:wg_t_01} presents the results for the temperature ratio 
$T_{0i}/T_{0e}=100$; \figref{fig:wg_t1} for the temperature ratio
$T_{0i}/T_{0e}=1$. 
The results confirm accurate performance by {\tt GS2} over the range of
parameters tested.

%==============================================================================

\subsection{Ion and Electron Heating} 
\label{sec:power}
An important goal of our nonlinear gyrokinetic simulations to be
presented in future papers is to calculate the ratio of ion to
electron heating in collisionless turbulence \citep[motivated by
issues that arise in the physics of accretion disks;
see][]{qua98,qua99}.  Using the heating \eqref{eq:heat}, the solutions
of the linear collisionless dispersion relation can be used to
calculate the heat deposited into the ions and electrons for a given
linear wave mode.  These results for ion to electron power are
verified against estimates of the heating from the hot-plasma
dispersion relation and compared to numerical results from {\tt GS2}
in \figref{fig:pipe_t_01}.  The power deposited into each species
$P_s$ is calculated by {\tt GS2} using both forms of the heating
\eqref{eq:heat} (neglecting interspecies collisions).  Here we have
plotted the ion to electron power for a temperature ratio of
$T_{0i}/T_{0e}=100$ and ion plasma beta values of $\beta_i=1,10$. The
{\tt GS2} results agree well with the linear gyrokinetic and linear
hot-plasma calculations over five orders of magnitude in the power
ratio.

%==============================================================================

\pseudofigureone{fig:dens_bi1}{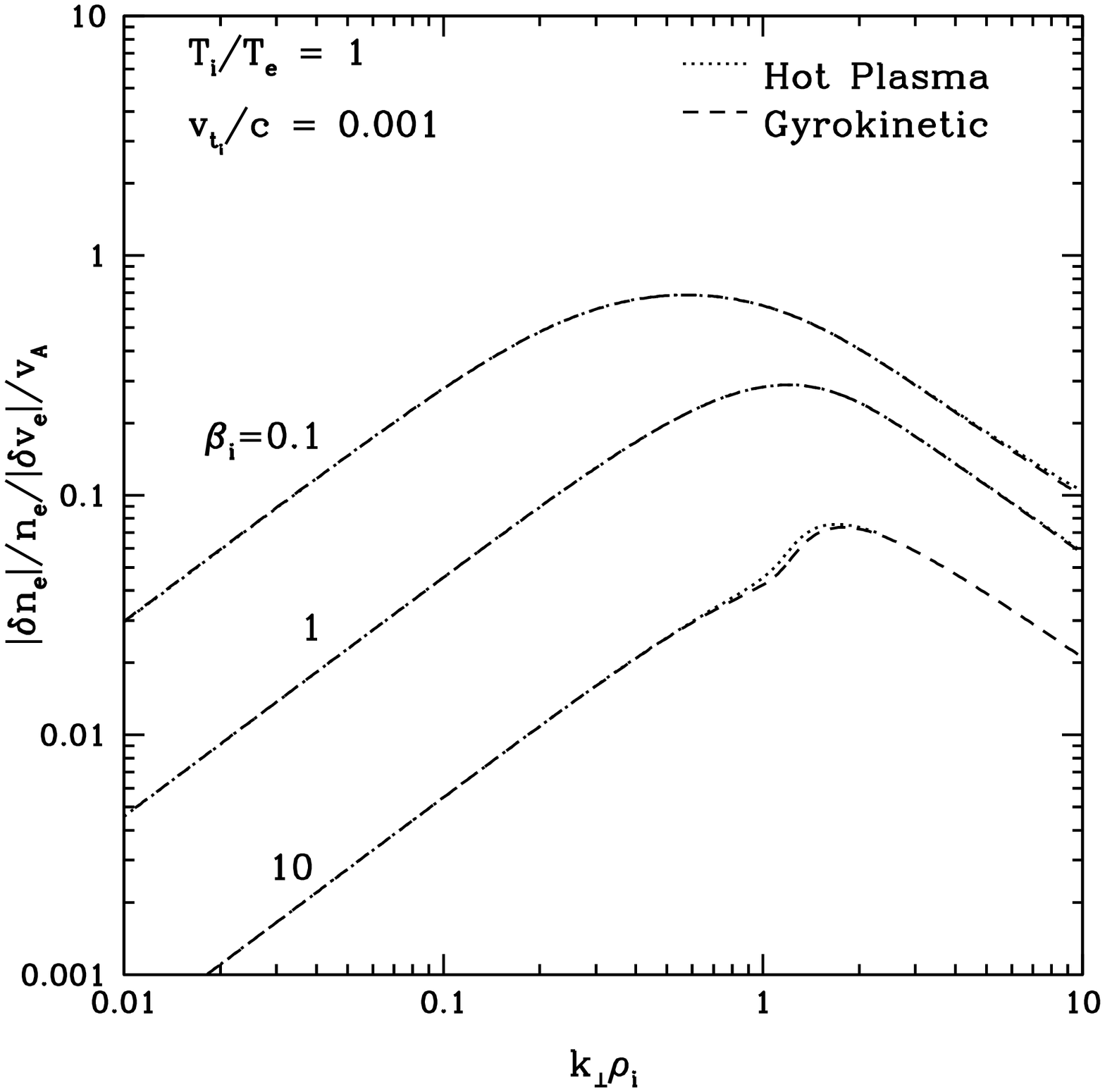}{Electron density fluctuations for 
a plasma with a temperature ratio 
of $T_{0i}/T_{0e}=1$ and values of ion plasma beta $\beta_i=0.1$, $1$, and $10$
using the gyrokinetic dispersion relation (dashed lines) and the hot-plasma 
dispersion relation (dotted lines). The fractional electron
density fluctuation $|\delta n_e|/ n_{0e}$ is normalized here by the
total electron velocity fluctuations relative to the Alfv\'en speed $|\delta
v_e|/ v_A$.}

\subsection{Density Fluctuations} 
\label{sec:density}
Alfv\'en waves in the MHD limit (at large scales) are incompressible,
with no motion along the magnetic field and no associated density
fluctuations. However, as Alfv\'en waves nonlinearly cascade to small scales
and reach $k_\perp \rho_i \sim 1$, finite-Larmor-radius effects give
rise to non-zero parallel motions, driving density fluctuations.
Here we compare the density fluctuations predicted by the linear
gyrokinetic dispersion relation with that from hot-plasma theory.
\figref{fig:dens_bi1} compares the density fluctuations for a plasma
with temperature ratio $T_{0i}/T_{0e}=1$ and values of ion plasma beta
$\beta_i=0.1$, $1$, and $10$, parameter values relevant to the
observations of interstellar scintillation in the ISM. These results
demonstrate that the fractional electron density fluctuations $|\delta
n_{0e}|/ n_{0e}$, normalized by the electron velocity fluctuation, 
peak near the ion Larmor radius as expected.

%==============================================================================
\subsection{Collisions} 
\label{sec:collisions}
Gyrokinetic theory is valid in both the collisionless and collisional
limits. To demonstrate the effect of 
collisionality---implemented in {\tt GS2} using a pitch-angle
scattering operator on each species with a coefficient
$\nu_s$---\figref{fig:collisions} presents the measured linear damping
rate in {\tt GS2} as the collision rate is increased.  Parameters for
this demonstration are $\beta_i=10$, $T_{0i}/T_{0e}=100$, and $k_\perp
\rho_i = 1.414$. The collision rates for both species are set to be
equal, $\nu=\nu_{ii}=\nu_{ee}$, and interspecies collisions are turned
off. The figure clearly demonstrates that in the collisionless limit,
$\nu \ll \omega$, the damping rate due to collisionless processes
becomes independent of $\nu$. As the collision rate is increased,
heating via collisionless Landau damping becomes less effective and
the measured damping rate decreases; this is expected because, in the
MHD limit where collisions dominate, Alfv\'en waves are undamped. As
discussed in \secref{sec:heating}, however, all heating is ultimately
collisional because collisions are necessary to smooth out the small-scale
structure in velocity space produced by wave-particle interactions. A
minimum collision rate must be specified for the determination of the
heating rate to converge. In the low velocity-space resolution
runs ($10 \times 8$ in velocity space) presented in
\figref{fig:collisions}, for $\nu/( k_\parallel v_A) < 0.01$, 
the heating rate did not converge accurately. Increasing the
velocity-space resolution lowers the minimum threshold on the
collision rate necessary to achieve convergence. 

\pseudofigureone{fig:collisions}{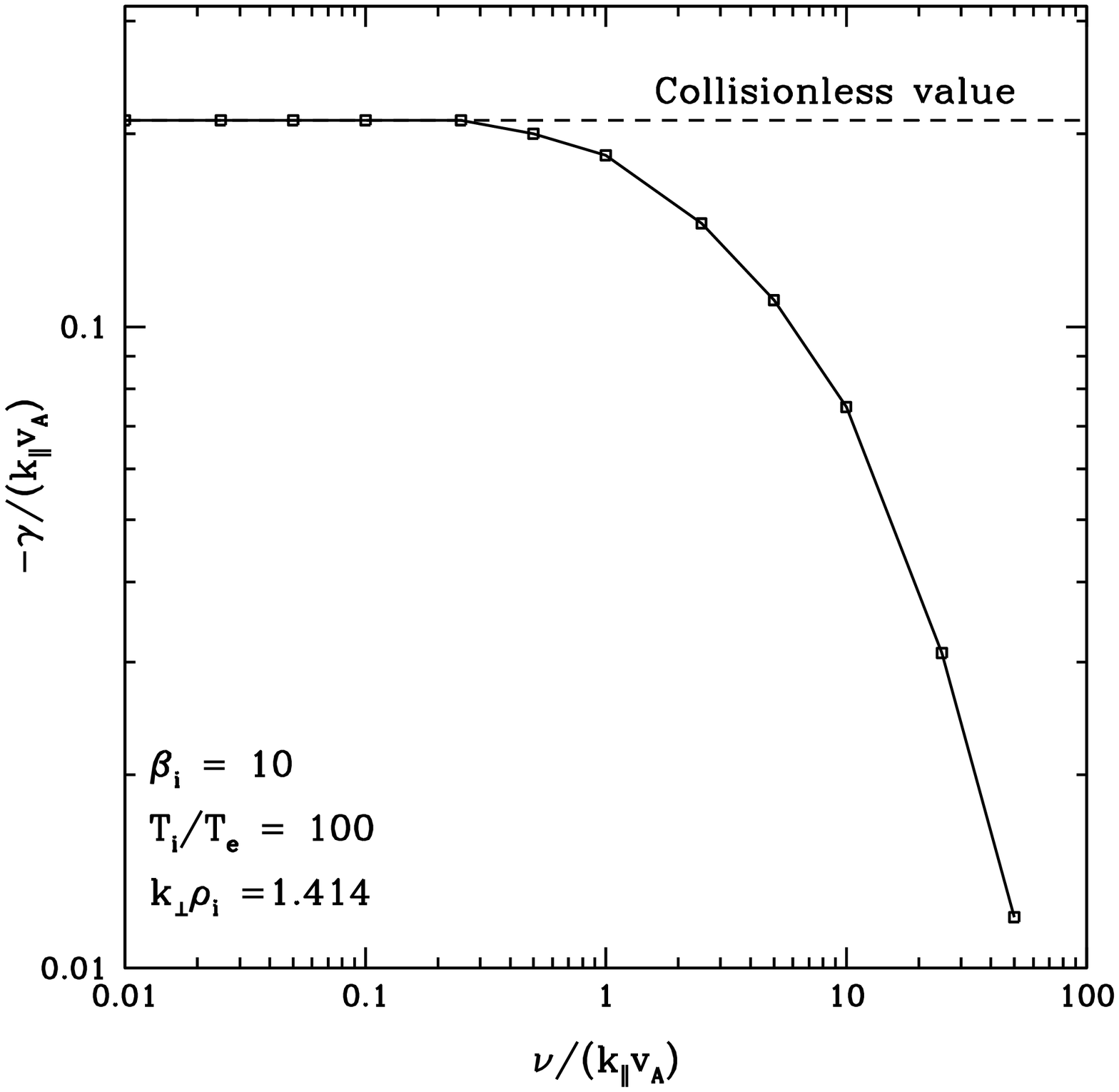}{Plot of the  damping rate $\gamma$ 
determined in linear runs of {\tt GS2} (open squares) vs.~the collision rate $\nu$
normalized by $k_\parallel v_A$. In the collisionless limit, $\nu
\ll \omega$, the damping rate is independent of the
collision rate as expected.  As the collision rate is increased,
collisionless processes for damping the wave are less effective and
the damping rate diminishes.}

%==============================================================================
\pseudofigureone{fig:parallel}{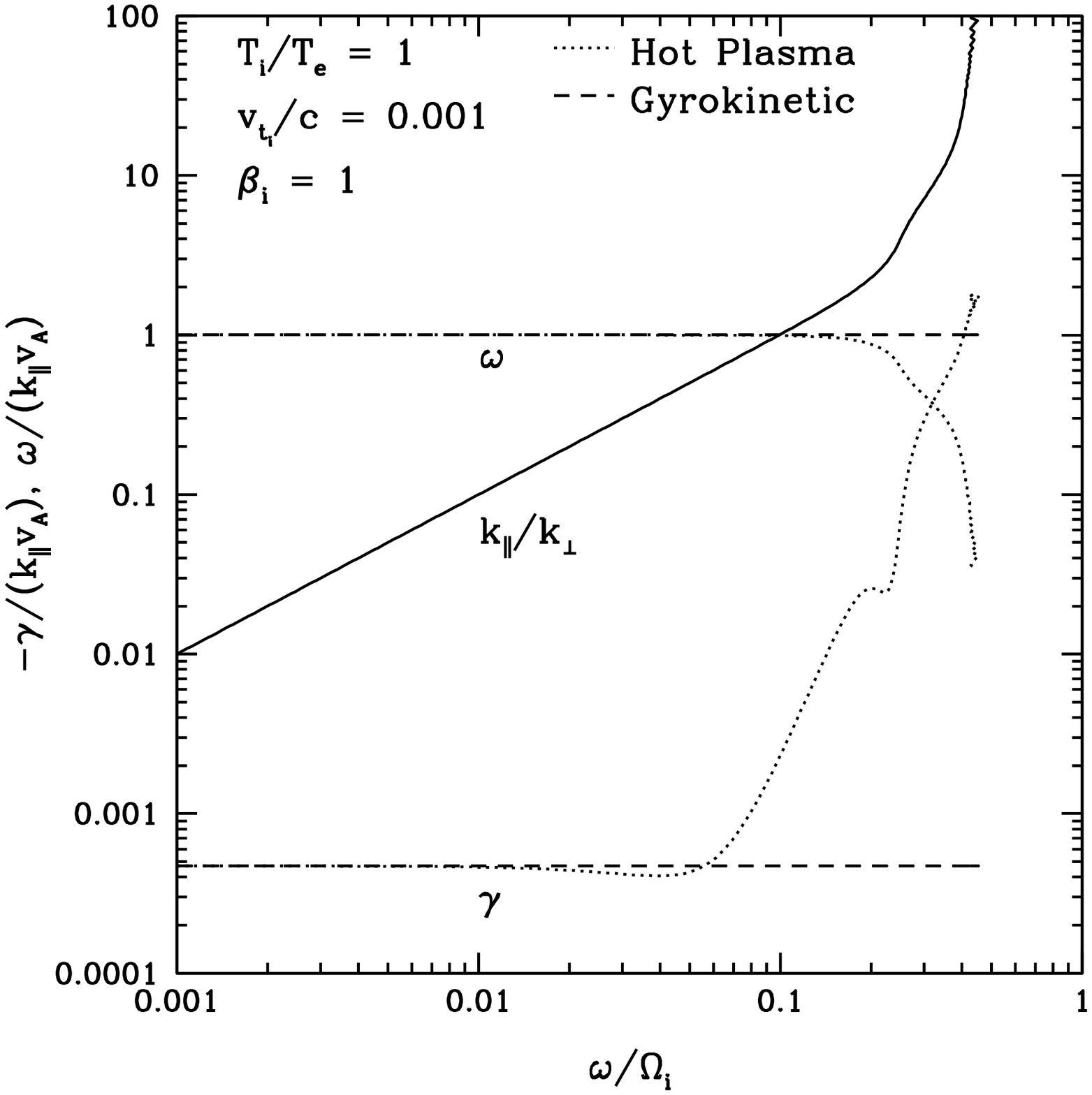}{For  a plasma with $T_{0i}/T_{0e}=1$, 
$\beta_i=1$, and $k_\perp \rho_i =0.1$, the limits of applicability of the gyrokinetic
solution (dashed line) are shown as the latter deviates from the hot-plasma 
solution (dashed line) when $\omega/\Omega_i \rightarrow 1$. 
Also plotted (solid line) is the value of $k_\parallel/k_\perp$
as a function of $\omega/\Omega_i$.}

\subsection{Limits of Applicability} 
\label{sec:limappl}
The gyrokinetic theory derived here is valid as long as three
important conditions are satisfied: (1) $k_\parallel \ll k_\perp$, 
(2) $\omega \ll \Omega_i$, and (3) $v_{th_s} \ll c$ (the
nonrelativistic assumption is not essential, but it is 
adopted in our derivation). 
As discussed at the end of \secref{sec:ordering}, the
gyrokinetic formalism retains the low-frequency dynamics of the slow
and Alfv\'en waves and collisionless dissipation via the Landau
resonance, but orders out the higher-frequency dynamics of the fast
MHD wave and cyclotron resonances.

A demonstration of the breakdown of the gyrokinetic approximation when
these limits are exceeded is provided in \figref{fig:parallel} for a
plasma with $T_{0i}/T_{0e}=1$, $\beta_i=1$, and $k_\perp \rho_i =0.1$.
Here, we increase the ratio of the parallel to perpendicular
wavenumber $k_\parallel/k_\perp$ from 0.001 to 100, and solve for the
frequency using the hot-plasma dispersion relation. The frequency
increases towards the ion cyclotron frequency as the wavenumber ratio
approaches unity. The frequency $\omega/k_\parallel v_A$
in gyrokinetics is independent of the value of $k_\parallel$, so we
compare these gyrokinetic values with the hot-plasma solution.
\figref{fig:parallel} plots the normalized real frequency
$\omega/k_\parallel v_A$ and damping rate $|\gamma|/k_\parallel v_A$ 
against the ratio of the real frequency to the ion cyclotron
frequency $\omega/\Omega_i$.  Also plotted is the value of
$k_\parallel/k_\perp$ at each $\omega/\Omega_i$. 
At $\omega/\Omega_i\sim 0.01$ and $k_\parallel/k_\perp \sim 0.1$, 
the damping rate deviates from the gyrokinetic solution as cyclotron damping 
becomes important. The real frequency departs from the gyrokinetic
results at $\omega/\Omega_i \sim 0.1$ and $k_\parallel/k_\perp \sim 1$. 
It is evident that gyrokinetic theory gives remarkably good
results even when $k_\parallel/k_\perp \sim 0.1$.

%==============================================================================
\section{CONCLUSION}
\label{sec:discuss}

This paper is the first in a series to study the properties of 
low-frequency anisotropic turbulence in collisionless 
astrophysical plasmas using
gyrokinetics.  Our primary motivation for investigating this problem
is that such turbulence appears to be a natural outcome of MHD
turbulence as energy cascades to small scales nearly perpendicular to
the direction of the local magnetic field (see \figref{fig:gk_fig}).
Gyrokinetic turbulence may thus be a generic feature of turbulent
astrophysical plasmas.

Gyrokinetics is an approximation to the full theory of collisionless
and collisional plasmas.  The necessary assumptions are that turbulent
fluctuations are anisotropic with parallel wavenumbers small compared
to perpendicular wavenumbers, $k_\parallel \ll k_\perp$, that
frequencies are small compared to the ion cyclotron frequency, $\omega
\ll \Omega_i$, and that fluctuations are small so that the typical
plasma or field line displacement is of order $\Order(k_\perp^{-1})$.
In this limit, one can average over the Larmor motion of particles,
simplifying the dynamics considerably.  Although gyrokinetics assumes
$\omega \ll
\Omega_i$, it allows for $k_\perp \rho_i \sim 1$, \emph{i.e.},
wavelengths in the direction perpendicular to the magnetic field can
be comparable to the ion Larmor radius.  On scales $k_\perp \rho_i
\lesssim 1$, the gyrokinetic approximation orders out the fast MHD
wave, but retains the slow wave and the Alfv\'en wave.  Gyrokinetics
also orders out cyclotron-frequency effects such as the cyclotron resonant
heating, but retains collisionless damping via the Landau resonance.\footnote{We 
are describing here the standard version of gyrokinetics. See \cite{qin00} for an 
extended theory that includes the fast wave and high frequency modes.}
It is worth noting that reconnection in the presence of a strong
guide field can be described by gyrokinetics, so the current sheets that
develop on scales of less than or equal to $\rho_i$ in a turbulent
plasma are self-consistently modeled in a nonlinear gyrokinetic
simulation. The enormous value of gyrokinetics as an approximation is
threefold: first, it considerably simplifies the linear and nonlinear
equations; second, it removes the fast cyclotron time scales and the
gyroangle dimension of the phase space; and third, it allows for a simple
physical interpretation in terms of the motion of charged rings.

In this paper, we have presented a derivation of the gyrokinetic
equations, including, for the first time, the equations describing
particle heating and global energy conservation.  The dispersion
relation for linear collisionless gyrokinetics is derived and its
physical interpretation is discussed. At scales $k_\perp\rho_i\ll1$, 
the familiar MHD Alfv\'en and slow modes are recovered. As  
scales such that $k_\perp \rho_i \sim 1$ are approached, 
there is a linear mixing of the Alfv\'en and slow modes. This
leads to effects such as collisionless damping of the Alfv\'en wave and
finite-density fluctuations in linear Alfv\'en waves. We have compared
the gyrokinetic results with those of hot-plasma kinetic theory,
showing the robustness of the gyrokinetic approximation. We also used
comparisons with the analytical results from both theories to verify
and demonstrate the accuracy of the gyrokinetic simulation code {\tt GS2} 
in the parameter regimes of astrophysical interest.  We note that
although our tests are linear, {\tt GS2} has already been used extensively for
nonlinear turbulence problems in the fusion research
\citep[e.g.,][]{dor00,jen00,rog00,jen01,jen01b,jen02,can04,ern04} 
and a program of nonlinear astrophysical turbulence simulations 
is currently underway.

In conclusion, we briefly mention some of the astrophysics problems
that will be explored in more detail in our future work:

\begin{enumerate}

\item Energy in Alfv\'enic turbulence is weakly damped until it cascades
to $k_\perp \rho_i \sim 1$. Thus, gyrokinetics can be used to calculate the
species by species heating of plasmas by low-frequency MHD turbulence
where the dominant heating is due to the Landau resonance. 
In future work, we will carry out gyrokinetic turbulent heating
calculations and apply them to particle heating in solar flares, the
solar wind, and hot radiatively inefficient accretion flows \citep[see
][ for related analytical and observational 
results]{qua98,gru98,qua99,lea98,lea99,lea00,cra03,cra05}. 
The use of the gyrokinetic formalism, which orders out 
high-frequency dynamics, such as fast MHD waves and 
cyclotron heating, is justified on the assumption that 
the turbulent cascade remains highly anisotropic down to 
scales of order ion Larmor radius, so that the fluctuation frequencies 
are well below the cyclotron frequency even when 
$k_\perp\rho_i\sim1$, rendering the cyclotron resonance unimportant 
to the plasma heating. 

\item {\it In situ} observations of the turbulent solar wind directly
measure the power spectra of the the magnetic field \citep{gol95b,lea98,lea99}
and the electric field \citep{bal05} down to scales smaller than
the ion gyroradius. 
Thus, detailed quantitative comparisons between
simulated power spectra and the measured ones are possible. It is
worth noting, however, that the solar wind is, in fact, sufficiently
collisionless that the gyrokinetic ordering here is not entirely appropriate
and the equilibrium distribution function $F_0$ may deviate from
a Maxwellian. Significant distortions of $F_0$, however, are tempered
by high-frequency kinetic mirror and firehose instabilities that may
play the role of collisions in smoothing out the distribution
function, so the solar wind plasma may not depart significantly
from gyrokinetic behavior. It must be kept in mind, though, that if 
these instabilities were effective in driving a cascade to higher $k_\parallel$
(and thus higher frequency), the gyrokinetic approximation would 
be violated and further analysis required to take account of the 
small-scale physics. 

\item In the interstellar medium of the Milky Way, 
the electron-density fluctuation power spectrum, inferred from
observations, is consistent with Kolmogorov turbulence over $10$
decades in spatial scale \citep[see][]{arm95}.  Observations suggest
the existence of an inner scale to the density fluctuations at
approximately the ion Larmor radius
\citep{spa90}. These observations may be probing the density
fluctuations associated with Alfv\'en waves on gyrokinetic scales (see
\S \ref{sec:density})---precisely the
regime that is best investigated by gyrokinetic simulations.

\end{enumerate}

\acknowledgements Much of this work was supported by the DOE Center for 
Multi-scale Plasma Dynamics, Fusion Science Center Cooperative
Agreement ER54785. G. W. H. is supported by NASA Grant NNG05GH78G and 
by U.S. DOE contract No.\ DE-AC02-76CH03073. 
E. Q. is also supported in part by NSF grant AST
0206006, NASA grant NAG5-12043, an Alfred P. Sloan Fellowship, and by the
David and Lucile Packard Foundation. A. A. S. is supported by 
a PPARC Advanced Fellowship and by King's College, Cambridge.

%==============================================================================
%==============================================================================
\appendix

%==============================================================================
\section{Derivation of the Gyrokinetic Equations}
\label{app:gk_deriv}

The nonlinear gyrokinetic equation in an inhomogeneous plasma was first 
derived by \citet{fri82}. 
In this Appendix, we derive the nonlinear gyrokinetic equation in the
simplest case where the equilibrium is a homogeneous plasma in a
constant magnetic field, \emph{i.e.}, $\nabla F_0 = 0$ and $\V{B}_0
= B_0\zhat$ in a periodic box. 
We begin with the Fokker--Planck equation and Maxwell's equations. 
The gyrokinetic ordering, which makes the expansion procedure 
possible was explained in \S\ref{sec:ordering}. 
The guiding center coordinates and 
the key mathematical operation --- ring averaging --- were 
introduced in \S\ref{sec:ring_ave}. 
In what follows, the Fokker--Planck equation is 
systematically expanded under the gyrokinetic ordering. 
The minus first, zeroth and first
orders are solved to determine both the form of the equilibrium
distribution function and the evolution equation for the perturbed
distribution function---the gyrokinetic equation 
(the slow evolution of the equilibrium enters in the second 
order and is worked out in \appref{app:gk_heat}). 
At each order, a ring average at constant guiding center position is 
employed to eliminate higher orders from the equation. 
Velocity integration of the perturbed distribution function yields the 
charge and current densities that appear in the gyrokinetic versions 
of Maxwell's equations.

%==============================================================================
\subsection{Maxwell's Equations and Potentials}

Let us start with Poisson's law. The equilibrium plasma is neutral, 
$\sum_s q_sn_{0s}=0$, so we have 
\begin{equation}
\nabla\cdot\delta\V{E} = 4\pi\sum_s q_s\delta n_s. 
\end{equation}
The left-hand side is $\Order(\epsilon B_0 v_{th_i}/c\rho_i)$ [see \eqref{eq:order_fluct}], 
the right-hand side is $\Order(\epsilon q_i n_{0i})$, 
so the ratio of the divergence of
the electric field to the charge density is $\Order(\beta_i^{-1}v_{th_i}^2/c^2)$. 
Therefore, in the limit of non-relativistic ions, 
the perturbed charge density is zero.\footnote{Alternatively, one can say that the divergence of 
the electric field
is small for gyrokinetic perturbations whose wavelengths are long
compared to the electron Debye length, $\lambda_{De}$.}
This establishes the condition of quasineutrality:
\begin{equation}
\sum_s q_s \delta n_{s}=0.
\label{eq:quasi0}
\end{equation}

In Faraday's law, 
\begin{equation}
c\nabla\times\delta\V{E} = -\frac{\partial\delta\V{B}}{\partial t},
\end{equation}
the left-hand side is $\Order(\epsilon \Omega_i B_0)$, whereas 
the right-hand side is $\Order(\epsilon^2 \Omega_i B_0)$.
Therefore, to dominant order the electric field satisfies 
$\nabla\times \delta \V{E}=0$, so the largest part of the electric field is
electrostatic. The inductive electric field does, however,
gives an important contribution to the parallel electric field. The
electric and magnetic fields are most conveniently written in terms of 
the scalar potential $\phi$ and vector potential $\V{A}$:
\begin{equation}
\delta \V{E} =  -\nabla \phi - \frac{1}{c}\frac{\partial \V{A}}{\partial t},
\quad \delta \V{B} =  \nabla \times \V{A}.
\label{eq:defpot}
\end{equation}
We choose the Coulomb gauge $\nabla\cdot {\bf A} = 0$.  Thus with
the gyrokinetic ordering, to $\Order(\epsilon^2)$, the vector potential is
\begin{equation}
\V{A} = A_\parallel \zhat + \V{A}_\perp = A_\parallel \zhat 
+ \nabla \lambda \times \zhat.
\label{eq:Adef}
\end{equation}
Hence, the perturbed magnetic field to $\Order(\epsilon^2)$ is given by
\begin{equation}
\delta \V{B} = \nabla A_\parallel \times \zhat - \nabla^2 \lambda \zhat 
= \nabla A_\parallel \times \zhat + \delta B_\parallel \zhat. 
\label{eq:bdef2}
\end{equation}
We will use the scalars $A_\parallel$ and $\delta B_\parallel$ (rather
than $\lambda$) in subsequent development.  

Consider next the Amp\`ere--Maxwell law: 
\begin{equation}
\nabla\times\delta\V{B} = \frac{4\pi}{c}\delta\V{j} + \frac{1}{c}\frac{\partial\delta\V{E}}{\partial t}.
\end{equation}
The left-hand side is $\Order (\epsilon B_0\Omega_i/v_{th_i})$,  
while the second term in the left-hand side (the displacement current)
is $\Order(\epsilon^2 B_0\Omega_i v_{th_i}/c^2)$. 
The ratio of the latter to the former is, therefore, 
$\Order(\epsilon v_{th_i}^2/c^2)$, so we can drop the displacement current 
and use the pre-Maxwell form of Amp\`ere's law:
\begin{equation}
\nabla \times \delta \V{B} = -\nabla^2 {\bf A} = - \nabla^2 A_\parallel \zhat 
+ \nabla \delta B_\parallel \times \zhat = \frac{ 4 \pi}{c} \delta \V{j}.
\label{eq:amp}
\end{equation}

%==============================================================================
\subsection{The Gyrokinetic Equation}
Let us start with the Fokker--Planck equation
\begin{equation}
\label{eq:fpeq}
\frac{d f_s}{d t} = 
\frac{\partial f_s}{\partial t} + \V{v}\cdot\nabla f_s
+ \frac{q_s}{m_s}\left(-\nabla\phi - \frac{1}{c}\frac{\partial\V{A}}{\partial t}  
+ \frac{\V{v}\times\V{B}}{c}\right)\cdot\frac{\partial f_s}{\partial \V{v}} 
= \left(\frac{\partial f_s}{\partial t}\right)_\text{coll} = C_{sr}(f_s,f_r) + C_{ss}(f_s,f_s),
\end{equation}
where the right-hand side is the standard Fokker--Planck
integro-differential collision operator \citep[e.g.,][]{hel02}.  
$C_{sr}(f_s,f_r)$ denotes the effect of collisions of species
$s$ on (the other) species $r$ and $C_{ss}(f_s,f_s)$ denotes like-particle
collisions. 
To reduce the clutter, we suppress the species label $s$ in this section  
and denote the entire collision term by $C(f,f)$. 

The distribution function is expanded in powers of $\epsilon$:
\begin{equation}
f = F_0 + \delta f, \quad 
\delta f = \delta f_1 + \delta f_2 + \cdots,
\end{equation}
where $\delta f_n \sim \Order(\epsilon^n F_0)$. 
With the ordering defined by equations (\ref{eq:order_fluct}-\ref{eq:order_kpar}), 
the terms in the Fokker--Planck are ordered as follows
\begin{eqnarray}
\nonumber
&&\underorder{\epsilon^2}{\frac{\partial F_0}{\partial t}}
+ \underorder{\epsilon}{\frac{\partial \delta f}{\partial t}}
+ \underorder{1}{\V{v}_\perp \cdot \nabla \delta f }
+ \underorder{\epsilon}{v_\parallel \zhat \cdot \nabla \delta f}
+ \frac{q}{m}\biggl(\underorder{1}{-\nabla\phi}
- \underorder{\epsilon}{\frac{1}{c}\frac{\partial \V{A}}{\partial t}}
+ \underorder{1}{\frac{\V{v} \times \delta \V{B}}{c}}
+ \underorder{1/\epsilon}{\frac{\V{v} \times \V{B}_0}{c}}
\biggr)\cdot \frac{\partial F_0}{\partial \V{v}}\\
&&\qquad
+\, \frac{q}{m}\biggl(\underorder{\epsilon}{-\nabla\phi}
- \underorder{\epsilon^2}{\frac{1}{c}\frac{\partial \V{A}}{\partial t}}
+ \underorder{\epsilon}{\frac{\V{v} \times \delta \V{B}}{c}}
+ \underorder{1}{\frac{\V{v} \times \V{B}_0}{c}}
\biggr)\cdot \frac{\partial \delta f}{\partial \V{v}} 
= \underorder{1}{C(F_0,F_0)}
+ \underorder{\epsilon}{C(F_0,\delta f)}
+ \underorder{\epsilon}{C(\delta f, F_0)}
+ \underorder{\epsilon^2}{C(\delta f, \delta f)},
\label{eq:fp_expanded}
\end{eqnarray}
where we label below each term its order relative to $F_0 v_{th_i}/l_0$. 
We now proceed with the formal expansion.

\subsubsection{Minus First Order, ${\Order(1/\epsilon)}$} 

From
\eqref{eq:fp_expanded}, in velocity variables transformed  from $\V{v}$ to
$(v, v_\perp,\theta)$, we obtain at this order
\begin{equation}
\frac{\partial  F_0}{\partial \theta}=0,
\label{eq:order0}
\end{equation}
so the equilibrium distribution function does not depend on gyrophase
angle, $F_0= F_0(v, v_\perp, t)$.

\subsubsection{Zeroth Order, $\Order(1)$} 

At this order, \eqref{eq:fp_expanded} becomes
\begin{equation}
\V{v}_\perp \cdot \nabla \delta f_1 + \frac{q}{m}\left( -\nabla \phi + 
\frac{\V{v} \times \delta \V{B}}{c}\right) \cdot \frac{\partial F_0}{\partial \V{v}}
- \Omega \frac{\partial \delta f_1}{\partial \theta} = C(F_0,F_0).
\label{eq:order1}
\end{equation}
At this stage both $F_0$ and $\delta f_1$ are unknown.  To eliminate
$\delta f_1$ from this equation (and thereby isolate information about
$F_0$), we multiply
\eqref{eq:order1} by $1+ \ln F_0$ and integrate over all space and
all velocities, making use of \eqref{eq:order0} and assuming that 
perturbed quantities spatially average to zero (this is exactly true 
in a periodic box). We find 
\begin{equation}
\int d^3\V{r}\int d^3\V{v}( \ln F_0) C(F_0,F_0) =0.
\end{equation}
It is known from the proof of Boltzmann's H Theorem 
that this uniquely constrains $F_0$ to be a Maxwellian:
\begin{equation}
 F_0= \frac{n_0}{\pi^{3/2}v_{th}^3} \exp \left( -
 \frac{v^2}{v_{th}^2}\right),
\label{eq:eqdist}
\end{equation}
where the mean plasma flow is assumed to be zero.  The temperature
$T_0(t) = (1/2)mv_{th}^2$ associated with this Maxwellian varies on the slow
(heating) time scale, $t_{heat} \sim \Order(\epsilon^{-2} l_0/v_{th})$, 
due to conversion of the turbulent energy into heat. 
In \appref{app:gk_heat}, we determine this heating in the 
second order of the gyrokinetic expansion. 
The density $n_0$ does not vary because the
number of particles is conserved. In most other derivations of
gyrokinetics, $F_0$ is not determined, although it is often assumed to
be a Maxwellian.

Substituting the solution for $F_0$ [\eqref{eq:eqdist}] into \eqref{eq:order1}
and using $C(F_0,F_0)=0$ yields
\begin{equation}
\V{v}_\perp \cdot \nabla \delta f_1 
- \Omega \frac{\partial \delta f_1}{\partial \theta} = -\V{v} \cdot
\nabla \left(\frac{q \phi}{T_0} \right) F_0.
\end{equation}
This inhomogeneous equation for $\delta f_1$ supports a particular
solution and a homogeneous solution.  Noting the particular solution
$\delta f_p = - (q \phi/T_0) F_0 + \Order(\epsilon^2 F_0)$, 
the first-order perturbation is written as $\delta f_1 =
- (q \phi/T_0) F_0 +h$, where the homogeneous solution $h$ satisfies
\begin{equation}
\V{v}_\perp \cdot \nabla  h 
- \Omega  \left(\frac{\partial h}{\partial \theta} \right)_\V{r} \
= \left( \frac{\partial h}{\partial \theta} \right)_\V{R}= 0,
\end{equation}
where we have transformed the $\theta$ derivative at constant position
$\V{r}$ to one at constant guiding center $\V{R}$.  Thus $h$ is
independent of the gyrophase angle at constant guiding center $\V{R}$
(but not at constant position $\V{r}$): $h=h(\V{R}, v, v_\perp, t)$. 
Therefore, the complete solution for the distribution function, 
after taking $1-q\phi / T_0 = \exp ( - q \phi / T_0) + \Order(\epsilon^2)$ 
and absorbing $\Order(\epsilon^2)$ terms into $\delta f_2$, is
\begin{equation}
f=F_0(v,\epsilon^2 t) \exp \left[-\frac{q \phi(\V{r},t)}{T_0}\right] 
+ h(\V{R},v, v_\perp, t) + \delta f_2 + \cdots
\label{eq:distfunc}
\end{equation}
The first term in the solution is the equilibrium distribution
function corrected by the Boltzmann factor.  Physically this arises
from the rapid (compared to the evolution of $\phi$) motion of
electrons along and ions across the field lines attempting to set up a
thermal equilibrium distribution.  However the motion of the particles
across the field is constrained by the gyration and the particles are
not (entirely) free to set up thermal equilibrium.  The second term is
the gyrokinetic distribution function that represents the response of
the rings to the perturbed fields.

\subsubsection{First Order, $\Order(\epsilon)$}

Plugging in the form of solution given in
\eqref{eq:distfunc} and transforming into guiding-center spatial
coordinates and velocity coordinates $(v, v_\perp,\theta)$, the
Fokker--Planck equation to this order becomes
\begin{equation}
\frac{\partial h}{\partial t} +\frac{d \V{R}}{d t} \cdot
\frac{\partial h}{\partial \V{R}} + \frac{q}{m}\left( -\nabla_\perp \phi + 
\frac{\V{v} \times \delta \V{B}}{c}\right) \cdot \left( \frac{\V{v}}{v} 
\frac{\partial h}{\partial v} 
+\frac{ \V{v}_\perp}{v_\perp} \frac{\partial h}{\partial v_\perp} 
\right) - C(h,F_0) - C(F_0,h) = 
\Omega \left(\frac{\partial \delta f_2}{\partial \theta} 
\right)_{\V{R}}
+ \frac{q}{T_0} \left( \frac{\partial \phi}{\partial t} -\frac{\V{v}}{c} \cdot
\frac{\partial \V{A}}{\partial t} \right) F_0,
\label{eq:order2}
\end{equation}
where 
\begin{equation}
\frac{d \V{R}}{dt} = v_\parallel\zhat + 
\frac{c}{B_0}\left(-\nabla\phi - \frac{1}{c}\frac{\partial\V{A}}{\partial t} 
+ \frac{\V{v}\times\delta\V{B}}{c}\right)\times\zhat.
\label{eq:vR_def}
\end{equation}
Note that the linearized collision operator $C(h,F_0) + C(F_0,h)$
involves $h$ and $F_0$ both of electrons and of ions.

To eliminate $\delta f_2$ from \eqref{eq:order2}, we ring average the
equation over $\theta$ at fixed guiding center $\V{R}$, taking
advantage of the fact that $\delta f_2$ must be periodic in $\theta$. 
The ring averaging also eliminates the third term on the left-hand  
side. Indeed, for an arbitrary function $a(\V{r})$, 
\begin{equation}
\left\langle\V{v}_\perp\cdot\nabla a\right\rangle_{\V{R}} = 
- \Omega \left\langle
\left(\V{v}\times\zhat\right)\cdot\left(\frac{\partial}{\partial\V{v}} 
\frac{\V{v}\times\zhat}{\Omega}\right)\cdot\nabla a
\right\rangle_{\V{R}} = 
\Omega \left\langle
\left(\V{v}\times\zhat\right)\cdot\left(\frac{\partial\V{r}}{\partial\V{v}}\right)_\V{R}
\cdot\nabla a \right\rangle_{\V{R}} = 
\Omega \left\langle
\left(\V{v}\times\zhat\right)\cdot\left(\frac{\partial a}{\partial\V{v}}\right)_\V{R}
\right\rangle_{\V{R}} =
- \Omega \left\langle \left(\frac{\partial a}{\partial\theta}\right)_\V{R}
\right\rangle_{\V{R}} = 0, 
\end{equation}
whence $\langle\V{v}\cdot\nabla_\perp\phi\rangle_\V{R} = 0$
and $\langle\V{v}_\perp\cdot(\V{v}\times\delta\V{B})\rangle_\V{R} 
= v_\parallel\langle\V{v}_\perp\cdot(\zhat\times\delta\V{B})\rangle_\V{R} 
= v_\parallel\langle\V{v}_\perp\cdot\nabla_\perp A_\parallel\rangle_\V{R} = 0$ 
[see \eqref{eq:bdef2}]. Thus, the ring-averaged \eqref{eq:order2} takes the form
\begin{equation}
\frac{\partial h}{\partial t} 
+\left\langle \frac{d \V{R}}{d t} \right\rangle_\V{R} \cdot
\frac{\partial h}{\partial \V{R}} 
- \left(\frac{\partial h}{\partial t}\right)_{\rm coll} = 
 \frac{q}{T_0} \frac{\partial \langle \chi \rangle_\V{R}}{\partial t} F_0,
\label{eq:gkprelim}
\end{equation}
where we have defined the gyrokinetic collision operator 
$({\partial h/\partial t})_{\rm coll} =
\langle C(h,F_0) + C(F_0,h)\rangle_{\V{R}}$ 
and the gyrokinetic potential $\chi=\phi-\V{v}\cdot\V{A}/c$.
Keeping only first-order contrubutions in \eqref{eq:vR_def}
and substituting for $\delta\V{B}$ from \eqref{eq:bdef2}, we find that 
the ring-averaged guiding center motion is given by
\begin{equation}
\left\langle \frac{d \V{R}}{d t} \right\rangle_\V{R} 
= v_\parallel \zhat  
- \frac{c}{B_0} \left\langle\nabla_\perp\phi\right\rangle_\V{R}\times\zhat 
+ \frac{v_\parallel}{B_0} \left\langle\nabla_\perp A_\parallel\right\rangle_\V{R}\times\zhat 
- \frac{1}{B_0}\left\langle\V{v}_\perp\delta B_\parallel\right\rangle_\V{R} 
= v_\parallel \zhat  
- \frac{c}{B_0}\frac{\partial \langle \chi \rangle_{\V{R}}}{\partial \V{R}} \times
\zhat,
\label{eq:Rmotion}
\end{equation}
where we have used the identity 
$\left\langle\V{v}_\perp\delta B_\parallel\right\rangle_\V{R}
= -\left\langle\nabla_\perp(\V{v}_\perp\cdot\V{A}_\perp)\right\rangle_\V{R}$. 
Substituting \eqref{eq:Rmotion} into \eqref{eq:gkprelim}, we obtain the
{\em gyrokinetic equation}:
\begin{equation}
\frac{\partial h}{\partial t} 
+v_\parallel \zhat  \cdot \frac{\partial h}{\partial \V{R}} 
+ \frac{c}{B_0} \left[ \langle \chi \rangle_\V{R} ,h \right]
- \left(\frac{\partial h}{\partial t}\right)_{\rm coll} = 
 \frac{q}{T_0} \frac{\partial \langle \chi \rangle_\V{R}}{\partial t} F_0,
\label{eq:gkequation}
\end{equation}
where the nonlinear effects enter via the Poisson bracket, defined by
\begin{equation}
\left[ \langle \chi \rangle_{\V{R}} ,h \right]= 
\left(\frac{\partial \langle \chi \rangle_{\V{R}} }{\partial \V{R}} \times
\zhat \right) \cdot \frac{\partial h}{\partial \V{R}} =
\frac{\partial \langle \chi \rangle_{\V{R}} }{\partial X} 
\frac{\partial h}{\partial Y} - 
\frac{\partial \langle \chi \rangle_{\V{R}} }{\partial Y} 
\frac{\partial h}{\partial X} .
\label{eq:poisbrack}
\end{equation}

The gyrokinetic \eqref{eq:gkequation} describes the time evolution of
$h$, the ring distribution function. The second term on the left-hand
side corresponds to the ring motion along $\V{B}_0$, the third term
to the ring motion across $\V{B}_0$, the fourth term to the effect of
collisions.  The source term on the right-hand side is the
ring-averaged change in the energy of the particles.  A more
detailed discussion of the physical aspects of this equation is given in
\secref{sec:gk_eqs}. The equilibrium distribution function $F_0$
changes only on the slow (heating) time scale and is thus formally fixed
(stationary) with respect to the time scale of \eqref{eq:gkequation}. 
The evolution of $F_0$ is calculated in \appref{app:gk_heat}.

%==============================================================================
\subsection{The Gyrokinetic Form of Maxwell's Equations}
To complete the set of gyrokinetic equations, we need to determine 
the electromagnetic field, which is encoded by $\chi$. 
To determine the three unknown scalars 
$\phi$, $A_\parallel$ and $\delta B_\parallel$ [which relates 
to $\V{A}_\perp$ via \eqsref{eq:defpot}{eq:bdef2}], 
we use the quasineutrality condition, \eqref{eq:quasi0}, 
and Amp\`ere's law, \eqref{eq:amp}, taken at $\Order(\epsilon)$ in the
gyrokinetic ordering.

%===============================================
\subsubsection{The Quasineutrality Condition}
The charge density needed in the quasineutrality condition,
\eqref{eq:quasi0}, can be determined by multiplying the distribution
function, \eqref{eq:distfunc}, expanded to first order $\Order(\epsilon)$, by
the charge $q_s$ and integrating over velocities.  Expanding the
exponential in the Boltzmann term and dropping terms of order
$\Order(\epsilon^2)$ and higher gives
\begin{equation}
\sum_s \left[- \frac{q_s^2 n_{0s}}{T_{0s}} \phi +  q_s
\int d^3\V{v} h_s\left(\V{r}+ \frac{\V{v} \times \zhat}{\Omega_s},\V{v},t\right) 
 \right]= 0.
\end{equation}
Note that the velocity integral must be
performed at fixed position $\V{r}$, because the charge must be
determined at fixed position $\V{r}$, not at fixed guiding center
$\V{R}$. Using the ring average at constant $\V{r}$ [\eqref{eq:avgr_def}], 
the quasineutrality condition can be written 
in the following form
\begin{equation}
\sum_s \left( - \frac{q_s^2 n_{0s}}{T_{0s}} \phi + q_s \int d^3 \V{v}
 \langle h_s \rangle_\V{r} \right)= 0.
\label{eq:ap_quasi}
\end{equation}

%===============================================
\subsubsection{The Parallel Amp\`ere's Law}
\label{sec:paramp}
The current density is calculated by multiplying the distribution function,
\eqref{eq:distfunc}, expanded to first order $\Order(\epsilon)$, 
by $q_s \V{v}$ and integrating over velocities. 
The Boltzmann part of the current is odd with respect to $v_\parallel$ 
and vanishes upon integration. 
The parallel component of Amp\`ere's law, \eqref{eq:amp}, is, therefore 
\begin{equation}
-\nabla_\perp^2 A_\parallel = \frac{ 4 \pi}{c} \delta j_\parallel  
= \sum_s \frac{ 4 \pi}{c} q_s
\int d^3 \V{v}  v_\parallel \langle h_s \rangle_\V{r},
\label{eq:ap_amp_par}
\end{equation}
where the ring average at fixed position $\V{r}$ appears in the same fashion 
as in \eqref{eq:ap_quasi}. 

%===============================================
\subsubsection{The Perpendicular Amp\`ere's Law}
The perpendicular component of Amp\`ere's law, \eqref{eq:amp}, is derived
in an analogous manner as the parallel component in \secref{sec:paramp}: 
Amp\`ere's law is crossed with $\zhat$, the
Boltzmann contribution vanishes upon integration over gyrophase angle
$\theta$, and a ring average at fixed position $\V{r}$ is performed.
The result is 
\begin{equation}
\nabla_\perp \delta B_\parallel = \frac{ 4 \pi}{c} \zhat \times \delta \V{j}
= \sum_s \frac{ 4 \pi}{c} q_s
\int d^3 \V{v} \langle\zhat \times \V{v}_\perp h_s \rangle_\V{r}.
\label{eq:ap_amp_perp}
\end{equation}
It is straightforward to show that \eqref{eq:ap_amp_perp} is the
gyrokinetic version of perpendicular pressure balance (no fast
magnetosonic waves). Integration by parts yields:
\begin{equation}
\nabla_\perp \frac{B_0\delta B_\parallel}{4\pi}
= - \nabla_\perp \cdot \V{\delta P}_\perp,
\label{eq:amp_perp}
\end{equation}
where the perpendicular pressure tensor is
\begin{equation}
 \V{\delta P}_\perp = \sum_s m_s 
\int d^3 \V{v} \left\langle\V{v}_\perp\V{v}_\perp h_s \right\rangle_\V{r}.
\label{eq:amp_perp2}
\end{equation}
%Note that in gyrokinetics the pressure tensor is not diagonal.  

In order to drive steady-state (non-decaying) turbulence, we introduce
additional externally driven antenna current $\V{j}_a$ to the right
hand sides of \eqref{eq:ap_amp_par} and \eqref{eq:ap_amp_perp}.

%==============================================================================
\section{Derivation of the Heating Equation}
\label{app:gk_heat}
While the gyrokinetic \eqref{eq:gkequation} determines the evolution
of the perturbation to the distribution function on the intermediate
time scale, we must go to second order, $\Order(\epsilon^2\omega)$, in
the gyrokinetic ordering to obtain the slow evolution of the
equilibrium distribution function $F_0$. This Appendix contains two
derivations of the heating \eqref{eq:heat}: the first is more
conventional, but longer (\S\ref{app:gk_heat_conv}); the second
employs entropy conservation and is, in a sense, more intuitive and
fundamental (\S\ref{app:entropy}).  We also discuss energy
conservation in driven systems and derive the power-balance
\eqref{eq:pb} (see \S\ref{app:driven}).

%===============================================
\subsection{Conventional Derivation of the Heating Equation}
\label{app:gk_heat_conv}

We begin by defining the medium-time average over a period $\Delta t$ 
long compared to the fluctuation time scale but short compared to
the heating time scale, $ 1/\omega \ll \Delta t \ll 1/\epsilon^2\omega$,
\begin{equation}
\overline{a}(t)=\frac{1}{\Delta t}\int^{t+\Delta t/2}_{t-\Delta t/2}dt' a(t').
\label{eq:timeave}
\end{equation}
The equilibrium distribution function $F_0$ is constant at times $\sim\Delta t$, 
so $\overline{F_0}=F_0$.

To determine the evolution of the equilibrium density and temperature
of a species $s$ on the heating (transport) time scale, we consider the
full (not ring averaged) Fokker--Planck \eqref{eq:fpeq}. 
To demonstrate that particle conservation implies 
that $n_{0s}$ is a constant, we 
integrate \eqref{eq:fpeq} over {\em all} space and velocity,
divide by system volume $V$, and discard all terms of order
$\Order(\epsilon^3)$ and higher: 
\begin{equation}
\int \frac{d^3\V{r}}{V} \int d^3\V{v} \frac{\partial f_s}{\partial t}= 
 \frac{d n_{0s}}{d t} + 
\frac{d}{d t}
\int \frac{d^3\V{r}}{V} \int d^3\V{v} \delta f_{2s}= 0.
\label{eq:fpeq2}
\end{equation}
Here we have used the conservation of particles by the collision
operator, the fact that the first-order perturbations spatially
average to zero, and an integration by parts over velocity 
to simplify the result. Performing
the medium-time average [\eqref{eq:timeave}] eliminates the
$\delta f_{2s}$ term, leaving
\begin{equation}
\frac{d  n_{0s}}{d t}= 0.
\end{equation}
Thus, for both species $s$, the density $n_{0s}$ is constant on the
heating (transport) time-scale.

The evolution of the temperature $T_{0s}$ is calculated similarly by multiplying
\eqref{eq:fpeq} by $m_s v^2/2$, integrating over all space
and velocity, and dividing by the system volume $V$.  Using 
the expansion of the distribution function, integration by parts in
velocity, we get
\begin{equation}
\frac{3}{2} n_{0s} \frac{d  T_{0s}}{d t} +
\frac{d}{d t}\int \frac{d^3\V{r}}{V} \int d^3\V{v} 
\frac{m_s v^2}{2} 
\delta f_{2s} = \int \frac{d^3\V{r}}{V} \int d^3\V{v} q_s (\V{v} \cdot \V{E}) f_s +
\int \frac{d^3\V{r}}{V} \int d^3\V{v} \frac{m_s v^2}{2} C_{sr}(f_s,f_r).
\label{eq:efp2}
\end{equation}
We have again assumed $\int d^3\V{r}\delta f_{1s}=0$ 
(first-order perturbations spatially average to zero). 
The first term on the right-hand side is the work done on the
particles by the fields, the second is the collisional energy
exchange between species. Note that collisions between like particles
do not produce a net loss of energy for a species and thus do not
appear in this expression (the operator $C_{ss}$ integrates to zero). 
At this order, the collisions between
species occur only between the Maxwellian equilibria of each
interacting species; the standard form of this collisional energy
exchange \citep[see, e.g.,][]{hel02} is given by
\begin{equation}
\int \frac{d^3\V{r}}{V} \int d^3\V{v} \frac{m_s v^2}{2} C_{sr}(f_s,f_r)=
n_{0s} \nu_{\E}^{sr} (T_{0r}-T_{0s}).
\label{eq:efp3}
\end{equation}
Expressions for the interspecies collision rate $\nu_{\E}^{sr}$ can be found in 
\citet{hel02}. The collisional energy exchange rate of ions on
electrons, $\nu_{\E}^{ie}$, is a factor $(m_e/m_i)^{1/2}$ times smaller than 
the ion-ion collision rate. It is, therefore, possible to sustain a temperature
difference between ions and electrons even though the plasma is
collisional enough to make $F_0$ Maxwellian for each species.

Splitting the electric field into potentials, 
$\V{E}=-\nabla \phi  - (1/c)\partial \V{A}/\partial t$, 
we can manipulate the scalar
potential part into a more useful higher order form. After some algebra and 
using \eqref{eq:efp3}, we find 
\begin{eqnarray}
\frac{3}{2} n_{0s}  \frac{d  T_{0s}}{d t}   + 
\frac{d }{d t}\int \frac{d^3\V{r}}{V} \int d^3\V{v} 
\left(\frac{m_s v^2}{2} \delta f_{2s} + q_s\phi \delta f_s\right) 
= \int \frac{d^3\V{r}}{V} \int d^3\V{v} q_s \frac{\partial}{\partial t}
\left(\phi - \frac{\V{v} \cdot \V{A}}{c}\right) \delta f_s 
 +  n_{0s} \nu_{\E}^{sr} (T_{0r}-T_{0s}),
\label{eq:efp6}
\end{eqnarray}
where $\delta f_s$ is the entire perturbed part of the distribution function. 
From this equation, we can check the order of the heating rate,
\begin{equation}
\frac{3}{2} n_{0s} \frac{d  T_{0s}}{d t} \sim
\Order(\epsilon^2 \omega T_{0s}).
\label{eq:heatorder}
\end{equation}
We see that the variation in the equilibrium quantities is, as
expected, on a time scale that is $\epsilon^2$ slower than the time
scale of the variations in the fluctuating quantities. Note that this
ordering is consistent with all the energy in an Alfv\'en wave cascade
becoming heat in a single cascade time at the driving scale.

Splitting the perturbed distribution function $\delta f_s$ 
into the Boltzmann and gyrokinetic parts [\eqref{eq:distfunc}], 
we obtain the following instantaneous form of the \emph{heating equation}
\begin{eqnarray}
\frac{3}{2} n_{0s} \frac{d  T_{0s}}{d t} +
\frac{d }{d t} \left[ \int \frac{d^3\V{r}}{V} \int d^3\V{v} 
\left( \frac{m_s v^2}{2} \delta f_{2s} + q_s \phi h_s \right) - 
\int \frac{d^3\V{r}}{V}  \frac{n_{0s}q_s^2 \phi^2}{2 T_{0s}}\right] 
= \int \frac{d^3\V{r}}{V} \int d^3\V{v} q_s 
\frac{\partial  \chi }{\partial t}
h_s  + n_{0s} \nu_{\E}^{sr} (T_{0r}-T_{0s}).
\label{eq:gkh}
\end{eqnarray}
To obtain the heating \eqref{eq:heat}, we take the 
medium-time average, defined by \eqref{eq:timeave}, 
of the above equation. The average of the
second term on left-hand side is zero; it does not contribute to the
average heating. Thus, we do not require the second-order perturbed 
distribution function $\delta f_{2s}$ to calculate the heating. 
The average of the first term on the right-hand side of \eqref{eq:gkh} 
is the desired heating term that relates the slow-time-scale 
evolution of the equilibrium to the solution of the the gyrokinetic equation. 
The $\V{r}$ integral is converted into the $\V{R}_s$ integral by noticing 
that $\int d^3\V{r}\int d^3\V{v} = \int d^3\V{v}\int d^3\V{R}_s$
(the velocity integration on the left is at constant $\V{r}$, 
while on the right it is at constant $\V{R}_s$). 
Therefore, 
\begin{equation}
\int \frac{d^3\V{r}}{V} \int d^3\V{v} 
q_s \frac{\partial\chi(\V{r})}{\partial t} h_s(\V{R}_s)  
= \int d^3\V{v} \int \frac{d^3\V{R}_s}{V} 
q_s\left[\frac{\partial}{\partial t}\chi\left(\V{R}_s-\frac{\V{v}\times\zhat}{\Omega_s}\right)\right]
h_s(\V{R}_s)  
= \int d^3\V{v} \int \frac{d^3\V{R}_s}{V} 
q_s \frac{\partial \langle \chi \rangle_{\V{R}_s}}{\partial t} h_s.
\label{eq:gkheat2}
\end{equation}

We must now demonstrate that the heating is ultimately collisional 
[the second equality in \eqref{eq:heat}]. 
To make this connection, we multiply the gyrokinetic \eqref{eq:gkequation}
by $T_{0s}h_s/F_{0s}$ and integrate over space (i.e., with respect to $\V{R}_s$)
and velocity to obtain the following equation 
\begin{equation}
\frac{d}{d t} \int d^3\V{v}\int \frac{d^3\V{R}_s}{V} 
\frac{T_{0s}}{2F_{0s}} h_s^2 
- \int d^3\V{v}\int \frac{d^3\V{R}_s}{V} 
\frac{T_{0s}}{F_{0s}} h_s \left(\frac{\partial h_s}{\partial t}\right)_{\rm coll} = 
\int d^3\V{v}\int \frac{d^3\V{R}_s}{V} 
q_s \frac{\partial \langle \chi \rangle_{\V{R}_s}}{\partial t} h_s
\label{eq:gkheat}
\end{equation}
(for reasons that will become apparent in \appref{app:entropy}, 
this will be referred to as the \emph{entropy-balance equation}).
Combining \eqsref{eq:gkh}{eq:gkheat} using
\eqref{eq:gkheat2} then gives the collisional form 
of the instantaneous heating equation:
\begin{eqnarray}
\frac{3}{2} n_{0s} \frac{d  T_{0s}}{d t} & + &
\frac{d }{d t} \left[ \int \frac{d^3\V{r}}{V} \int d^3\V{v} 
\left( \frac{m_s v^2}{2} \delta f_{2s} + q_s \phi \langle h_s\rangle_{\V{r}} 
- \frac{T_{0s}}{2F_{0s}} \langle h_s^2\rangle_{\V{r}} \right) - 
\int \frac{d^3\V{r}}{V}  \frac{n_{0s}q_s^2 \phi^2}{2 T_{0s}}\right] \nonumber \\
&=& - \int \frac{d^3\V{r}}{V} \int d^3\V{v} \frac{T_{0s}}{F_{0s}} 
\left\langle h_s \left(\frac{\partial h_s}{\partial t}\right)_{\rm coll} \right\rangle_{\V{r}}
+ n_{0s} \nu_{\E}^{sr} (T_{0r}-T_{0s}),
\label{eq:heating_instant}
\end{eqnarray}
where we have used $\int d^3\V{r}\int d^3\V{v} a(\V{R}_s) = \int
d^3\V{r}\int d^3\V{v} \langle a\rangle_{\V{r}}$ (this manipulation is
done purely for notational cleanliness: the expressions under the
integrals are now explicitly functions of $\V{r}$ and $\V{v}$, not of
$\V{R}_s$).  Under medium-time averaging, the second term on the left
hand side of \eqref{eq:heating_instant} again vanishes and the second
equality in \eqref{eq:heat} is obtained:
\begin{equation}
\frac{3}{2} n_{0s}  \frac{d  T_{0s}}{d t} = 
- \int \frac{d^3\V{r}}{V} \int d^3\V{v} \frac{T_{0s}}{F_{0s}} 
\overline{\left\langle h_s 
\left(\frac{\partial h_s}{\partial t}\right)_{\rm coll}\right\rangle_{\V{r}}}
+ n_{0s} \nu_{\E}^{sr} (T_{0r}-T_{0s}).
\label{eq:heating}
\end{equation}
The term that has averaged out 
does not contribute to the net (slow-time-scale) heating because 
it represents the sloshing of energy back and forth 
between particles and fields (on the fluctuation time scale). 
On the right-hand side, the collisional term is negative definite for like-particle 
and pitch-angle collision operators [these are the only relevant cases:
for $s=i$, the ion-ion collisions dominate; for $s=e$, the dominant
terms are electron-electron collisions and the pitch-angle scattering
of the electrons off the ions; all other parts of the collision
operator are subdominant by at least one factor of
$(m_e/m_i)^{1/2}$]. 
 
The \eqref{eq:heating} for the heating on the slow time scale 
is sign-definite, so it is, in practice, easier to
average numerically than the instantaneous heating 
\eqref{eq:gkh}: unlike in 
\eqref{eq:gkh}, calculating the average heating does not require 
precisely capturing the effect of near-cancellation of the
intermediate-time-scale oscillations of the instantaneous energy transfer 
between particles and waves. The net heating is always 
collisional, regardless of the collision rate---when collisions are small, 
$h_s$ develops small
scales in velocity space, typically $\Delta {\bf v}\sim
\Order(\nu^{1/2})$, so that the heating is independent of
the collision rate $\nu$. As we shall see below, \eqref{eq:heating}
relates heating to the collisional entropy production. 

%===============================================
\subsection{Entropy Argument to Derive the Heating Equation}
\label{app:entropy}

Ignoring the ion-electron collisions, whose rate is 
$(m_e/m_i)^{1/2}$ times smaller than that of the ion-ion collisions, 
Boltzmann's $H$ Theorem gives the time evolution 
of the entropy of the ions $S_i$ as follows 
\begin{eqnarray}
\frac{dS_i}{dt} = 
-\frac{d}{dt}\int\frac{d^{3}{\bf r}}{V}\int d^{3}{\bf v}  f_i\ln f_i 
= -\int\frac{d^{3}{\bf r}}{V}\int d^{3}{\bf v} \ln f_i C_{ii}(f_i,f_i).
\label{eq:htheorem}
\end{eqnarray}
It can be easily shown that the right-hand side is 
non-negative \citep[see, e.g.,][]{lif81} and, therefore, that entropy always increases.
It can also be shown that the entropy increase is zero if, and only
if, the distribution function is a Maxwellian.  Expanding $f_i$ 
about the Maxwellian $F_{0i}$, we obtain, to order $\Order(\epsilon^2)$,
\begin{eqnarray}
\frac{dS_i}{dt} =  
-\frac{d}{dt}\int\frac{d^{3}{\bf r}}{V}\int d^{3}{\bf v}  
\left[F_{0i}\ln F_{0i} + (1 + \ln F_{0i})\delta f_{2i} 
+ \frac{\delta f_{1i}^2}{2F_{0i}}\right] 
= -\int\frac{d^{3}{\bf r}}{V}\int d^{3}{\bf v} 
\frac{\delta f_{1i}}{F_{0i}} 
\left(\frac{\partial\delta f_{1i}}{\partial t}\right)_{\rm coll},
\label{eq:htheorem2}
\end{eqnarray}
where we have made use of the energy conservation properties of
ion-ion collisions and of the fact that $\delta f_{1i}$ 
spatially averages to zero. 
Evaluating the the zeroth-order (Maxwellian) part of the 
integral on the left-hand side and splitting $\delta f_{1i}$ 
into the Boltzmann and gyrokinetic parts [\eqref{eq:distfunc}], 
we obtain the slow evolution of temperature
\begin{eqnarray}
\frac{3}{2} n_{0i} \frac{1}{T_{0i}} 
\frac{d  T_{0i}}{d t} + 
\frac{d }{d t} \left[ \int \frac{d^3\V{r}}{V} \int d^3\V{v} 
\left(\frac{m_i v^2}{2T_{0i}} \delta f_{2i} 
+ \frac{q_i \phi}{T_{0i}} h_i 
- \frac{ h_i^2}{2F_{0i}} \right) - 
\int \frac{d^3\V{r}}{V}  \frac{n_{0i}q_i^2 \phi^2}{2 T_{0i}^2}\right] 
= - \int \frac{d^3\V{r}}{V} \int d^3\V{v} \frac{1}{F_{0i}} 
\left\langle h_i \left(\frac{\partial h_i}{\partial t}\right)_{\rm coll}
\right\rangle_{\V{r}}.
\label{eq:heating_entropy}
\end{eqnarray}
This result is the same as \eqref{eq:heating_instant}; the heating is
now manifestly expressed as the irreversible entropy production. Under 
medium-time average, the second term on the left-hand side of 
\eqref{eq:heating_entropy} again vanishes, so 
the heating \eqref{eq:heating} is recovered.

%===============================================
\subsection{Energy Conservation in Driven Systems}
\label{app:driven}

To determine an equation for the conservation of energy in
gyrokinetics, we use Poynting's theorem
\begin{equation}
\frac{d }{d t}  \int d^3\V{r} \left( \frac{E^2}{8 \pi}+
\frac{B^2}{8 \pi} \right) + \frac{c}{4 \pi} \oint d\V{S} \cdot( \V{E} \times \V{B}) 
= - \int d^3\V{r} \left( \V{j} + \V{j}_a \right) \cdot \V{E},
\end{equation}
where $\V{j}_a$ is the current in the antenna driving the system 
and $\V{j}$ is the plasma current. 
We shall drop the surface term (the Poynting flux)
--- this is justified, e.g., in a numerical box 
with periodic boundary conditions. 

From \secref{sec:ordering}, we know that 
$|\delta \V{E}|^2 \sim \Order (\epsilon^2 B_0^2v_{th_i}^2/c^2)$ 
and $|\delta \V{B}|^2\sim \Order (\epsilon^2 B_0^2)$ [\eqref{eq:order_fluct}].  
Thus, in the non-relativistic limit,
the magnetic energy dominates and we may neglect the electric field
energy (this is consistent with neglecting the displacement current in
the non-relativistic ordering). We are left with
\begin{equation}
\frac{d }{d t}  \int d^3\V{r} \frac{|\delta\V{B}|^2}{8 \pi} 
= - \int d^3\V{r} \left( \V{j} + \V{j}_a \right) \cdot \V{E}.
\label{eq:jdote}
\end{equation}
Under medium-time averaging, the left-hand side vanishes and we 
are left with the steady-state balance: 
\begin{equation}
\int d^3\V{r}\, \overline{\left( \V{j} + \V{j}_a \right) \cdot \V{E}} = 0.
\end{equation}

On the other hand, using \eqsref{eq:efp2}{eq:heating_instant} to calculate 
$\int d^3\V{r}(\V{j} \cdot \V{E}) = \int d^3\V{r}\sum_s\int d^3\V{v}q_s(\V{v}\cdot\V{E})f_s$, 
we can convert \eqref{eq:jdote} into the following instantaneous \emph{power-balance equation}
\begin{eqnarray}
\frac{d}{d t}   \int \frac{d^3\V{r}}{V} \left[
\sum_s \int d^3\V{v} \frac{T_{0s}}{2F_{0s}} 
\left( h_s - \frac{q_s \phi}{T_{0s}}F_{0s} \right)^2  
+ \frac{|\delta\V{B}|^2}{8 \pi} 
\right] 
= \sum_s \int \frac{d^3\V{r}}{V} \int d^3\V{v} \frac{T_{0s}}{F_{0s}} 
\left\langle h_s \left(\frac{\partial h_s}{\partial t}\right)_{\rm coll}\right\rangle_{\V{r}}
- \int \frac{d^3\V{r}}{V} \V{j}_a  \cdot \V{E}.
\label{eq:energy_entropy}
\end{eqnarray}
The first term on the right-hand side is the nonnegative-definite 
collisional entropy production [see \eqref{eq:heating_entropy}], 
the second term is the external energy input. 
The left-hand side is the time derivative of the fluctuation energy 
(kinetic plus magnetic): 
\begin{equation}
\delta W = \int \frac{d^3\V{r}}{V} 
\left[\sum_s \int d^3\V{v} \frac{T_{0s}\delta f_s^2}{2F_{0s}}
+ \frac{|\delta\V{B}|^2}{8 \pi} \right].
\end{equation}
Note that in the large-scale limit, appropriate for MHD, we have 
Alfv\'en waves, for which $h_s \simeq q_s\left\langle \phi \right\rangle_{\V{R}_s} /T_{0s}$ 
--- the kinetic energy then becomes the $\V{E} \times \V{B}$ velocity squared as it should be 
in MHD. 

If we medium-time average \eqref{eq:energy_entropy}, the left-hand
side vanishes and we have the power balance between external energy
injection and collisional dissipation:
\begin{equation}
\sum_s \int \frac{d^3\V{r}}{V} \int d^3\V{v} \frac{T_{0s}}{F_{0s}} 
\overline{\left\langle h_s \left(\frac{\partial h_s}{\partial t}\right)_{\rm coll}\right\rangle_{\V{r}}}
= \int \frac{d^3\V{r}}{V} \overline{\V{j}_a  \cdot \V{E}}.
\end{equation}
%Using \eqref{eq:heating}, we can rewrite this power balance as 
%\begin{equation}
%\sum_s \frac{3}{2} n_{0s}  \frac{d  T_{0s}}{d t}
%= -\int \frac{d^3\V{r}}{V} \overline{\V{j}_a  \cdot \V{E}},
%\end{equation}
%where we have omitted the interspecies collisions. 

%==============================================================================
\section{Derivation of the Linear Collisionless Dispersion Relation} 
\label{app:disprel}
The dispersion relation for a linear, collisionless gyrokinetic system
is derived beginning with the 
linearized collisionless 
version of the gyrokinetic \eqref{eq:gkequation1}, 
\begin{equation}
\frac{\partial h_s}{\partial t} 
+v_\parallel  \frac{\partial h_s}{\partial z} =
\frac{q_s}{T_{0s}} F_{0s} 
\frac{\partial \langle \chi \rangle_{\V{R}_s}}{\partial t},
\label{eq:gkequation_lin}
\end{equation}
and the field equations (\ref{eq:ap_quasi}),
(\ref{eq:ap_amp_par}), and (\ref{eq:ap_amp_perp}). First, 
the electromagnetic fields and the gyrokinetic distribution function
are expanded in plane waves, allowing the ring averages appearing in the
equations to be written as multiplications by Bessel functions. The
gyrokinetic equation is then solved algebraically for the distribution
function.  Next, this solution is
substituted into equations~(\ref{eq:ap_quasi}), (\ref{eq:ap_amp_par}),
and (\ref{eq:ap_amp_perp}), and the integration over velocity is
performed using the plasma dispersion function to simplify the
parallel velocity integrals and modified Bessel functions to express
the perpendicular velocity integrals.  The condition for the existence
of a solution to the resulting set of algebraic equations 
is the dispersion relation. In this Appendix,
the plasma species subscript $s$ is suppressed when unnecessary.

%==============================================================================
\subsection{Solving for the Distribution Function}
First, we decompose the electromagnetic potentials into plane wave
solutions of the form $a(\V{r},t)= \hat{a}\exp[ i(\V{k} \cdot
\V{r}-\omega t)]$ (where $a$ denotes $\chi$, $\phi$, or $\V{A}$) and
the gyrokinetic distribution function into solutions of the form
$h(\V{R},v,v_\perp,t)= \hat{h}\exp [i(\V{k} \cdot \V{R}-\omega
t)]$.  To solve \eqref{eq:gkequation_lin} for $\hat{h}$, we need to
express $\langle \chi \rangle_{\V{R}}$ in algebraic terms.  Under the
plane-wave decomposition, ring averages reduce to multiplications by
Bessel functions.  For example, the terms with the scalar potential
$\phi$ in the definition of $\chi=\phi - \V{v}\cdot\V{A}/c$, yield
\begin{equation}
\langle \phi(\V{r},t) \rangle_\V{R}= \hat{\phi} 
e^{i(\V{k} \cdot \V{R}-\omega t)}
\frac{1}{2 \pi} \oint d \theta \exp\left(i\frac{k_\perp v_\perp}{\Omega} \cos \theta\right)
=J_0\left(\frac{k_\perp v_\perp}{\Omega}\right) \hat{\phi} 
e^{i(\V{k} \cdot \V{R}-\omega t)}.
\label{eq:phibessguid}
\end{equation}
Here we have used the definition of the zeroth-order Bessel function
\citep{abr72} and the relation between the position and guiding center,
\eqref{eq:posguid}, noting that $\V{k} \cdot [\zhat\times\V{v}/\Omega] = 
(k_\perp v_\perp/\Omega) \cos \theta$, where $\theta$ is the angle 
between $\V{k}_\perp$ and the particle's instantaneous 
Larmor radius $\zhat\times\V{v}/\Omega$.  After further
algebraic manipulations of this kind, the ring averaged potential
$\langle \chi \rangle_{\V{R}}$ can be written in terms of zeroth- and
first-order Bessel functions,
\begin{equation}
\langle \chi \rangle_\V{R}= \left[ J_0\left(\frac{k_\perp v_\perp}{\Omega}\right) 
\left(\hat{\phi} - \frac{v_\parallel \hat{A}_\parallel }{c}\right) +
\frac{J_1\left({k_\perp v_\perp/\Omega}\right)} {{k_\perp
v_\perp/\Omega}}
\frac{m v^2_\perp}{q} \frac{\delta \hat{B}_\parallel}{B_0} \right]
 e^{i(\V{k} \cdot \V{R}-\omega t)},
\label{eq:gkpotential}
\end{equation}
where we have used the definition  $\delta \hat{B}_\parallel = i(
\mathbf{k}_\perp \times \hat{\V{A}}_\perp ) \cdot \zhat$.
Similarly, the ring average of the distribution function at constant
$\V{r}$ becomes
\begin{equation}
\langle h(\V{R},v,v_\perp,t) \rangle_\V{r}= 
J_0\left(\frac{k_\perp v_\perp}{\Omega}\right) 
\hat{h} e^{i(\V{k} \cdot \V{r}-\omega t)}.
\label{eq:gbesspos}
\end{equation}

The linearized gyrokinetic equation can now be solved for the
distribution function:
\begin{equation}
\hat{h}= \frac{q F_{0}}{T_0} 
\frac{\omega}{\omega - k_\parallel v_\parallel} 
\langle \hat{\chi} \rangle_{\V{R}}.
\label{eq:h_sol2}
\end{equation}
Using \eqref{eq:gkpotential}, this can be written in terms of the
potentials as follows
\begin{equation}
\hat{h}= \frac{q F_{0}}{T_0} \left\{ J_0\left(\frac{k_\perp v_\perp}{\Omega}\right) 
\frac{\omega \hat{A}_\parallel}{k_\parallel c} +  
\frac{\omega}{\omega - k_\parallel v_\parallel} \left[
J_0\left(\frac{k_\perp v_\perp}{\Omega}\right) 
\left( \hat{\phi}-\frac{\omega\hat{A}_\parallel}{k_\parallel c} \right) 
+ \frac{J_1\left({k_\perp v_\perp/\Omega}\right)}{{k_\perp v_\perp/\Omega}} 
\frac{2v_\perp^2}{v_{th}^2}\frac{T_0}{q}\frac{\delta \hat{B}_\parallel}{B_0} 
\right] \right\}.
\label{eq:h_solution}
\end{equation}

%==============================================================================
\subsection{Performing the Integration over Velocity}
Because the solution for the distribution function
\eqref{eq:h_solution} is a product of functions of $v_\parallel$ and
$v_\perp$, the integrals over velocity space, $\int
d^3\V{v}=\int_{-\infty}^{\infty}  dv_\parallel \int_0^\infty v_\perp
dv_\perp \int_0^{2\pi} d\theta$, in
\eqsmoreref{eq:ap_quasi}{eq:ap_amp_perp} can be expressed in terms of plasma
dispersion functions and modified Bessel functions. 
%Hence, we can
%derive an algebraic set of equations for the potentials 
%$\hat{\phi}$, $\hat{A}_\parallel$, and
%$\delta \hat{B}_\parallel$ from the quasineutrality condition
%\eqref{eq:ap_quasi}, the parallel Amp\`ere's law \eqref{eq:ap_amp_par}, and the
%perpendicular Amp\`ere's law \eqref{eq:ap_amp_perp}. 

Integrals over $v_\parallel$, when not immediately completed, are
written in terms of the plasma dispersion function \citep{fri61}
\begin{equation}
Z(\xi) = \frac{1}{\sqrt{\pi}} \int_L dx \frac{e^{-x^2}}{x-\xi},
\label{eq:plasmadisp}
\end{equation}
where $\xi = \omega/|k_\parallel| v_{th}$
and the integral is performed over the Landau contour from $-\infty$
to $+\infty$ below the pole at $x=\xi$ in the complex plane.
Using this definition, we can write
\begin{equation}
\frac{1}{\sqrt{\pi}}\int \frac{d v_\parallel}{v_{th}}\,e^{-v_\parallel^2/v_{th}^2}
\frac{\omega}{\omega-k_\parallel v_\parallel} 
= - \xi Z(\xi).
\end{equation}

Integrations over $v_\perp$ can 
be written in terms of modified Bessel functions. Three such integrals arise:
\begin{eqnarray}
\Gamma_0 (\alpha)& =& \int_0^\infty \frac{2 v_\perp d v_\perp}{v_{th}^2} 
\left[J_0\left(\frac{k_\perp v_\perp}{\Omega}\right)\right]^2 e^{- v_\perp^2/v_{th}^2}  
=  I_0(\alpha) e^{-\alpha}, \nonumber \\
\Gamma_1(\alpha) &=& \int_0^\infty \frac{2 v_\perp d v_\perp}{v_{th}^2} 
\frac{2 v_\perp^2}{v_{th}^2} 
\frac{ J_0\left({k_\perp v_\perp/\Omega}\right) 
J_1\left({k_\perp v_\perp/\Omega}\right)}{{k_\perp v_\perp/\Omega}}
e^{- v_\perp^2/v_{th}^2} = [I_0(\alpha)-I_1(\alpha)] e^{-\alpha}, \nonumber \\
\Gamma_2 (\alpha)&=& \int_0^\infty \frac{2 v_\perp d v_\perp}{v_{th}^2} 
\frac{4 v_\perp^4}{v_{th}^4} 
\left[\frac{ J_1\left({k_\perp v_\perp/\Omega}\right)}
{k_\perp v_\perp/\Omega}\right]^2
e^{- v_\perp^2/v_{th}^2} =2 \Gamma_1(\alpha),
\label{eq:gam0def}
\end{eqnarray}
where $I_0$ and $I_1$ are the modified Bessel functions, 
$\alpha =k_\perp^2 \rho^2/2$,
and we have used the relation \citep{wat66},
\begin{equation}
\int^{\infty}_{0} dx\, x J_n(px) J_n(qx) e^{-a^2 x^2} =
\frac{1}{2a^2} I_n\left(\frac{pq}{2a^2}\right)
e^{-(p^2+q^2)/4a^2}. 
\label{eq:bessint}
\end{equation}

%==============================================================================
\subsection{Quasineutrality Condition}
Beginning with the gyrokinetic quasineutrality condition, 
\eqref{eq:ap_quasi}, we  write the ring average of the distribution 
function at constant position $\V{r}$ as a multiplication by a Bessel
function [\eqref{eq:gbesspos}] and substitute $\hat{h}$ from 
\eqref{eq:h_solution} into the velocity integral. This gives
\begin{eqnarray} 
\sum_s\frac{q_s^2n_{0s}}{T_{0s}}\,\hat{\phi} &=& \sum_s 
2 \pi q_s \int_{-\infty}^{\infty} dv_\parallel \int_0^\infty v_\perp dv_\perp 
J_0\left(\frac{k_\perp v_\perp}{\Omega_s}\right) \hat{h}\nonumber\\ 
&=& \sum_s \frac{q_s^2 n_{0s}}{T_{0s}} \left[ 
\Gamma_{0}(\alpha_s)\frac{\omega \hat{A}_\parallel}{k_\parallel c}
-\Gamma_{0}(\alpha_s) \xi_s Z(\xi_s)
\left(\hat{\phi} - \frac{\omega \hat{A}_\parallel}{k_\parallel c} \right)
- \Gamma_{1}(\alpha_s) \xi_s Z(\xi_s) 
\frac{T_{0s}}{q_s} \frac{\delta \hat{B}_\parallel}{B_0}\right],
\label{eq:quaslin}
\end{eqnarray}
where, in performing the velocity integrals, we have used the definitions 
given in the previous section.

%==============================================================================
\subsection{Parallel Amp\`ere's Law}
Following a similar sequence of steps for the parallel Ampr\`ere's law, 
\eqref{eq:ap_amp_par}, we have
\begin{eqnarray}
k_\perp^2 \hat{A}_\parallel &=& \frac{4 \pi}{c}\sum_s 
2 \pi q_s\int_{-\infty}^{\infty} dv_\parallel \int_0^\infty v_\perp dv_\perp 
J_0\left(\frac{k_\perp v_\perp}{\Omega_s}\right) v_\parallel \hat{h}\nonumber\\ 
&=& -\frac{4\pi\omega}{ck_\parallel}\sum_s \frac{q_s^2 n_{0s}}{T_{0s}}
\left[1+\xi_s Z(\xi_s)\right]\left[\Gamma_{0}(\alpha_s) 
\left(\hat{\phi} - \frac{\omega \hat{A}_\parallel}{k_\parallel c} \right)
+ \Gamma_{1}(\alpha_s) \frac{T_{0s}}{q_s} \frac{\delta \hat{B}_\parallel}{B_0}
\right].
\label{eq:paramplin}
\end{eqnarray}

%==============================================================================
\subsection{Perpendicular Amp\`ere's Law}
It is convenient to take the divergence of the perpendicular Amp\`ere's law, 
\eqref{eq:ap_amp_perp}, before processing it in the same way as the two other 
field equations. This gives
\begin{eqnarray}
\frac{\delta \hat{B}_\parallel}{B_0} &=& - \frac{4\pi}{B_0^2}
\sum_s 2\pi T_{0s}\int_{-\infty}^{\infty} dv_\parallel \int_0^\infty v_\perp dv_\perp 
\frac{2 v_\perp^2}{v_{th_s}^2}\frac{ J_1\left({k_\perp v_\perp/\Omega_s}\right)}
{{k_\perp v_\perp/\Omega_s}}\hat{h}
\nonumber\\
&=& - \frac{4\pi}{B_0^2} \sum_s q_s n_{0s} \left[ \Gamma_{1}(\alpha_s)
\frac{\omega \hat{A}_\parallel}{k_\parallel c} 
-\Gamma_{1}(\alpha_s) \xi_s Z(\xi_s) 
\left(\hat{\phi} - \frac{\omega \hat{A}_\parallel}{k_\parallel c} \right)
- 2\Gamma_{1}(\alpha_s) \xi_s Z(\xi_s) 
\frac{T_{0s}}{q_s} \frac{\delta \hat{B}_\parallel}{B_0} \right].
\label{eq:perpamplin}
\end{eqnarray}

%==============================================================================
\subsection{Dispersion Relation}
Before combining the three field equations derived above
to produce the dispersion relation, we
specify a hydrogen plasma, 
allowing us to take $n_{0i}=n_{0e}$
and $q_i=-q_e=e$.  
We now divide \eqref{eq:quaslin} by $q_i^2 n_{0i}/T_{0i}$, 
\eqref{eq:paramplin} by $(4\pi\omega/ck_\parallel)(q_i^2 n_{0i}/T_{0i})$, 
and \eqref{eq:perpamplin} by $(4\pi/B_0^2)q_i n_{0i}$.
Noting two manipulations,
\begin{equation}
\frac{k_\perp^2 k_\parallel^2 c^2}{4 \pi \omega^2}
\frac{T_{0i}}{q_i^2 n_{0i} }= 
\frac{k_\perp^2 \rho_i^2 }{2}  \frac{k_\parallel^2 v_A^2}{\omega^2}
\quad\text{and}\quad 
\frac{B_0^2}{4 \pi n_{0i} q_i} =  \frac{2}{\beta_i} \frac{T_{0i}}{q_i},
\end{equation}
we arrive at an algebraic linear system of equations that
can be written succinctly in matrix form as
\begin{equation}
\left( \begin{array}{ccc}
A & A-B & C \\
A-B & A-B-\alpha_i/\overline{\omega}^2 & C+E \\
C & C+E & D- 2/\beta_i 
\end{array} \right) 
\left( \begin{array}{c}
\hat{\phi} \\
-{\omega \hat{A}_\parallel/k_\parallel c}\\
(T_{0i}/q_i) {\delta \hat{B}_\parallel/B_0}
\end{array} \right)=0,
\label{eq:matrix}
\end{equation}
where $\overline{\omega}=\omega/|k_\parallel|v_A$ and 
the definitions of the coefficients $A$, $B$, $C$, $D$, $E$ are given in
equations~(\ref{eq:defa}-\ref{eq:defe}).  
Setting the determinant of
this matrix equal to zero gives the dispersion relation for linear,
collisionless gyrokinetics [\eqref{eq:disprel}]
\begin{equation}
\left(\frac{\alpha_i A}{\overline{\omega}^2} -A B + B^2 \right)
\left(  \frac{2A}{\beta_i}- AD +  C^2 \right)
=\left( AE +  BC \right)^2.
\label{eq:disprel_app}
\end{equation}
The left-hand side of \eqref{eq:disprel_app} contains two factors, the first
corresponding to the Alfv\'en-wave branch and the second to the slow-wave 
branch; the right-hand side represents a finite-Larmor-radius
coupling between the Alfv\'en and slow modes that occurs 
as $k_\perp\rho_i$ approaches unity.

%==============================================================================
\section{Analytical Limits of the Dispersion Relation} 
\label{app:limits}

The linear, collisionless dispersion relation \eqref{eq:disprel}
harbors the plasma dispersion functions $Z(\xi_s)$ and the integrals
of the Bessel functions over the perpendicular velocity
$\Gamma_{0}(\alpha_s)$ and $\Gamma_{1}(\alpha_s)$. These functions can be 
expanded for large and small arguments, allowing an
analytical form of the dispersion relation to be derived in these
limits. The arguments in which the expansions will be made are 
\begin{equation} 
\xi_i = \frac{\omega}{|k_\parallel|v_{th_i}} = \frac{\overline{\omega}}{\sqrt{\beta_i}}, \quad
\xi_e = \frac{\omega}{|k_\parallel|v_{th_e}} =
\xi_i\left(\frac{m_e}{m_i}\right)^{1/2} \left(\frac{T_{0i}}{T_{0e}}\right)^{1/2},\quad 
\alpha_i = \frac{(k_\perp \rho_i)^2}{2},\quad
\alpha_e = \frac{(k_\perp \rho_e)^2}{2} = \alpha_i\frac{m_e}{m_i} \frac{T_{0e}}{T_{0i}}.
\end{equation}
Thus, the natural
subsidiary expansion parameters in the dispersion relation are
$\alpha_i$, $\beta_i$, $m_e/m_i$ and $T_{0i}/T_{0e}$. The limit of
long perpendicular wavelength, $\alpha_i \ll 1$, discussed in
\secref{sec:lowk}, illuminates the physical meaning of each of the
factors in the dispersion relation through a connection to the
MHD Alfv\'en and slow modes. 
In the short-wavelength limit, $\alpha_i\gg1$, discussed in \secref{sec:highk}, 
kinetic Alfv\'en waves replace the MHD modes.
In this Appendix, we examine the limits of high and low 
$\beta_i$, while keeping $\alpha_i$ finite. This allows us to 
connect the large- and small-wavelength asymptotics.

%==============================================================================
\subsection{High Beta Limit, $\beta_i \gg 1$}
\label{app:weak}

For $\beta_i\gg1$, we use the small-argument 
expansion of the plasma dispersion functions, $Z(\xi_s)\simeq i\sqrt{\pi}$, 
because $\xi_i=\overline{\omega}/\sqrt{\beta_i}\ll1$ 
and $\xi_e=(m_e/m_i)^{1/2} (T_{0i}/T_{0e})^{1/2}\xi_i\ll1$. 
To ensure the latter to be true, we need
$T_{0i}/T_{0e} \ll (m_i/m_e)\beta_i$, which is not at all 
very restrictive. 
We can also take $\alpha_e\ll1$ because $m_e/m_i\ll1$. 
The coefficients of the gyrokinetic 
dispersion relation become
\begin{eqnarray}
A &\simeq& 1 + \frac{T_{0i}}{T_{0e}} + i\sqrt{\pi}\xi_i
\left[\Gamma_0(\alpha_i) + \left(\frac{T_{0i}}{T_{0e}}\right)^{3/2}\left(\frac{m_e}{m_i}\right)^{1/2}\right],\\
B &\simeq& 1 - \Gamma_0(\alpha_i),\\
C &\simeq& i\sqrt{\pi}\xi_i
\left[\Gamma_1(\alpha_i) - \left(\frac{T_{0i}}{T_{0e}}\frac{m_e}{m_i}\right)^{1/2}\right],\\
D &\simeq& 2i\sqrt{\pi}\xi_i
\left[\Gamma_1(\alpha_i) + \left(\frac{T_{0e}}{T_{0i}}\frac{m_e}{m_i}\right)^{1/2}\right] 
\equiv 2i\sqrt{\pi}\xi_i G(\alpha_i),\\
E &\simeq& \Gamma_1(\alpha_i) - 1,
\end{eqnarray}
where we have dropped all terms of order $1$ and higher in $m_e/m_i$. 
The auxiliary function $G(\alpha_i)$ 
introduced in the expression for $D$ will be useful below. 
We will see that there are two interesting 
limits: $k_\perp \rho_i \sim \Order(\beta_i^{-1/4})$, $\overline{\omega}\sim \Order(1)$ 
and $k_\perp \rho_i \sim \Order(1)$, $\overline{\omega}\sim \Order(\beta_i^{-1/2})$ 
(the ordering of $\overline{\omega}$ is assumed \emph{a priori} and 
verified by the result, in the usual fashion).

\subsubsection{The limit $k_\perp \rho_i \sim 1/\beta_i^{1/4}$}

In this ordering, $\alpha_i\sim\xi_i\sim\Order(\beta_i^{-1/2})$. 
Expanding $\Gamma_0(\alpha_i)\sim 1-\alpha_i$ and $\Gamma_1(\alpha_i)\sim1-(3/2)\alpha_i$, 
we find that $A\sim\Order(1)$ and $B,C,D,E\sim\Order(\beta_i^{-1/2})$.
The dispersion relation becomes
\begin{equation}
-\left(\frac{\alpha_i}{\overline{\omega}^2} - B\right) D = E^2,
\end{equation}
where $B\simeq\alpha_i$, $E\simeq-(3/2)\alpha_i$ and 
$D\simeq2i\overline{\omega}\sqrt{\pi/\beta_i}$. 
This is a quadratic equation for $\overline{\omega}$. Its solution is 
\begin{equation}
\overline{\omega} = - i \frac{9}{16}\sqrt{\frac{\beta_i}{\pi}}\,\alpha_i \pm
\sqrt{1-\left(\frac{9}{16}\sqrt{\frac{\beta_i}{\pi}}\,\alpha_i\right)^2},
\label{eq:highbeta_sol1}
\end{equation}
which agrees with the \emph{a priori} ordering $\overline{\omega}\sim\Order(1)$.
 
In the limit $k_\perp \rho_i \ll \beta_i^{-1/4}$, we recover, as expected, the
Alfv\'en wave with weak damping [see \eqsref{eq:disprel_alf}{eq:damp_alf}]: 
\begin{equation}
\label{eq:damped_aw}
\overline{\omega} = \pm 1 - i\frac{9}{16}\frac{k_\perp^2 \rho_i^2 }{2}\sqrt{\frac{\beta_i}{\pi}}.
\end{equation}

For $\alpha_i > (16/9)\sqrt{\pi/\beta_i}$, the frequency is purely imaginary. 
In the intermediate asymptotic limit $\beta_i^{-1/4} \ll k_\perp \rho_i \ll 1$, 
we have 
\begin{eqnarray}
\label{eq:overlap_weak}
\overline{\omega} &=& 
- i\frac{8}{9}\left(\frac{k_\perp^2 \rho_i^2 }{2}\right)^{-1}\sqrt{\frac{\pi}{\beta_i}},\quad
{\rm weakly~damped},\\
\label{eq:overlap_strong}
\overline{\omega} &=& 
- i\frac{9}{8}\frac{k_\perp^2 \rho_i^2 }{2}\sqrt{\frac{\beta_i}{\pi}},\quad
\qquad\ \ {\rm strongly~damped}.
\end{eqnarray}

\subsubsection{The limit $k_\perp \rho_i \sim 1$}

In this ordering, $\alpha_i\sim\Order(1)$, $\xi_i\sim\Order(\beta_i^{-1})$. 
Then $A,B,E\sim\Order(1)$, $C,D\sim\Order(\beta_i^{-1})$. The dispersion 
relation now is 
\begin{equation}
\frac{\alpha_i}{\overline{\omega}^2}\left(\frac{2}{\beta_i}-D\right) = E^2.
\end{equation}
Since $D\simeq2i\overline{\omega}\sqrt{\pi/\beta_i} G(\alpha_i)$, 
this is again a quadratic equation for $\overline{\omega}$. 
Its solution is 
\begin{equation}
\overline{\omega} = - i \sqrt{\frac{\pi}{\beta_i}}\frac{\alpha_i G(\alpha_i)}{[\Gamma_1(\alpha_i)-1]^2} \pm
\sqrt{\frac{2}{\beta_i}\frac{\alpha_i}{[\Gamma_1(\alpha_i)-1]^2} - 
\left(\sqrt{\frac{\pi}{\beta_i}}\frac{\alpha_i G(\alpha_i)}{[\Gamma_1(\alpha_i)-1]^2}\right)^2},
\label{eq:highbeta_sol2}
\end{equation}
which agrees with the \emph{a priori} ordering $\overline{\omega}\sim\Order(\beta_i^{-1/2})$.

In the limit $k_\perp \rho_i \ll 1$,
the two solutions are
\begin{eqnarray}
\label{eq:sw_ap}
\overline{\omega} &=& -\frac{i}{\sqrt{\pi\beta_i}},\\ 
\overline{\omega} &=& - i\frac{8}{9}\left(\frac{k_\perp^2 \rho_i^2 }{2}\right)^{-1}\sqrt{\frac{\pi}{\beta_i}}.
\end{eqnarray}
The first solution is the damped slow mode [\eqref{eq:sw}]
the second solution matches the weakly damped Alfv\'en mode 
in the intermediate limit [\eqref{eq:overlap_weak}]. 

In the limit $k_\perp \rho_i \gg 1$, 
$\Gamma_1(\alpha_i)\to0$, $G(\alpha_i)\to(T_{0e}/T_{0i})^{1/2}(m_e/m_i)^{1/2}$, 
and \eqref{eq:highbeta_sol2} reproduces the $\beta_i \gg 1$ limit of
kinetic Alfv\'en waves [see \eqsref{eq:kaw}{eq:kaw_damp}]:
\begin{equation}
\overline{\omega} = \pm \frac{k_\perp\rho_i}{\sqrt{\beta_i}} - i\frac{k_\perp^2\rho_i^2}{2} 
\sqrt{\frac{\pi}{\beta_i}}\left(\frac{T_{0e}}{T_{0i}}\frac{m_e}{m_i}\right)^{1/2}.
\label{eq:kaw_ap}
\end{equation}

\subsubsection{Summary}

Thus, at $k_\perp\rho_i\sim\beta^{-1/4}$, 
the low-frequency weakly damped Alfv\'en waves [\eqref{eq:damped_aw}] 
are converted 
into two aperiodic modes, one weakly, one strongly damped 
[\eqsref{eq:overlap_weak}{eq:overlap_strong}]. 
At $k_\perp\rho_i\sim1$, 
the weakly damped Alfv\'en mode and the weakly damped slow mode 
[\eqref{eq:sw_ap}] are converted into two weakly damped kinetic Alfv\'en waves 
[\eqref{eq:kaw_ap}]. These are finally damped at 
$k_\perp\rho_e\sim1$. Note that the slow mode we are referring to 
is the weakest-damped of many modes into which the 
two MHD slow waves and the entropy mode are converted 
when their parallel wavelengths exceed the ion mean free path.

\begin{figure}[t]
\plottwo{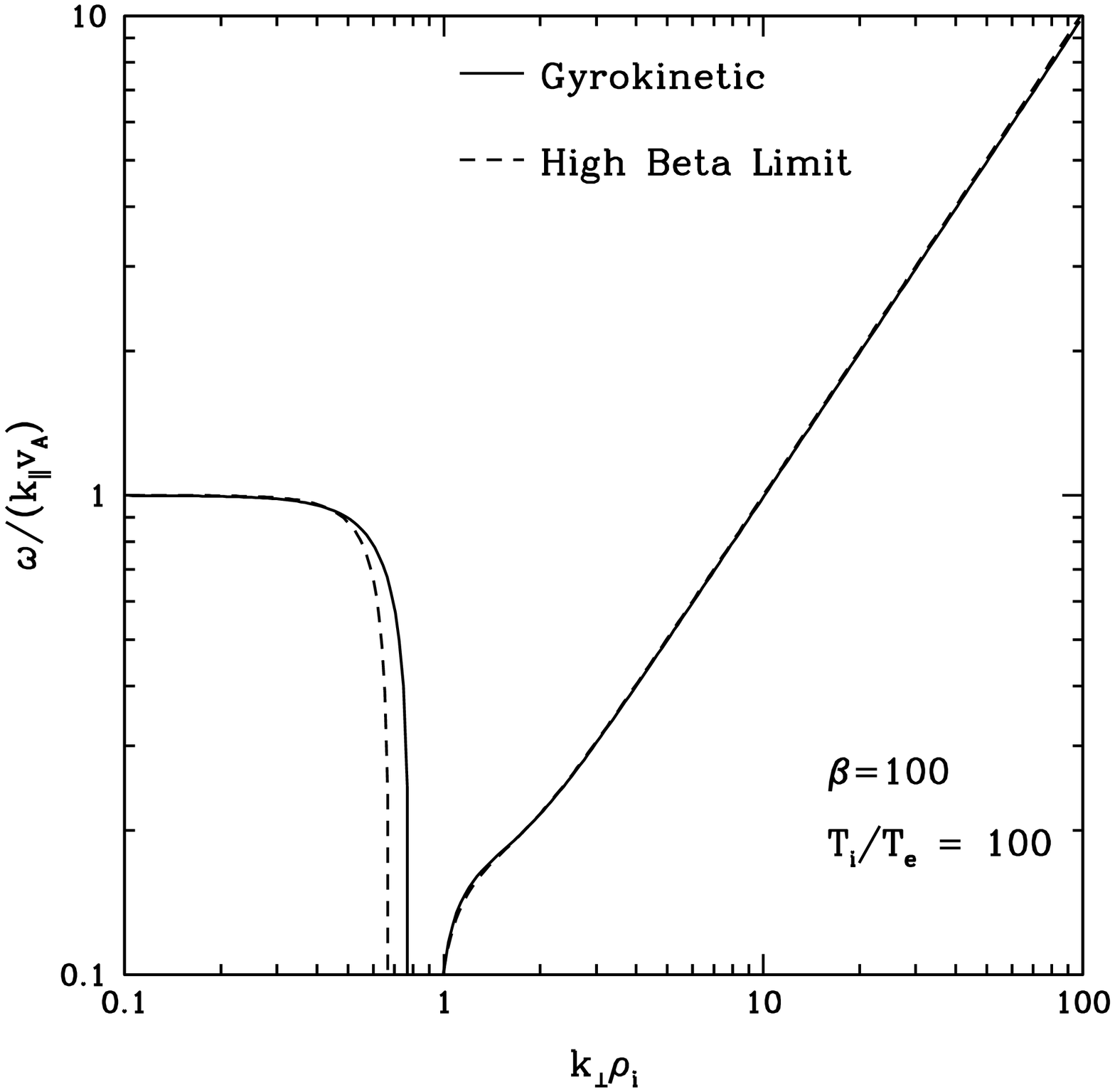}{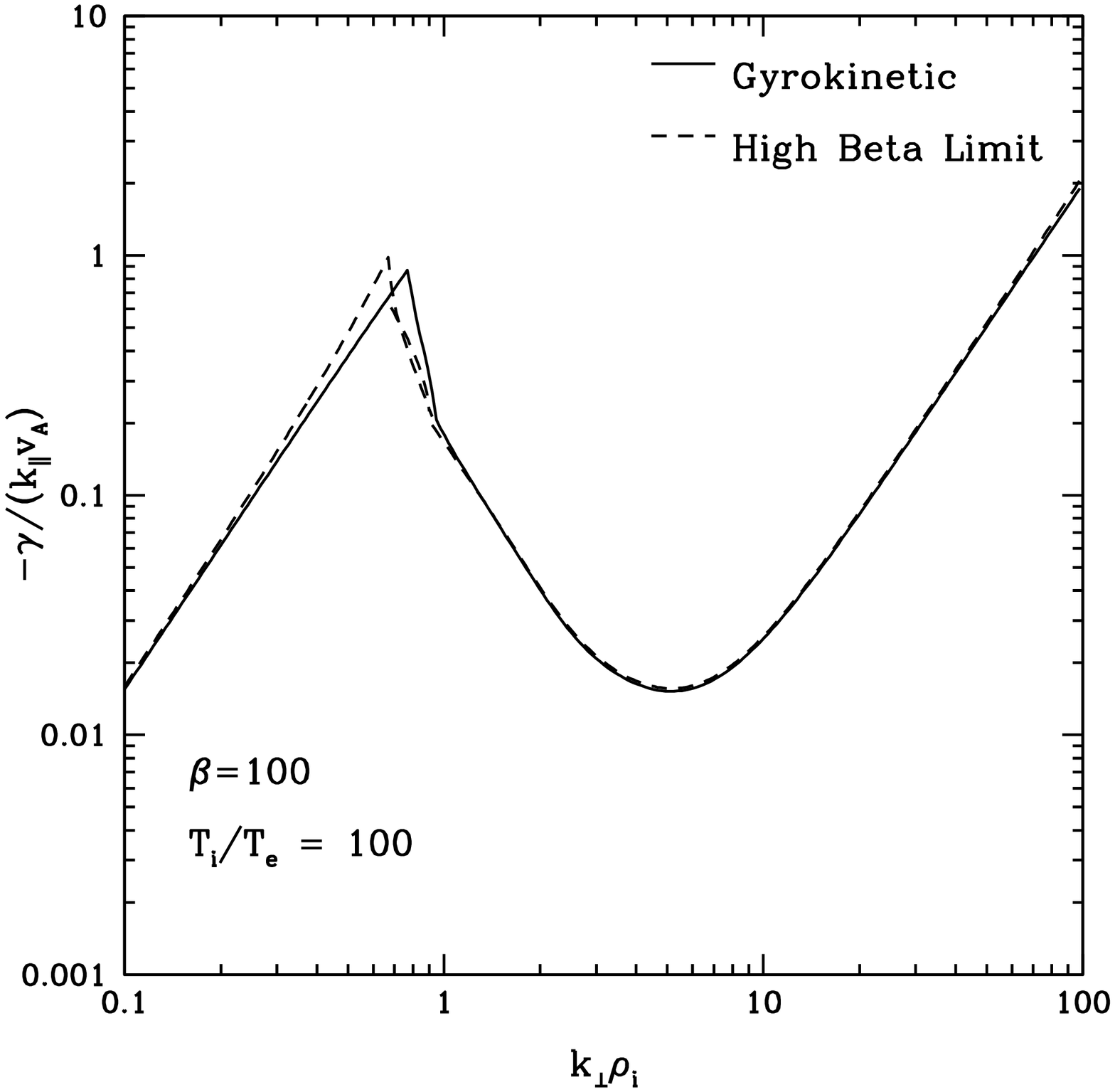}
\caption{The real frequency (left) and damping rate 
(right) of the weakly damped Alfv\'en mode 
derived by numerical solution of the linear, collisionless gyrokinetic
dispersion relation (solid line) compared to the high-beta analytical
limit (dashed line). The approximate solution consists of 
two solutions, valid for $k_\perp \rho_i \sim \Order(\beta_i^{-1/4})$
and $k_\perp \rho_i \sim \Order(1)$ [the $+$ branch of \eqref{eq:highbeta_sol1} 
and the $-$ branch of \eqref{eq:highbeta_sol2}, respectively]. 
In the intermediate limit  
$\beta_i^{-1/4} \ll k_\perp \rho_i \ll 1$, both solutions are plotted
to confirm that they match.}
\label{fig:weakb}
\end{figure}

The real frequency and damping rate for $\beta_i=100$ and
$T_{0i}/T_{0e}= 100$ for the branch corresponding to 
the weakly damped Alfv\'en mode are plotted in
\figref{fig:weakb}.

%==============================================================================
\begin{figure}[t]
\plottwo{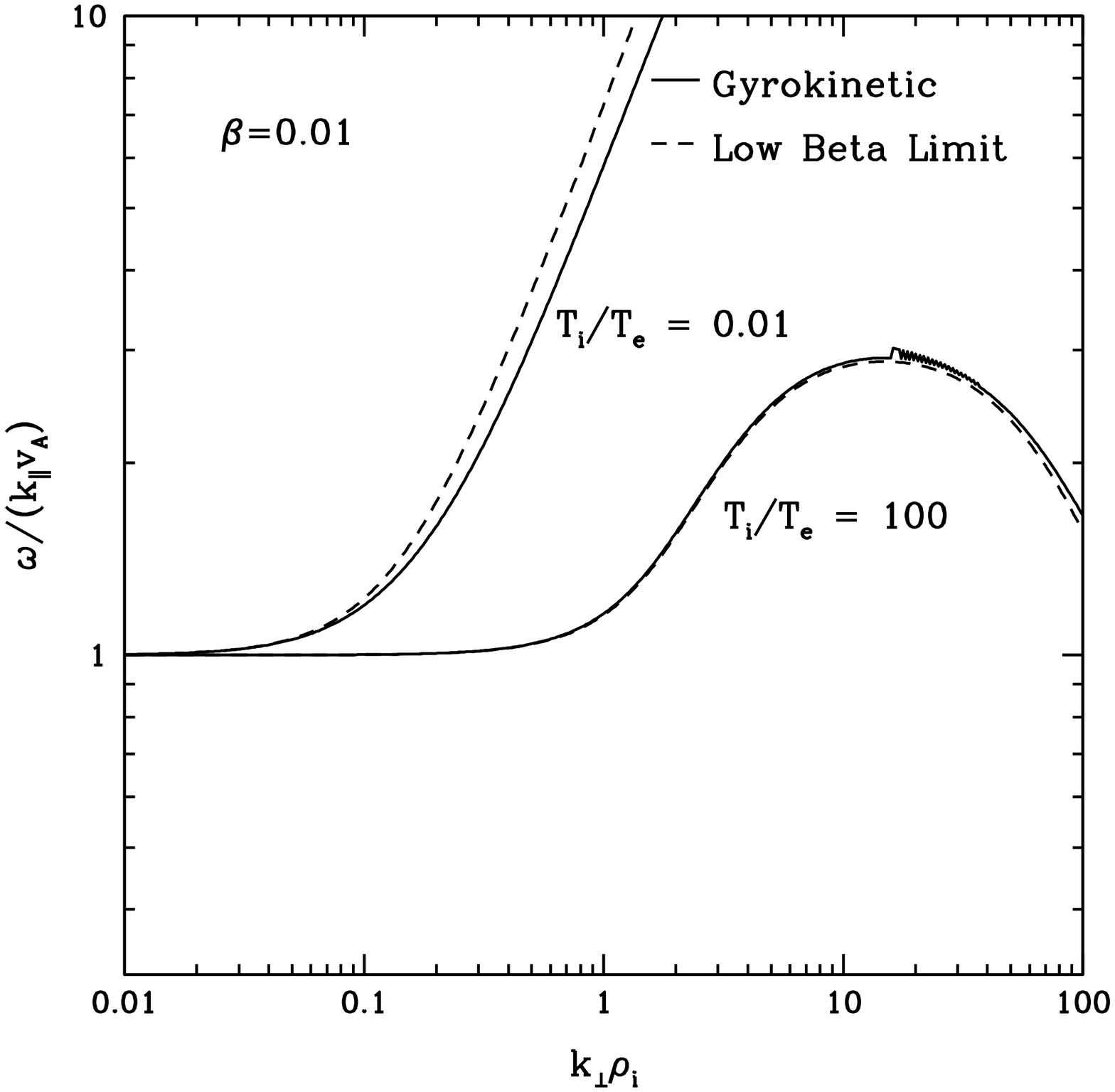}{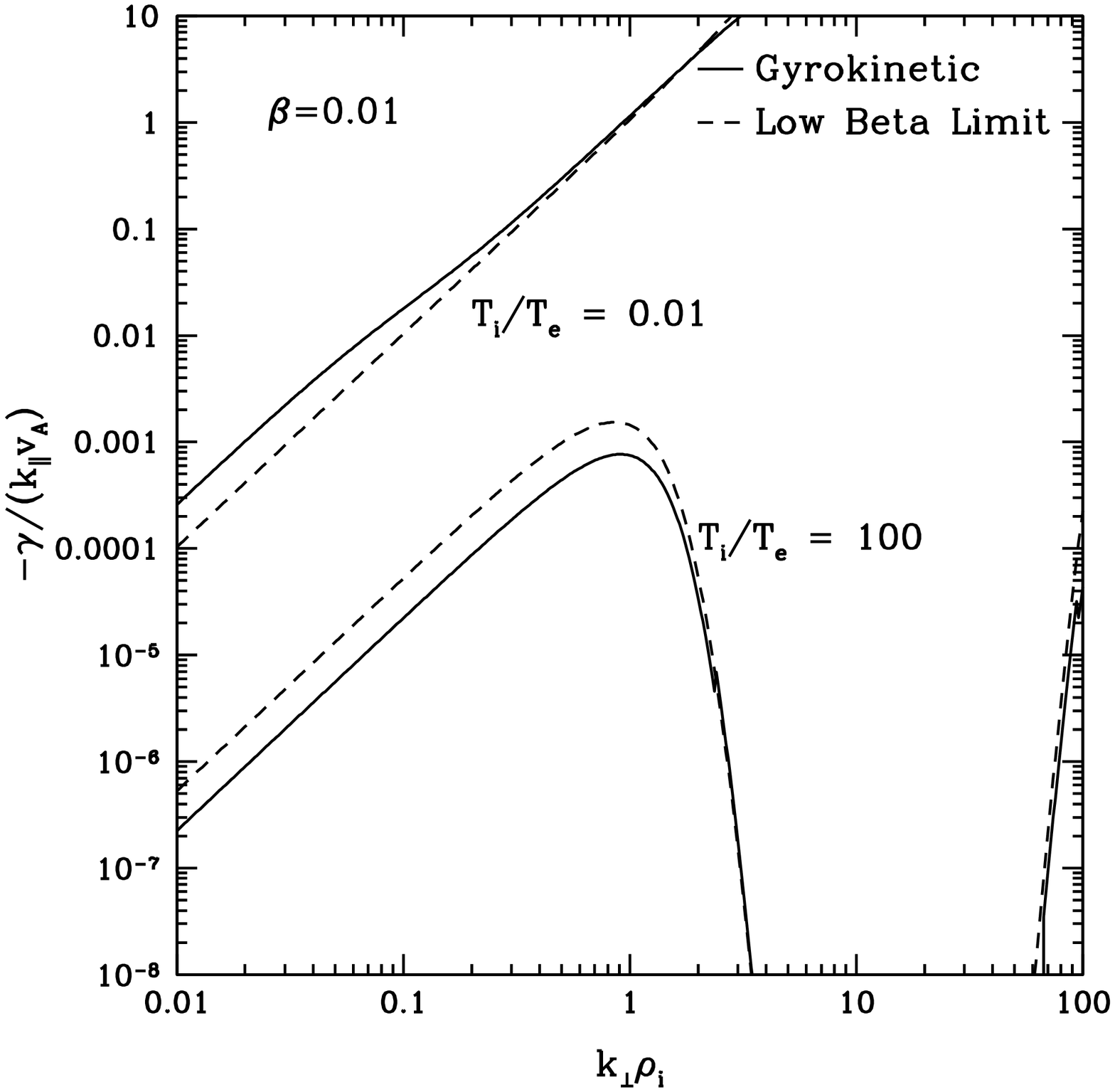}
\caption{The real frequency (left) and damping rate (right) of the 
Alfv\'en mode derived by numerical solution of the linear, collisionless gyrokinetic
dispersion relation (solid line) compared to the low-beta analytical
limits $T_{0i}/T_{0e}\gg1$ [\eqsref{eq:lowbeta2_omega}{eq:lowbeta2_gamma}]
and $T_{0i}/T_{0e}\ll(m_i/m_e)\beta_i$ [\eqsref{eq:aw_lowbeta}{eq:damp_lowbeta}].} 
\label{fig:strongb}
\end{figure}

\subsection{Low Beta Limit, $\beta_i \ll 1$}
\label{app:strong}

For $\beta_i\ll1$, it turns out that $A,B,C,D,E\sim\Order(1)$, so the 
gyrokinetic dispersion relation reduces to
\begin{equation}
\left(\frac{\alpha_i A}{\overline{\omega}^2} - AB + B^2\right)\frac{2A}{\beta_i} = 0.
\end{equation}
We will focus on the first factor, which corresponds to Alfv\'en modes 
[the long-wavelength limit of the second factor gives the 
ion acoustic wave, see \S\ref{sec:lowk}]. 
We order $\overline{\omega}\sim\Order(1)$ and consider 
two interesting limits: $(m_e/m_i)(T_{0i}/T_{0e})\ll\beta_i\ll1$ 
and $\beta_i\sim m_e/m_i\ll1$, $T_{0i}/T_{0e}\gg1$.
The solutions in these two limits are presented below. 
These solutions are plotted together with 
numerical solutions of the full dispersion relation in
\figref{fig:strongb}.

\subsubsection{The limit $(m_e/m_i)(T_{0i}/T_{0e})\ll\beta_i\ll1$.}

In this limit, $\xi_i=\overline{\omega}/\sqrt{\beta_i}\gg1$ 
and $\xi_e=(m_e/m_i)^{1/2}(T_{0i}/T_{0e})^{1/2}\xi_i\ll1$ (slow ions, fast electrons). 
Expanding the ion and electron plasma dispersion functions in large 
and small argument, respectively, we get 
\begin{equation}
A \simeq 1 - \Gamma_0(\alpha_i) + \frac{T_{0i}}{T_{0e}} 
+ i\overline{\omega}\sqrt{\frac{\pi}{\beta_i}}\left[\Gamma_0(\alpha_i) \exp\left(-\frac{\overline{\omega}^2}{\beta_i}\right) + 
\left(\frac{T_{0i}}{T_{0e}}\right)^{3/2}\left(\frac{m_e}{m_i}\right)^{1/2}\Gamma_0(\alpha_e)\right].
\end{equation}
The resulting dispersion relation is 
\begin{equation}
\frac{T_{0i}}{T_{0e}} B\Gamma_0(\alpha_e)\overline{\omega}^2 
- \alpha_i\left[1-\Gamma_0(\alpha_i) + \frac{T_{0i}}{T_{0e}} \right] 
= - i(B\overline{\omega}^2 - \alpha_i){\rm Im}(A),
\end{equation}
where the right-hand side is small, so we can solve perturbatively 
for real frequency and small damping. The result is 
\begin{eqnarray}
\label{eq:aw_lowbeta}
\overline{\omega} &=& \pm\sqrt{\frac{\alpha_i[1-\Gamma_0(\alpha_i) + T_{0i}/T_{0e}]}{(T_{0i}/T_{0e})B\Gamma_0(\alpha_e)}},\\ 
\overline{\gamma} &=& 
- \frac{\alpha_i}{2[(T_{0i}/T_{0e})\Gamma_0(\alpha_e)]^2}
\sqrt{\frac{\pi}{\beta_i}}\left[\Gamma_0(\alpha_i) \exp\left(-\frac{\overline{\omega}^2}{\beta_i}\right) + 
\left(\frac{T_{0i}}{T_{0e}}\right)^{3/2}\left(\frac{m_e}{m_i}\right)^{1/2}\Gamma_0(\alpha_e)\right],
\label{eq:damp_lowbeta}
\end{eqnarray}
where $\overline{\gamma}={\rm Im}(\omega)/|k_\parallel|v_A$. 
Note that $\overline{\omega}\sim\Order(1)$, as promised at the outset. 
In the limit $\alpha_i,\alpha_e\ll1$, 
\eqref{eq:aw_lowbeta} reduces to the Alfv\'en wave solution, 
$\overline{\omega}=\pm1$.

\subsubsection{The limit $\beta_i\sim m_e/m_i\ll1$, $T_{0i}/T_{0e}\gg1$.}

In this limit, $\xi_i\sim (m_i/m_e)^{1/2}\gg1$ and, therefore, 
$\xi_e\sim (T_{0i}/T_{0e})^{1/2}\gg1$ (both ions and electrons are slow). 
Expanding all plasma dispersion functions in large arguments, we get 
\begin{equation}
A \simeq B - \frac{\Gamma_0(\alpha_e)}{2\overline{\omega}^2}\frac{m_i}{m_e}\beta_i 
+ i\overline{\omega}\sqrt{\frac{\pi}{\beta_i}}\left[\Gamma_0(\alpha_i) \exp\left(-\frac{\overline{\omega}^2}{\beta_i}\right) +
\left(\frac{T_{0i}}{T_{0e}}\right)^{3/2}\left(\frac{m_e}{m_i}\right)^{1/2}\Gamma_0(\alpha_e)
\exp\left(-\frac{T_{0i}}{T_{0e}}\frac{m_e}{m_i}\frac{\overline{\omega}^2}{\beta_i}\right)\right].
\end{equation}
The resulting dispersion relation is
\begin{equation}
\frac{\alpha_i\Gamma_0(\alpha_e)}{2\overline{\omega}^2}\frac{m_i}{m_e}\beta_i 
- B \left[\alpha_i + \frac{\Gamma_0(\alpha_e)}{2}\frac{m_i}{m_e}\beta_i\right] 
= - i(B\overline{\omega}^2 - \alpha_i){\rm Im}(A),
\end{equation}
where again the right-hand side is small and solving perturbatively gives
\begin{eqnarray}
\label{eq:lowbeta2_omega}
\overline{\omega} &=& \pm\sqrt{\frac{\alpha_i\Gamma_0(\alpha_e)(m_i/m_e)\beta_i}{\left[2\alpha_i 
+ \Gamma_0(\alpha_e)(m_i/m_e)\beta_i\right]B}},\\
\overline{\gamma} &=& -\frac{2\alpha_i^3\Gamma_0(\alpha_e)(m_i/m_e)\beta_i}{\left[2\alpha_i 
+ \Gamma_0(\alpha_e)(m_i/m_e)\beta_i\right]^3 B^2}
\sqrt{\frac{\pi}{\beta_i}}\left[\Gamma_0(\alpha_i) \exp\left(-\frac{\overline{\omega}^2}{\beta_i}\right) +
\left(\frac{T_{0i}}{T_{0e}}\right)^{3/2}\left(\frac{m_e}{m_i}\right)^{1/2}\Gamma_0(\alpha_e)
\exp\left(-\frac{T_{0i}}{T_{0e}}\frac{m_e}{m_i}\frac{\overline{\omega}^2}{\beta_i}\right)\right].
\label{eq:lowbeta2_gamma}
\end{eqnarray}

%==============================================================================

%==============================================================================

\end{document}